\documentclass[useAMS,usenatbib]{mn2e}

\usepackage{graphicx}
\usepackage{subfigure}
\usepackage{paralist}
\usepackage{amssymb}
\usepackage{bm}
\usepackage{bbding}
\usepackage[fleqn]{amsmath}
\usepackage{multirow}
\usepackage[colorlinks,
            linkcolor=blue, 
            anchorcolor=blue, 
            citecolor=blue, 
            ]{hyperref}
\title[Dynamical structures]{Dynamical structures of misaligned circumbinary planets under hierarchical three-body systems}
\author[H. Lei, Y. Gong]{
Hanlun Lei$^{1,2}$\thanks{leihl@nju.edu.cn}, Yan-Xiang Gong$^{3}$\thanks{yxgong@tsu.edu.cn}\\
$^{1}$School of Astronomy and Space Science, Nanjing University, Nanjing 210023, China\\
$^{2}$Key Laboratory of Modern Astronomy and Astrophysics in Ministry of Education, Nanjing University, Nanjing 210023, China\\
$^{3}$College of Physics and Electronic Engineering, Taishan University, Taian 271000, China
}

\begin{document}

\date{Accepted. Received; in original form}

\pagerange{\pageref{firstpage}--\pageref{lastpage}} \pubyear{2024}

\maketitle
\label{firstpage}

\begin{abstract}
All circumbinary planets (CBPs) currently detected are located in almost co-planar configurations with respect to the binary orbit, due to the fact that CBPs with higher misalignment are more difficult to detect. However, observations of polar circumbinary gas and debris disks in recent years and long-term orbital stability of inclined planets indicate that it is possible to form misaligned CBPs around eccentricity binaries (even polar CBPs). In this work we focus on the dynamical structures of CBPs in a wide range of parameters in order to provide a guidance for the space where the binary can host planets for a long enough time. To this end, the dynamical model is approximated as a hierarchical three-body problem, and the secular approximation is formulated up to the hexadecapolar order in semimajor axis ratio. Dynamical maps show that there are complex structures in the parameter space. A web of secular resonances is produced in the entire parameter space and it can well explain those numerical structures arising in dynamical maps. Based on perturbative treatments, an adiabatic invariant is introduced and thus dynamical structures can be explored by analysing phase portraits. It is found that (a) the quadrupole-order resonance (nodal resonance) is responsible for the distribution of V-shape region, and high-order and secondary resonances dominate those structures inside or outside V-shape region, and (b) the secondary 1:1 resonance is the culprit causing symmetry breaking of dynamical structures inside polar region.
\end{abstract}


\section{Introduction}
\label{Sect1}

Stellar binaries or multiples are popular in the Universe and, for massive stars, the fraction can reach up to 70\% \citep{tokovinin1997multiplicity}. There are about a hundred of planets detected in multiple-star systems \citep{martin2018populations}. In particular, planets in binary systems may be located in either circumstellar (S-type) or circumbinary (P-type) configurations. A dominant fraction of planets in binary systems are found in the S-type configurations. There is about more than a
dozen of circumbinary planets (CBPs) in the P-type configurations observed by the Kepler and
TESS space telescopes \citep{georgakarakos2024empirical,kostov2020toi,kostov2016kepler,kostov2021tic}. The lack of P-type planets may be ascribed to observational biases and dynamical processes \citep{cuello2019planet}. All the CBPs currently detected are located in almost co-planar configurations with respect to the binary orbit, which is a consequence of observational bias instead of a representative of the underlying population \citep{czekala2019degree}. 

In recent years, misaligned gas or debris disks have been observed around stellar binaries, such as 99 Herculis \citep{kennedy201299}, IRS 43 \citep{brinch2016misaligned}, GG Tau \citep{cazzoletti2017testing}, HD 142527 \citep{verhoeff2011complex}, HD 98800 B \citep{kennedy201299} and AC Her \citep{martin2023ac}, suggesting that misaligned planet formation is possible. As for the stability of P-type planets, polar orbits are stable over a wide range of binary parameters \citep{chen2020polar,cuello2019planet}. Based on $N$-body simulations, it is shown that terrestrial planet formation around an eccentric binary is more likely in a polar configuration than in a coplanar configuration \citep{childs2021formation}. \citet{martin2017polar} showed that misaligned discs should be more common around long-period and eccentric stellar binaries. Therefore, misaligned (even polar) orbits provide appropriate dynamical environments for planet formation. The reason why there are no highly misaligned or polar CBPs detected so far is that misaligned CBPs are usually difficult to detect due to long orbital periods as well as complex spectra of stellar binaries \citep{martin2014planets}. 

Misaligned terrestrial CBPs could form in the presence of a misaligned circumbinary gas disk, which can help to nodally align planetary orbits and maintain planetary inclination during the formation process \citep{childs2022misalignment}. As the first observational evidence of a polar circumbinary planet, the post-AGB star binary system (AC Her) is found to hold a dust disk where the inner edge is much farther out than the tidal truncation radius \citep{martin2023ac}. It strongly supports that there is likely a highly misaligned circumbinary planet which is located inside the disk around AC Her. A promising technique to detect misaligned CBPs is to measure binary eclipse timing variations (ETVs) \citep{sybilski2010detecting}. Applications of ETVs show that binary systems KIC 5095269, KIC 07177553 and KIC 7821010 hold inclined planetary mass objects around them \citep{borkovits2016comprehensive, getley2017evidence}. With the improvement of ETVs, it is believed that additional highly misaligned (even polar) CBPs around eccentric binaries or multiples with longer periods may be detected by taking advantage of such a technique based on future observations \citep{zhang2019distinguishing}.

In this work, we study dynamical structures of misaligned CBPs around eccentric binaries in a large parameter space (producing a global picture for the dynamics of CBPs) and then provide analytical understanding about those structures. The purpose of this work is to understand the dynamics of misaligned orbits under different circumbinary systems in order to make predictions about parameter space where the inner eccentric binary can host planets as long as possible. We approximate the dynamical model for describing CBPs as a hierarchical three-body problem, where the planet is treated as an outer test particle \citep{li2014analytical,de2019inverse,vinson2018secular,naoz2017eccentric}. The secular dynamics under hierarchical three-body configurations has been investigated extensively in the literature, where the eccentric von Zeipel-Lidov-Kozai (ZLK) mechanism is effective in understanding different astrophysical phenomena \citep{ford2000secular, katz2011long, lithwick2011eccentric, naoz2013secular, li2014analytical, naoz2016eccentric}. Those short-range forces, such as general relativity effect, could significantly influence the eccentric ZLK effects \citep{naoz2013secular,naoz2017eccentric,liu2015suppression, lepp2022radial}. For the P-type configuration, it is known that the inverse ZLK resonance appears at the hexadecapolar-level Hamiltonian \citep{de2019inverse,vinson2018secular}. Therefore, in the present work the secular approximation is formulated up to the hexadecapolar order in semimajor axis ratio in order to cover the effects caused by inverse ZLK resonance. Dynamical maps in the entire parameter space are produced by taking the second-derivative-based indicator $||\Delta D||$ as index and complex structures can be found in dynamical maps. Analytical study based on perturbative treatments is performed to understand numerical structures. 

The remaining part of this work is organised as follows. In Sect. \ref{Sect2}, the Hamiltonian model at the hexadecapolar-level approximation is formulated and validated. Varieties of stability conditions are briefly discussed in Sect. \ref{Sect3} and dynamical maps are produced in Sect. \ref{Sect4}. In Sect. \ref{Sect5}, dynamics under the quadrupole-order Hamiltonian is investigated and applications of resonance curves to numerical structures are performed in Sect. \ref{Sect6}. In Sect. \ref{Sect7}, dynamics of secondary resonances is studied by means of perturbative treatments. Conclusions are summarised in Sect. \ref{Sect8}.

\section{Hamiltonian model}
\label{Sect2}

\begin{figure}
\centering
\includegraphics[width=0.8\columnwidth]{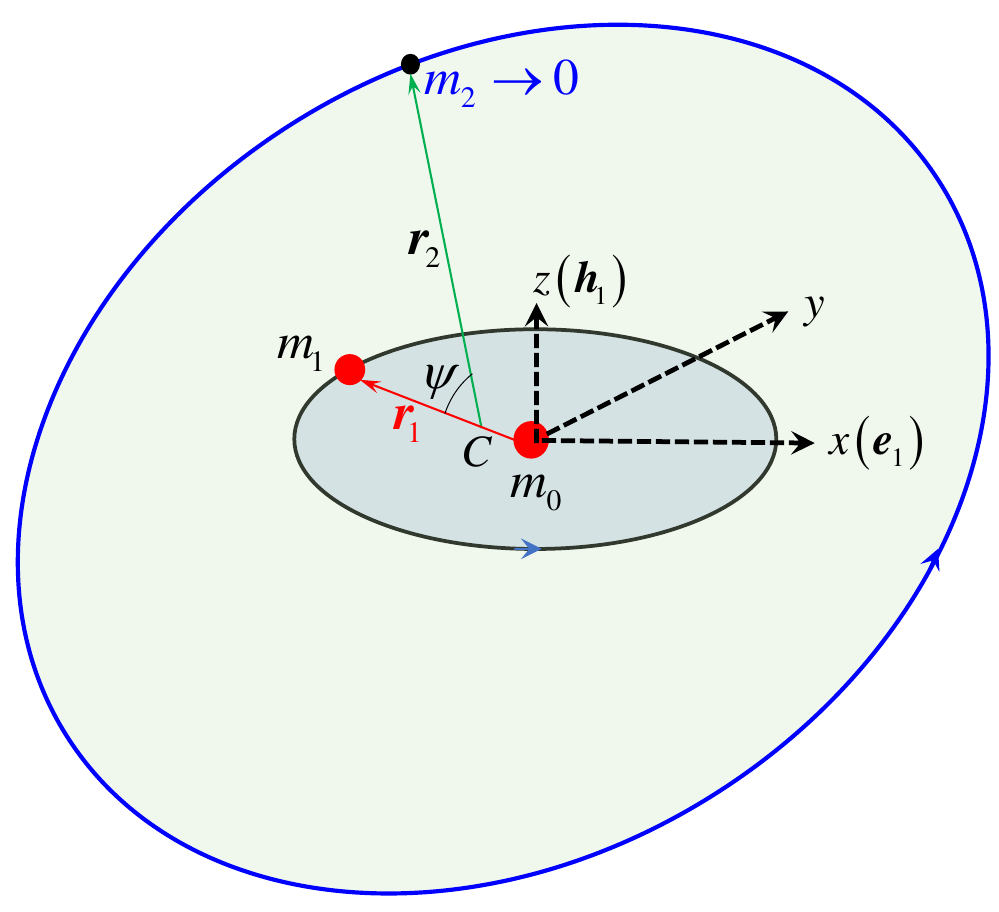}
\caption{Schematic diagram for the geometric configuration of circumbinary planetary system and definitions of variables adopted in this work. The circumbinary planet is approximated as a test particle (i.e., $m_2 \to 0$), whose long-term dynamics is governed by the gravitational attraction from the inner binary with mass $m_0$ and $m_1$. The barycentre of the inner binary is denoted by C. The invariant plane-based right-handed coordinate system $O$-$xyz$ is defined with the origin at the location of $m_0$, the $x$-axis aligned with the pericentre vector of $m_1$ and the $z$-axis along with its angular momentum vector. The position vector from $m_0$ towards $m_1$ is denoted by $\bm r_1$, the position vector from C towards $m_2$ is denoted by $\bm r_2$ (Jacobi coordinates), and their relative angle is $\psi$.}
\label{Fig1}
\end{figure}

In this work, we consider secular dynamics of CBPs. To this end, planets are approximated as exterior (massless) test particles in a hierarchical three-body configuration, where the inner binary is composed of a central object with mass $m_0$ and an inner perturber with mass $m_1$. The gravitational influence upon the binary stars coming from the test particle is negligible (i.e., the so-called test-particle approximation), thus the inner binary moves around their barycentre on a fixed Keplerian orbit. Please refer to Fig. \ref{Fig1} for the schematic diagram of dynamical model. Similar configuration has been considered in previous works for studying secular dynamics of exterior test particles \citep{de2019inverse,naoz2017eccentric,vinson2018secular}. 

As usual, we adopt the Jacobi coordinates to describe the motion of the perturber and test particles, i.e., the inner perturber $m_1$ moves around the central object $m_0$ and the test particle moves around the barycentre of the inner binary. To characterise their orbits, the following orbital elements are used: the semimajor axis $a_{1,2}$, eccentricity $e_{1,2}$, inclination $i_{1,2}$, longitude of ascending node $\Omega_{1,2}$, argument of pericentre $\omega_{1,2}$ and mean anomaly $M_{1,2}$. Unless otherwise stated, the subscript `1' stands for the variables of the perturber and the subscript `2' represents the variables of test particles. 

Considering the fact that the perturber's orbit corresponds to the invariant plane of system, we choose an inertial $m_0$-centred coordinate reference frame with the invariant plane as the $x$-$y$ plane and the angular momentum vector of system as the $z$-axis, leading to $i_1 = 0$. Without loss of generality, the $x$-axis is directed toward the perturber's pericentre, which means $\varpi_1 (=\omega_1 + \Omega_1)= 0$. Please refer to Fig. \ref{Fig1} for the definition of coordinate system.

Under the test-particle assumption, the perturber's orbit remains stationary. Thus, the orbital elements $a_1$ and $e_1$ are constant. Under hierarchical configurations, the semimajor axis ratio $\alpha = a_1/a_2$ is a small parameter, thus the disturbing function from the inner perturber can be expanded as a power series in $\alpha$ (i.e., Legendre expansion of disturbing function). As a result, the (unit mass) Hamiltonian, governing the evolution of test particles, can be written as \citep{harrington1968dynamical,harrington1969stellar}
\begin{equation}\label{Eq1}
\begin{aligned}
{\cal H} =&  - \frac{{{\cal G}\left( {{m_0} + {m_1}} \right)}}{{2{a_2}}} - \frac{{{\cal G}{m_0}{m_1}}}{{{a_2}(m_0 + m_1)}}\sum\limits_{n = 2}^ N  {{{\left( {\frac{{{a_1}}}{{{a_2}}}} \right)}^n}} \\
& \times \frac{{m_0^{n - 1} - {{\left( { - {m_1}} \right)}^{n - 1}}}}{{{{\left( {{m_0} + {m_1}} \right)}^{n-1}}}}{\left( {\frac{{{r_1}}}{{{a_1}}}} \right)^n}{\left( {\frac{{{a_2}}}{{{r_2}}}} \right)^{n + 1}}{P_n}\left( {\cos \psi } \right)
\end{aligned}
\end{equation}
where $\cal G$ is the Universal gravitational constant, $r_1$ is the distance between $m_0$ and $m_1$, $r_2$ is the distance between the test particle and the barycentre of the inner binary and the angle $\psi$ stands for the relative angle between the radius vectors of the test particle and the perturber (see Fig. \ref{Fig1} for details). ${P_n}\left( {\cos \psi } \right)$ is the Legendre polynomial of order $n$. In particular, the Hamiltonian represented by equation ($\ref{Eq1}$) is referred to as quadrupole-level approximation when the truncated order is taken as $N=2$, as octupole-level approximation when $N=3$, as hexadecapolar-level approximation when $N=4$, and so on.

In long-term evolutions, the short-term influences associated with the mean motion of the test particle and the inner perturber can be filtered out by means of double-averaging technique \citep{naoz2016eccentric}. Such a similar phase-averaged or secular approximation can be found in \citet{lei2020dynamical} and \citet{lei2022secular} for studying secular dynamics of navigation satellites. Performing double averages for the Hamiltonian and truncating it at the hexadecapolar order, we can get the resulting double-averaged Hamiltonian as
\begin{equation}\label{Eq2}
{\cal H} =  - {{\cal C}_0}\left( {{F_2} + {\varepsilon _1}{F_3} + {\varepsilon _2}{F_4}} \right),
\end{equation}
where the constant terms have been removed. In the averaged Hamiltonian, the mean anomalies $M_{1,2}$ become cyclic variables and thus the semimajor axes $a_{1,2}$ are constants during the long-term evolution. In equation (\ref{Eq2}), the coefficient ${\cal C}_0$ is given by
\begin{equation*}
{{\cal C}_0} = \frac{{{\cal G}{m_0}{m_1}}}{{16\left( {{m_0} + {m_1}} \right)}}\frac{{a_1^2}}{{a_2^3}},
\end{equation*}
which remains unchanged during the long-term evolution, the quadrupole-order term is
\begin{equation*}
\begin{aligned}
{F_2} =& \frac{1}{{{{\left( {1 - e_2^2} \right)}^{3/2}}}}\left[ {\left( {2 + 3e_1^2} \right)\left( {3{{\theta}^2} - 1} \right)} \right.\\
&\left. { + 15e_1^2\left( {1 - {{\theta}^2}} \right)\cos \left(2{\Omega _2}\right)} \right]
\end{aligned}
\end{equation*}
the octupole-order term is
\begin{equation*}
\begin{aligned}
{F_3} =&  - \frac{{15}}{{32}}\frac{{{e_1}{e_2}}}{{{{\left( {1 - e_2^2} \right)}^{5/2}}}}\left\{ {\left( {4 + 3e_1^2} \right) \times } \right.\\
&\left[ {\left( { - 1 + 11\theta  + 5{\theta ^2} - 15{\theta ^3}} \right)\cos \left( {{\Omega _2} - {\omega _2}} \right)} \right.\\
&\left. { + \left( { - 1 - 11\theta  + 5{\theta ^2} + 15{\theta ^3}} \right)\cos \left( {{\Omega _2} + {\omega _2}} \right)} \right]\\
&+ 35e_1^2\left[ {\left( {1 - \theta  - {\theta ^2} + {\theta ^3}} \right)\cos \left( {3{\Omega _2} - {\omega _2}} \right)} \right.\\
&\left. {\left. { + \left( {1 + \theta  - {\theta ^2} - {\theta ^3}} \right)\cos \left( {3{\Omega _2} + {\omega _2}} \right)} \right]} \right\}
\end{aligned}
\end{equation*}
and the hexadecapolar-order term is
\begin{equation*}
\begin{aligned}
{F_4} =& \frac{{15}}{{64}}\frac{{2 + 3e_2^2}}{{{{\left( {1 - e_2^2} \right)}^{7/2}}}}\left\{ {{d_1} - 11 - 18{\theta ^2} + 21{\theta ^4}} \right.\\
& - 4{d_2}\left( {1 - 8{\theta ^2} + 7{\theta ^4}} \right)\cos \left( {2{\Omega _2}} \right)\\
&\left. { + 7{d_3}\left( {1 - 2{\theta ^2} + {\theta ^4}} \right)\cos \left( {4{\Omega _2}} \right)} \right\}\\
& + \frac{{15}}{{64}}\frac{{e_2^2}}{{{{\left( {1 - e_2^2} \right)}^{7/2}}}}\left\{ { - 6\left( {1 - 8{\theta ^2} + 7{\theta ^4}} \right)\cos \left( {2{\omega _2}} \right)} \right.\\
& + 4{d_2}\left( {1 + 5\theta  - 6{\theta ^2} - 7{\theta ^3} + 7{\theta ^4}} \right)\cos \left( {2{\Omega _2} - 2{\omega _2}} \right)\\
& + 4{d_2}\left( {1 - 5\theta  - 6{\theta ^2} + 7{\theta ^3} + 7{\theta ^4}} \right)\cos \left( {2{\Omega _2} + 2{\omega _2}} \right)\\
& + 7{d_3}\left( {1 - 2\theta  + 2{\theta ^3} - {\theta ^4}} \right)\cos \left( {4{\Omega _2} - 2{\omega _2}} \right)\\
&\left. { + 7{d_3}\left( {1 + 2\theta  - 2{\theta ^3} - {\theta ^4}} \right)\cos \left( {4{\Omega _2} + 2{\omega _2}} \right)} \right\}
\end{aligned}
\end{equation*}
where $\theta = \cos{i_2}$ and the constant coefficients $d_{1,2,3}$ in the expression of $F_4$ are given by
\begin{equation*}
\begin{aligned}
{d_1} =& \frac{{64}}{5}{\rm{ + }}24e_1^2 + 9e_1^4,\\
{d_2} =& \frac{{21}}{8}\left( {2e_1^2 + e_1^4} \right),\\
{d_3} =& \frac{{63}}{8}e_1^4.
\end{aligned}
\end{equation*}
The factors $\varepsilon _1$ and $\varepsilon _2$ stand for the significance of the octupole-order and hexadecapolar-order contributions compared to the quadrupole-order term, given by
\begin{equation*}
{\varepsilon _1} = \frac{{{m_0} - {m_1}}}{{{m_0} + {m_1}}}\frac{{{a_1}}}{{{a_2}}},\;{\varepsilon _2} = \frac{{m_0^3 + m_1^3}}{{{{\left( {{m_0} + {m_1}} \right)}^3}}}\frac{{a_1^2}}{{a_2^2}}
\end{equation*}
It is observed that $\varepsilon _1$ is proportional to the semimajor axis ratio $\alpha = a_1/a_2$ and $\varepsilon _2$ is proportional to $\alpha^2$. In particular, when the inner binary is of equal mass (i.e., $m_0 = m_1$), the octupole-order term disappears because of ${\varepsilon _1} = 0$, and the hexadecapolar-order term takes its minimum.

One remark is made here. The Hamiltonian at the hexadecapolar-level approximation given here is confirmed to be in agreement with the ones presented in \citet{vinson2018secular} and \citet{de2019inverse} and the octupole-order version is in agreement with the one given in \citet{naoz2017eccentric}, but with different expressions.

As the coefficient ${\cal C}_0$ is a constant during the long-term evolution, it is convenient to normalise the Hamiltonian by ${\cal C}_0$ as follows\footnote{Normalisation of Hamiltonian is equivalent to re-scaling the unit of time.}:
\begin{equation}\label{Eq3}
\begin{aligned}
{\cal H} & = - {F_2} - {\varepsilon _1}{F_3} - {\varepsilon _2}{F_4}\\
& = {{\cal H}_0}\left( {{e_2},{i_2},{\Omega _2}} \right) + \varepsilon {{\cal H}_1}\left( {{e_2},{i_2},{\Omega _2},{\omega _2}} \right)
\end{aligned}
\end{equation}
where ${\cal H}_0 = -F_2$ and $\varepsilon {\cal H}_1 = -{\varepsilon _1}{F_3}-{\varepsilon _2}{F_4}$. From the viewpoint of perturbation theory, the Hamiltonian ${\cal H}_0$ (quadrupole-level approximation) describes the unperturbed dynamics and the Hamiltonian ${\cal H}_1$ plays a role of perturbation which is dependent on ${\varepsilon _1}$ and ${\varepsilon _2}$.

To study the long-term evolution from the viewpoint of Hamiltonian, we introduce the normalised Delaunay variables:
\begin{equation*}
\begin{aligned}
{g_2} &= {\omega _2},\quad {G_2} = \sqrt {1 - e_2^2}, \\
{h_2} &= {\Omega _2},\quad {H_2} = {G_2}\cos {i_2}.
\end{aligned}
\end{equation*}
In terms of Delaunay variables, the normalised Hamiltonian is expressed as
\begin{equation}\label{Eq4}
{\cal H} = {{\cal H}_0}\left( {{G_2},{H_2},{h_2}} \right) + \varepsilon {{\cal H}_1}\left( {{G_2},{g_2},{H_2},{h_2}} \right).
\end{equation}
The Hamiltonian canonical relation leads to the equations of motion \citep{morbidelli2002modern}
\begin{equation}\label{Eq5}
\begin{aligned}
\frac{{{\rm d}g_2}}{{{\rm d}\tau }} &= \frac{{\partial {\cal H}}}{{\partial G_2}},\quad \begin{array}{*{20}{c}}
{}&{}
\end{array}\frac{{{\rm d}h_2}}{{{\rm d}\tau }} = \frac{{\partial {\cal H}}}{{\partial H_2}},\\
\frac{{{\rm d}G_2}}{{{\rm d}\tau }} &=  - \frac{{\partial {\cal H}}}{{\partial g_2}},\quad \begin{array}{*{20}{c}}
{}&{}
\end{array}\frac{{{\rm d}H_2}}{{{\rm d}\tau }} =  - \frac{{\partial {\cal H}}}{{\partial h_2}},
\end{aligned}
\end{equation}
and the associated Lagrange planetary equations in terms of orbit elements can be written as \citep{murray1999solar}
\begin{equation}\label{Eq6}
\begin{aligned}
\frac{{{\rm d}{e_2}}}{{{\rm d}\tau }} &= \frac{\eta }{{{e_2}}}\frac{{\partial {\cal H}}}{{\partial {\omega _2}}},\\
\frac{{{\rm d}{i_2}}}{{{\rm d}\tau }} &= \frac{{\csc {i_2}}}{\eta }\left( {\frac{{\partial {\cal H}}}{{\partial {\Omega _2}}} - \cos {i_2}\frac{{\partial {\cal H}}}{{\partial {\omega _2}}}} \right),\\
\frac{{{\rm d}{\Omega _2}}}{{{\rm d}\tau }} &=  - \frac{{\csc {i_2}}}{\eta }\frac{{\partial {\cal H}}}{{\partial {i_2}}},\\
\frac{{{\rm d}{\omega _2}}}{{{\rm d}\tau }} &= \frac{{\cot {i_2}}}{\eta }\frac{{\partial {\cal H}}}{{\partial {i_2}}} - \frac{\eta }{{{e_2}}}\frac{{\partial {\cal H}}}{{\partial {e_2}}}
\end{aligned}
\end{equation}
where $\eta = \sqrt{1-e_2^2}$. The relationship between the time $\tau$ under the normalised Hamiltonian and the time $t$ under the non-normalised Hamiltonian is given by
\begin{equation*}
\begin{aligned}
\tau  =& \frac{{{{\cal C}_0}}}{{\sqrt {{\cal G}\left( {{m_0} + {m_1}} \right){a_2}} }}t\\
=& \frac{{{\cal G}{m_0}{m_1}}}{{16\left( {{m_0} + {m_1}} \right)}}\frac{{a_1^2}}{{a_2^3}} \frac{{{1}}}{{\sqrt {{\cal G}\left( {{m_0} + {m_1}} \right){a_2}} }}t.
\end{aligned}
\end{equation*}

\begin{figure*}
\centering
\includegraphics[width=\columnwidth]{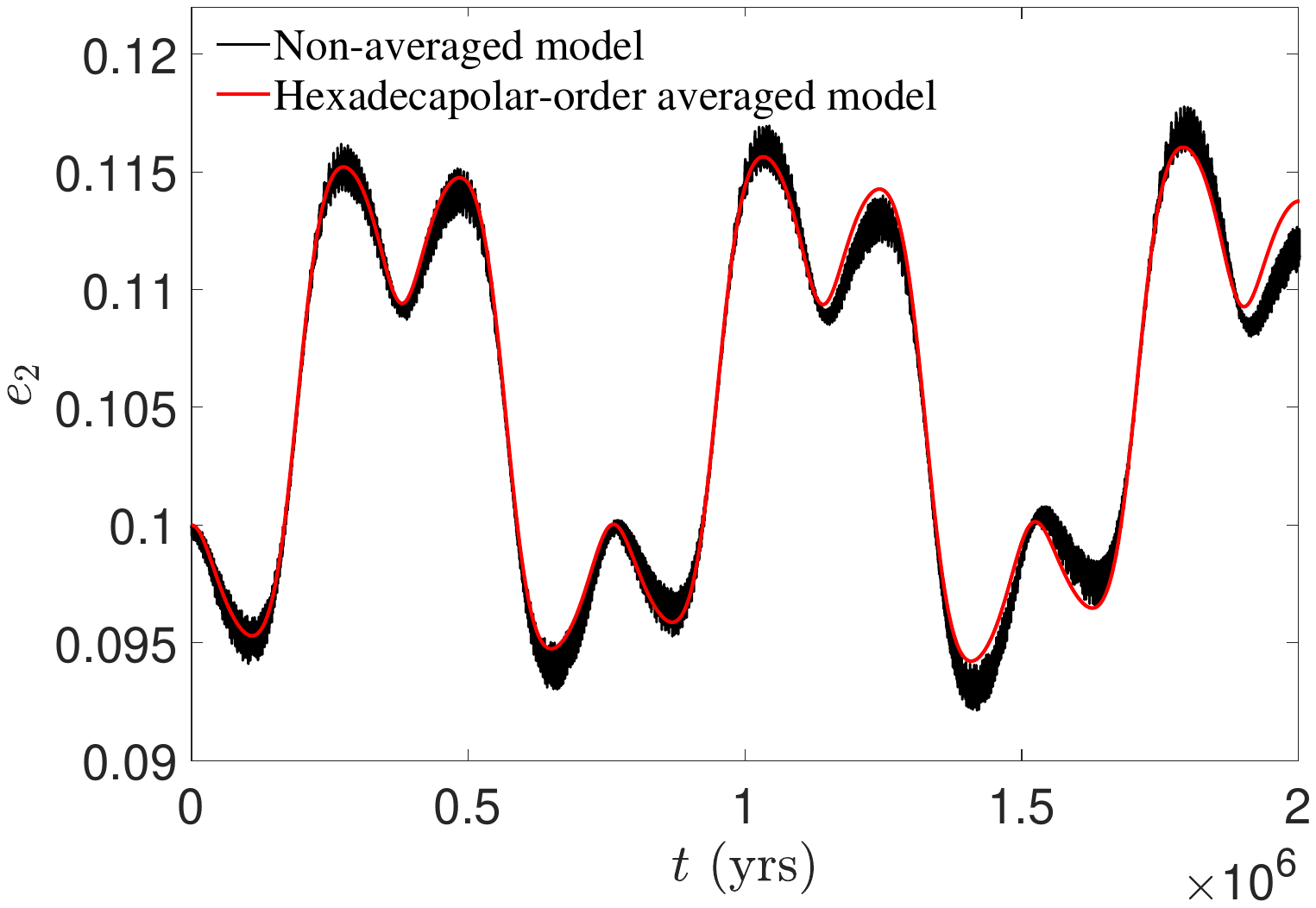}
\includegraphics[width=\columnwidth]{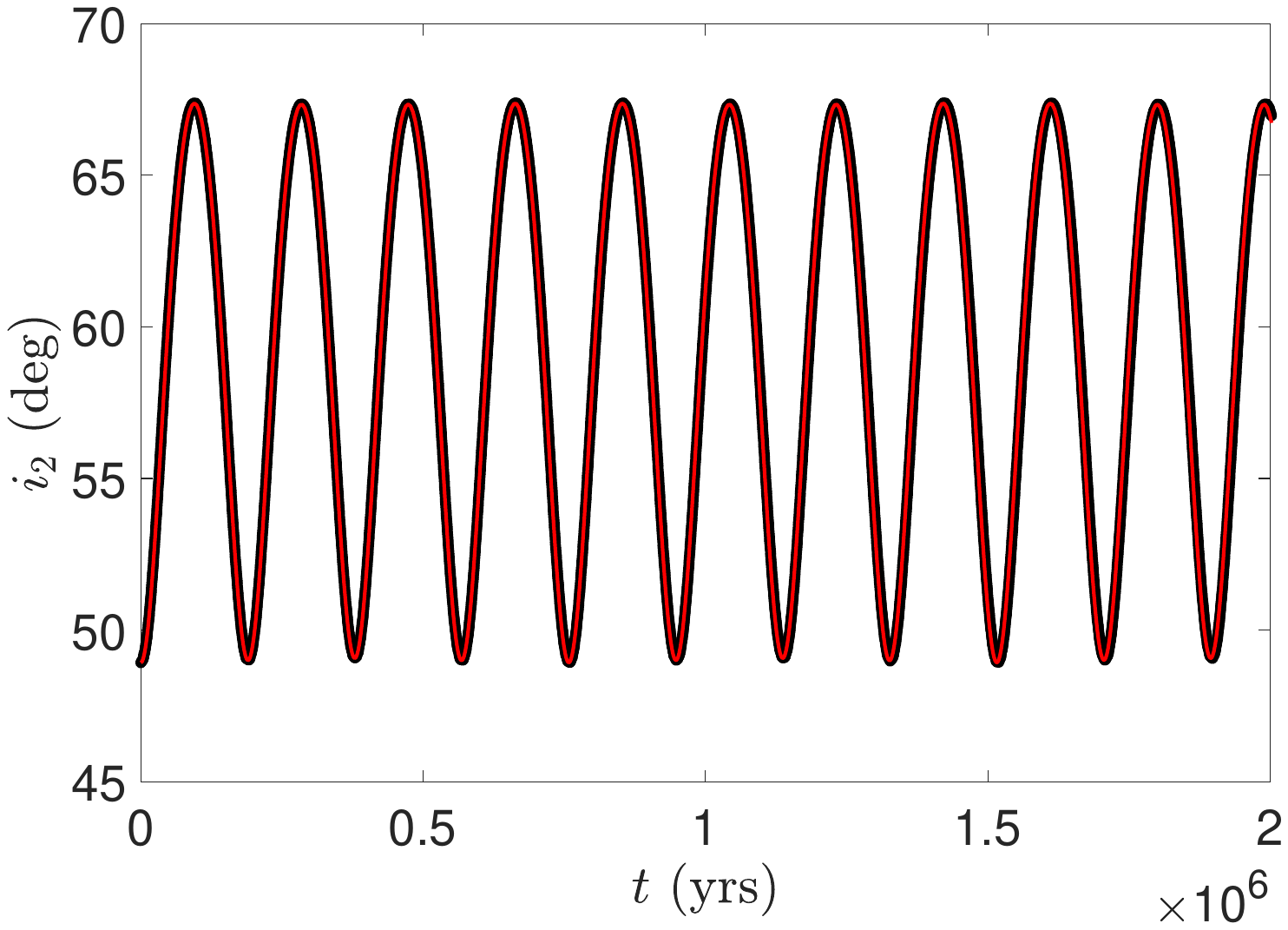}
\caption{Comparison between the N-body integration results (black lines) and the secular approximation at the hexadecapolar order (red lines) over $2 \times 10^6$ $\rm yrs$. The left panel is for the evolution of eccentricity $e_2$ and the right panel is for the evolution of inclination $i_2$. In practical simulations, the system parameters are taken as $m_0 = 1.0 m_{\odot}$, $m_1 = 0.5 m_{\odot}$, $a_1 = 3 \rm au$ and $a_2 = 40 \rm au$; in this case it holds $C_0 = 2.9296875 \times 10^{-6}$, $\varepsilon_1 = 2.5 \times 10^{-2}$ and $\varepsilon_2 = 1.875 \times 10^{-3}$. Under the test-particle approximation, the eccentricity of the inner body remains constant and it is taken as $e_1 = 0.3$ in practice. The initial conditions of both trajectories are set as $e_{2,0}=0.1$, $i_{2,0}=48.925^{\circ}$ and $\Omega_{2,0}=\omega_{2,0}=\pi/2$ (the slight difference between the `mean' and `osculating' elements at the initial instant is ignored). Good agreement can be observed between the trajectories propagated under the non-averaged model and the averaged model truncated at the hexadecapolar order.}
\label{Fig2}
\end{figure*}

In Fig. \ref{Fig2}, time histories of eccentricity and inclination are compared for the trajectories numerically propagated under the non-averaged dynamical model (direct N-body integration) and under the double-averaged model up to the hexadecapolar order (secular approximation). Please refer to the caption for detailed setting of system parameters as well as initial condition. It is observed that (a) the evolution under the non-averaged model exhibits short-term oscillations, while the short-term oscillations are filtered out under the double-averaged Hamiltonian model, (b) variation of inclination presents good periodicity as the evolution of inclination is dominated by the integrable quadrupole-order Hamiltonian, and (c) there are good agreements between the N-body integration results and secular approximations for the long-term evolution. As a result, secular approximation up to hexadecapolar order is effective to predict long-term evolution for CBPs.

Without otherwise stated, it is assumed that the central object holds the mass of $m_0 = 1.0\, m_{\odot}$ and the semimajor axes of the inner and outer orbits are taken as $a_1 = 3\, \rm au$ and $a_2 = 40\, \rm au$ in the entire work. It means that the semimajor axis ratio is fixed at $\alpha = 3/40$.

\section{Stability condition}
\label{Sect3}

It is of significance to perform stability analysis of multi-body systems in order to answer the question: in what regions of parameter space can CBPs persist for a long enough time? About this topic, \citet{dvorak1989stability} numerically derived stability limits of outer planetary orbits (P-types) in binary systems, then \citet{holman1999long} provided an empirical expression for the critical semimajor axis $a_{2,c}$ as a function of binary mass faction $\mu_b$ and eccentricity $e_1$. Extending to 3D space, \citet{doolin2011dynamics} performed a numerical investigation about the stability of test-particle CBPs in circumbinary phase space as a function of binary eccentricity $e_1$ and mass fraction $\mu_b$ and discovered that circumbinary orbits are surprisingly stable throughout the parameter space $(e_1,\mu_b)$. With a given pair of $(e_1,\mu_b)$, complex structures can be further observed in the $(a_2,i_2)$ space. \citet{chen2019orbital} and \citet{chen2020polar} extended the study of \citet{doolin2011dynamics} to the non-zero mass case and they systematically explored the dependence of stability of nearly circular bots upon binary eccentricity $e_1$, binary mass fraction $\mu_b$, planet mass $m_2$, planet semimajor axis $a_2$, and planet inclination $i_2$ based on a large number of numerical simulations. Similarly, complex structures are observed in the $(a_2,i_2)$ space. Based on $\sim$150 million full N-body simulations, \citet{quarles2018stability} developed numerical tools for quick, easy, and accurate determination of stability limit in the $(e_1,\mu_b)$ space and, in particular, they provided open-source python software to make use of their simulations. Recently, \citet{georgakarakos2024empirical} revisited the problem of dynamical stability of 3D and eccentric circumbinary planetary orbits by performing more than $3 \times 10^8$ numerical simulations and they provided empirical expressions in the form of multidimensional and parameterised fits for stability limits.

There are many works in studying long-term stability of planets moving in binary systems, and varieties of stability conditions can be found in previous literature under different assumptions. In this section, we briefly review them under the unified notation system adopted in this work (see Fig. \ref{Fig1} for the configuration) and then make a direct comparison with applications to CBPs. 

For co-planar and nearly circular planetary orbits, \citet{dvorak1989stability} provided a stability criterion where the semimajor axis of the lower critical orbit is a function of $a_1$ and $e_1$ in the following form:
\begin{equation}\label{Eq7}
{a_{2,c}} = {a_1}\left( {2.37 + 2.76{e_1} - 1.04e_1^2} \right)
\end{equation}
which is independent on the mass parameter of the inner binary $\mu_b=m_1/(m_0 + m_1)$, planetary eccentricity and inclination ($e_2$ and $i_2$). If $a_2$ is greater than the critical value $a_{2,c}$, the associated system is unstable. 

The influence of the mass parameter of binary ($\mu_b$) is considered in \citet{holman1999long}, who presented a widely used stability condition for P-type configurations as follows:
\begin{equation}\label{Eq8}
\begin{aligned}
{a_{2,c}} =& {a_1}\left( {1.6 + 5.1{e_1} - 2.22e_1^2 + 4.12{\mu _b} - 4.27{e_1}{\mu _b}} \right.\\
&\left. { - 5.09\mu _b^2 + 4.61e_1^2\mu _b^2} \right)
\end{aligned}
\end{equation}
which is updated by \citet{quarles2018stability} in the following form (see their Fit 1):
\begin{equation}\label{Eq9}
\begin{aligned}
{a_{2,c}} =& {a_1}\left( {1.48 + 3.92{e_1} - 1.41e_1^2 + 5.14{\mu _b} + 0.33{e_1}{\mu _b}} \right.\\
&\left. { - 7.95\mu _b^2 - 4.89e_1^2\mu _b^2} \right).
\end{aligned}
\end{equation}
Both expressions in \citet{holman1999long} and \citet{quarles2018stability} are independent on planetary eccentricity and inclination ($e_2$ and $i_2$), thus they are suitable for low-inclination and low-eccentricity planetary orbits under P-type configurations.

Considering the influence of planetary eccentricity $e_2$ upon stability condition, \citet{adelbert2023stability} provided a stability criterion as follows:
\begin{equation}\label{Eq10}
\begin{aligned}
{a_{2,c}} =& \frac{{{a_1}}}{{1 - {e_2}}}\left[ {1.36 + 5.79{e_1} - 5.87e_1^2 + 1.99{\mu _b}} \right.\\
&\left. { - 3.14\mu _b^2 + \left( {1.85 - 2.1e_1^2 + 3{e_1}{\mu _b}} \right){e_2}} \right]
\end{aligned}
\end{equation}
where the influence of planetary inclination $i_2$ is not considered. Thus, the stability condition given by equation (\ref{Eq10}) is suitable for low-inclination planetary orbits.

Furthermore, considering the influence of planetary eccentricity $e_2$ and inclination $i_2$, \citet{mardling2001tidal} presented a simple stability criterion (in the test-particle approximation with $m_2 \to 0$),
\begin{equation}\label{Eq11}
{a_{2,c}} = 2.8{a_1}\frac{{{{\left( {1 + {e_2}} \right)}^{2/5}}}}{{{{\left( {1 - {e_2}} \right)}^{6/5}}}}\left( {1 - 0.3\frac{{{i_2}}}{\pi }} \right)
\end{equation}
and recently \citet{georgakarakos2024empirical} provided a generalised stability criterion which is suitable for a large-range parameter space $(e_2 < 0.8)$ as follows:
\begin{equation}\label{Eq12}
\begin{aligned}
{\log _{10}}{a_{2,c}} =& {\log _{10}}{a_1} + 0.23612 - 0.29377\left( {{{\log }_{10}}{\mu _b}} \right) + 0.2271{i_2} \\
& + 1.06753{e_1}+ 0.62109{e_2} - 0.21512{\left( {{{\log }_{10}}{\mu _b}} \right)^2} \\
& - 0.06648i_2^2 - 1.52936e_1^2- 0.4748e_2^2 - 0.31329{e_1}{i_2} \\
& - 0.00869{e_2}{i_2} + 0.11846{e_1}i_2^2- 0.03932{\left( {{{\log }_{10}}{\mu _b}} \right)^3}\\
&- 0.00933i_2^3 + 0.87506e_1^3 + 1.25895e_2^3.
\end{aligned}
\end{equation}
It is noted that planetary inclination $i_2$ in both Eqs. (\ref{Eq11}) and (\ref{Eq12}) is given in unit of radian. The stability condition of \citet{mardling2001tidal} does not consider the influence of mass parameter $\mu_b$ and binary's eccentricity $e_1$. From the stability condition given by \citet{georgakarakos2024empirical}, it is observed that the critical semimajor axis $a_{2,c}$ is a function of $a_1$, $\mu_b$, $e_1$, $e_2$ and $i_2$, which is available for varieties of configurations. Please refer to \citet{georgakarakos2024empirical} for the stability criterion about arbitrary planetary eccentricities and its applications to those CBPs observed so far.

\begin{figure*}
\centering
\includegraphics[width=\columnwidth]{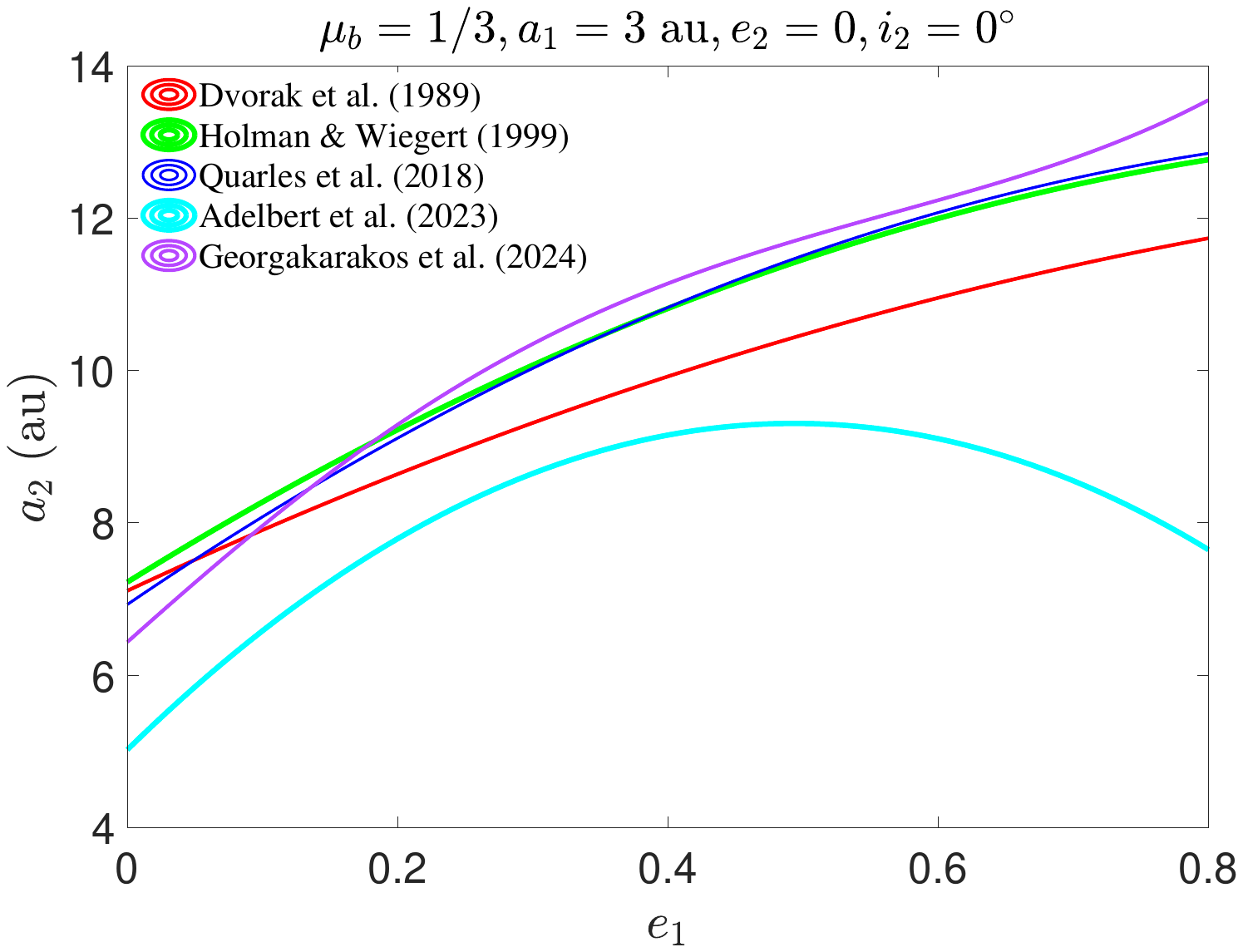}
\includegraphics[width=\columnwidth]{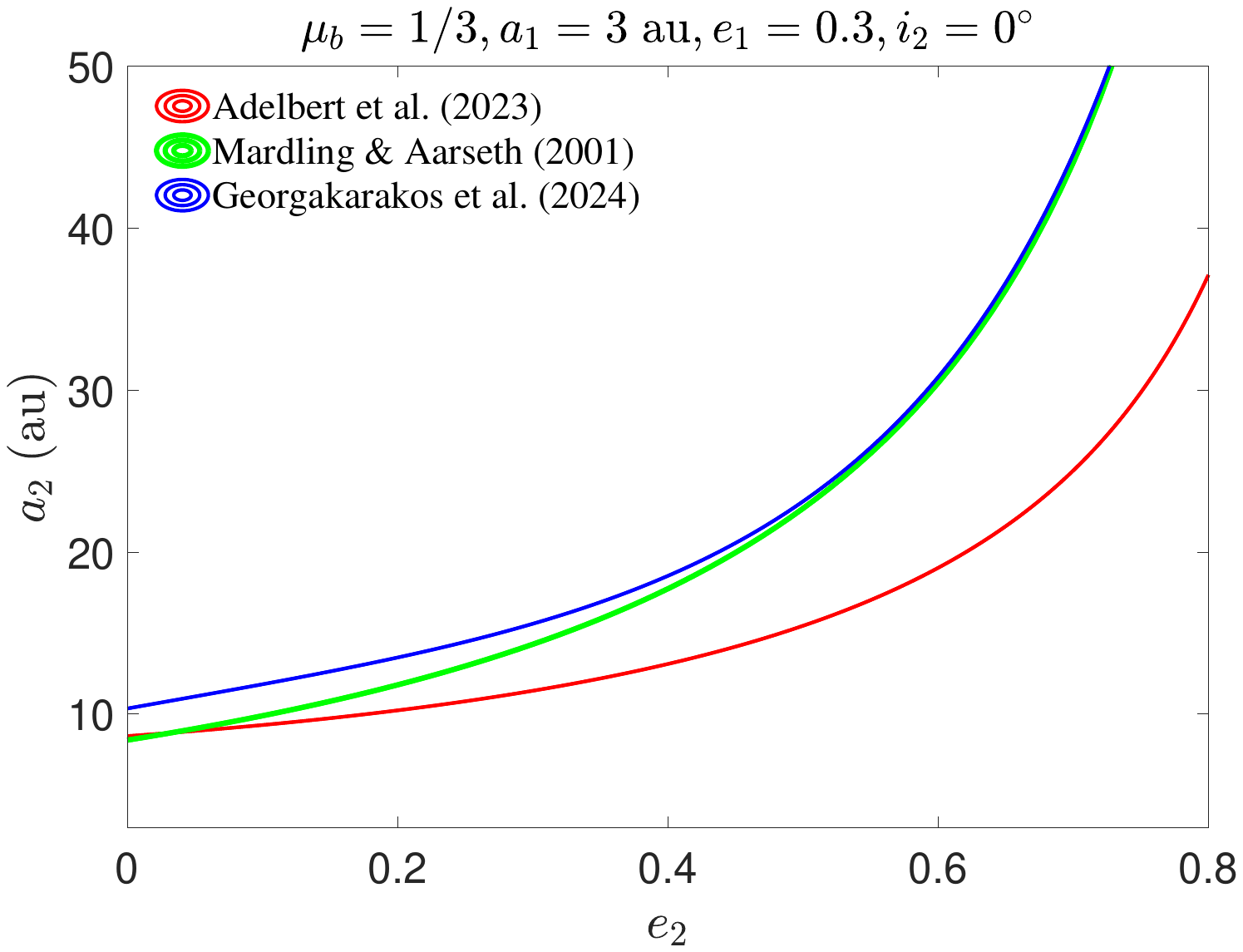}\\
\includegraphics[width=\columnwidth]{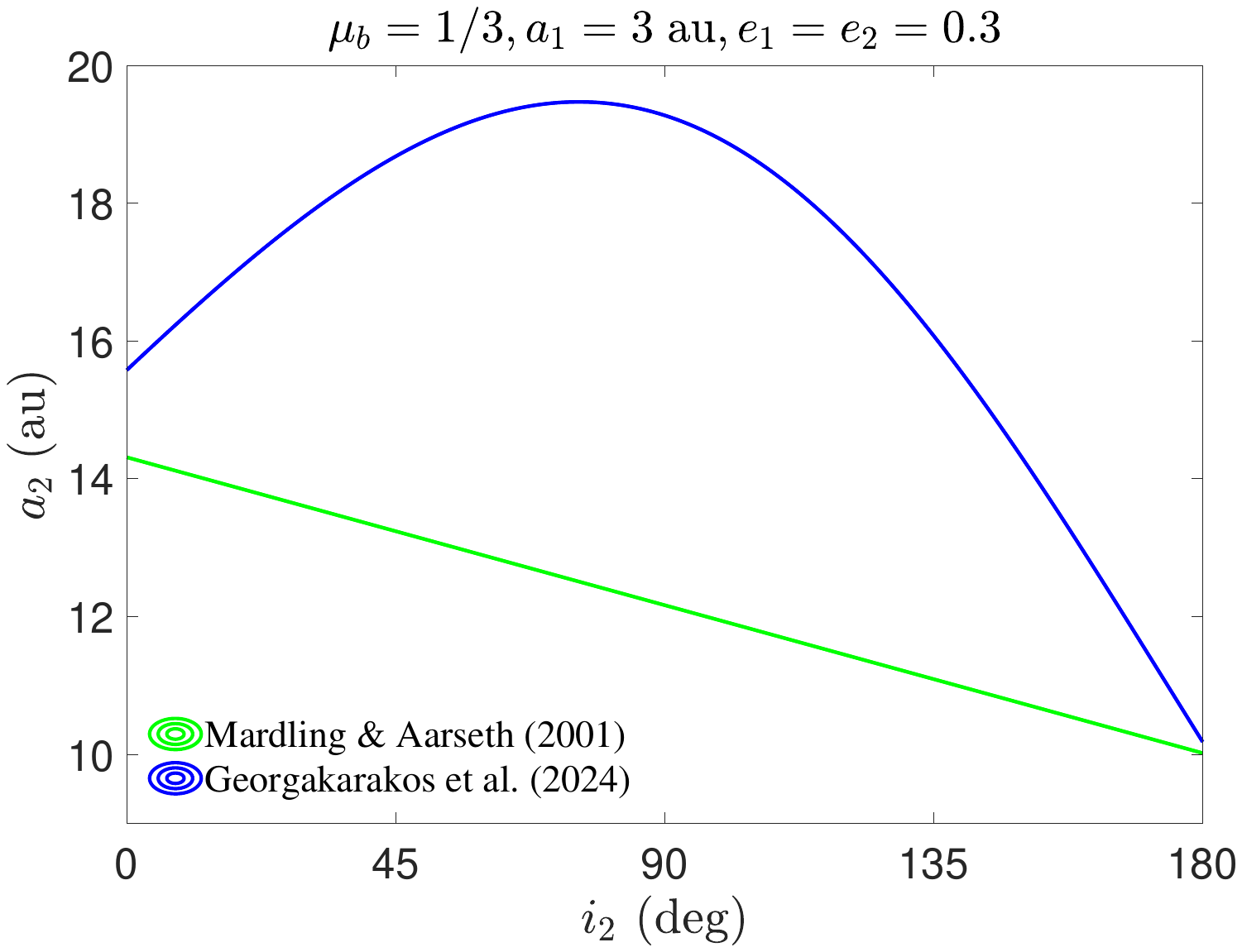}
\includegraphics[width=\columnwidth]{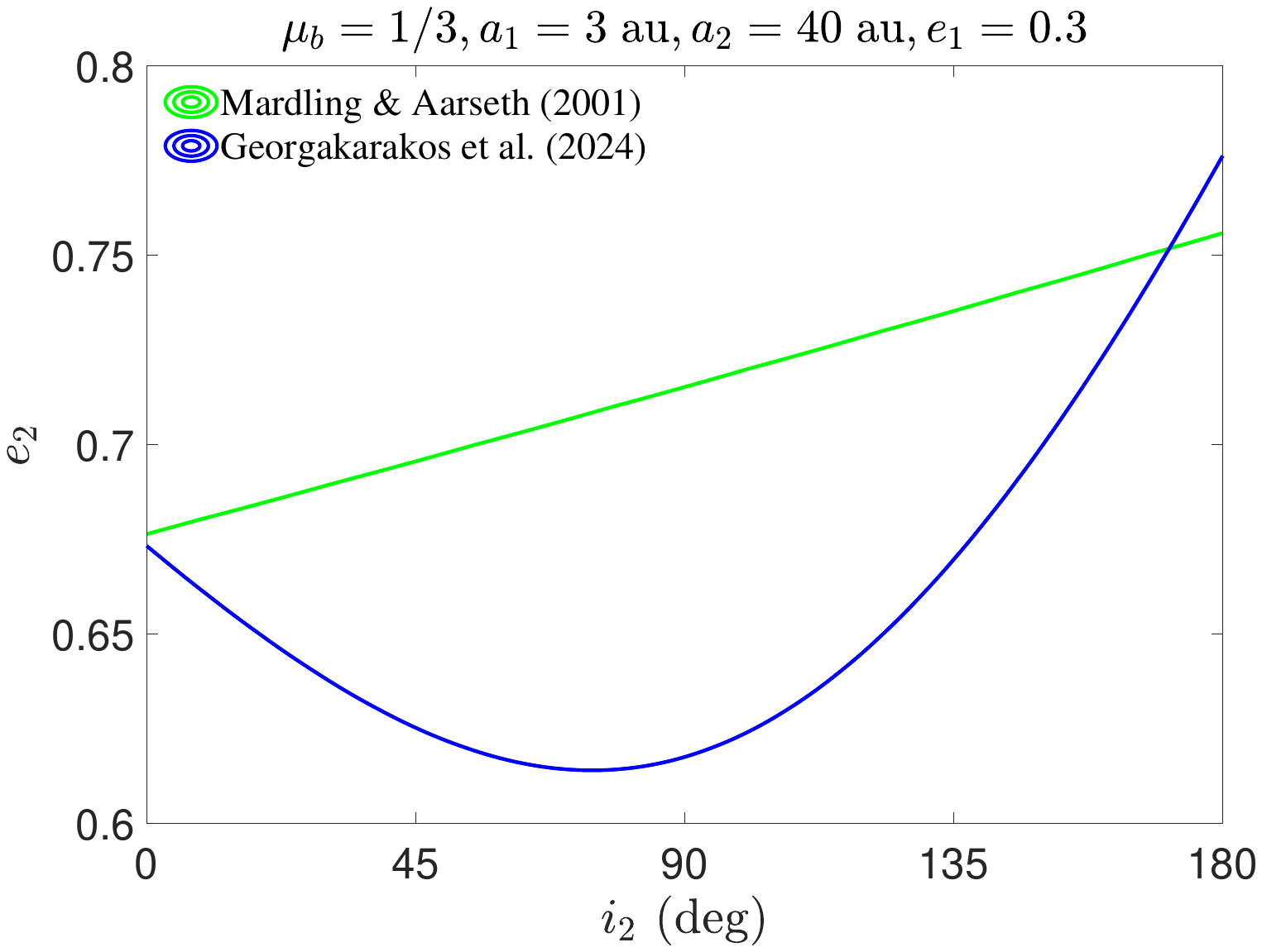}
\caption{Comparisons among critical curves of stability. The points located below the critical curves are stable. The stability criteria shown here come from \citet{dvorak1989stability}, \citet{holman1999long}, \citet{mardling2001tidal}, \citet{quarles2018stability}, \citet{adelbert2023stability} and \citet{georgakarakos2024empirical}.}
\label{Fig3}
\end{figure*}

To make a direct comparison among different versions of stability condition, we take a representative circumbinary planetary system with $m_0 = 1.0 m_{\odot}$ and $m_1 = 0.5 m_{\odot}$ as an example. The results are shown in Fig. \ref{Fig3}. The detailed setting of other parameters are provided on the top of each figure. In the $(e_1, a_2)$ space, $a_{2,c}$ is an increasing function of $e_1$ except for the curve given by \citet{adelbert2023stability}. In the $(e_2,a_2)$ space, the curve of \citet{mardling2001tidal} agrees well with that of \citet{adelbert2023stability} in the low-eccentricity space and agrees with that of \citet{georgakarakos2024empirical} in the high-eccentricity space. Only \citet{mardling2001tidal} and \citet{georgakarakos2024empirical} can provide the stability dependence on planetary inclination $i_2$, and the last two panels of Fig. \ref{Fig3} provide their stability curves in the $(i_2,a_2)$ and $(i_2,e_2)$ space. It is observed that the critical curves of \citet{mardling2001tidal} are linear functions of inclination $i_2$, while the curves of \citet{georgakarakos2024empirical} present nonlinear behaviours: (a) $a_{2,c}$ increases first and then decreases with $i_2$ and (b) $e_{2,c}$ deceases first and then increases with $i_2$.

In the following simulations, we need to ensure that the system parameters satisfy the stability condition. In practice, we adopt the stability limit given by \citet{mardling2001tidal}. Numerical simulations are stopped when stability condition is violated. Results remain qualitatively unchanged if a different stability condition is taken.

\section{Dynamical structures}
\label{Sect4}

In this section, phase-space structures of CBPs under hierarchical three-body configurations are numerically explored by taking advantage of dynamical maps, which provide a global view about the dynamics. The indicator we adopt for producing dynamical maps is the second-derivative-based index $||\Delta D||$ where $D$ is the action diameter of an orbit. This indicator was recently developed by \citet{Daquin2023Detection} and it is a sensitive and robust index for detecting separatrices, resonant centre and chaotic motion in the phase space. Especially, this indicator has a strong ability for distinguishing minute structures caused by high-order or secondary resonances \citep{lei2024dynamical,huang2024dynamical}. The introduction of $||\Delta D||$ follows from the same authors' recent developments on Lagrangian descriptors and arc-length of orbits \citep{daquin2022global}. Compared to other Chaotic index (e.g., fast Lyapunov indicator), the computation of $||\Delta D||$ is free of tangent dynamics \citep{guzzo2002numerical,Daquin2023Detection}. 

As for the current problem, the diameter metric of an orbit is defined by
\begin{equation}\label{Eq13}
D\left( {{e_{2,0}},{i_{2,0}},{\Omega _{2,0}},{\omega _{2,0}}} \right) = {\left\| {\left( {\delta {e_2},\delta {i_2}} \right)} \right\|_\infty}
\end{equation}
where $\left( {{e_{2,0}},{i_{2,0}},{\Omega _{2,0}},{\omega _{2,0}}} \right)$ is the initial state, $\delta {e_2}$ and $\delta {i_2}$ represent the amplitude of the eccentricity and inclination over a given period of time $t_{\max}$, determined by
\begin{equation*}
\begin{aligned}
\delta {e_2}\left( {{e_{2,0}},{i_{2,0}},{\Omega _{2,0}},{\omega _{2,0}}} \right) &= \mathop {\max }\limits_{0 \le t  \le {t_{\max }}} {e_2}\left( {{e_{2,0}},{i_{2,0}},{\Omega _{2,0}},{\omega _{2,0}}, t } \right)\\
& - \mathop {\min }\limits_{0 \le t  \le {t_{\max }}} {e_2}\left( {{e_{2,0}},{i_{2,0}},{\Omega _{2,0}},{\omega _{2,0}},t } \right)
\end{aligned}
\end{equation*}
and
\begin{equation*}
\begin{aligned}
\delta {i_2}\left( {{e_{2,0}},{i_{2,0}},{\Omega _{2,0}},{\omega _{2,0}}} \right) &= \mathop {\max }\limits_{0 \le t  \le {t_{\max }}} {i_2}\left( {{e_{2,0}},{i_{2,0}},{\Omega _{2,0}},{\omega _{2,0}}, t } \right)\\
& - \mathop {\min }\limits_{0 \le t  \le {t_{\max }}} {i_2}\left( {{e_{2,0}},{i_{2,0}},{\Omega _{2,0}},{\omega _{2,0}}, t } \right)
\end{aligned}
\end{equation*}
Based on the diameter $D$, \citet{Daquin2023Detection} defined a second-derivative-based index $||\Delta D||$ as follows:
\begin{equation}
\left\| {\Delta D} \right\| = \left|\frac{{{\partial ^2}D}}{{\partial e_{2,0}^2}}\right| + \left|\frac{{{\partial ^2}D}}{{\partial i_{2,0}^2}}\right| + \left|\frac{{{\partial ^2}D}}{{\partial \Omega _{2,0}^2}}\right| + \left|\frac{{{\partial ^2}D}}{{\partial \omega _{2,0}^2}}\right|.
\end{equation}

For the current problem, it is found that the last two terms have small contribution to $||\Delta D||$. To save computational cost, $||\Delta D||$ is simplified to be
\begin{equation}
\left\| {\Delta D} \right\| = \left|\frac{{{\partial ^2}D}}{{\partial e_{2,0}^2}}\right| + \left|\frac{{{\partial ^2}D}}{{\partial i_{2,0}^2}}\right|,
\end{equation}
where the second-order derivative $\frac{{{\partial ^2}D}}{{\partial {x^2}}}$ can be numerically calculated by means of central difference
\begin{equation*}
\frac{{{\partial ^2}D}}{{\partial {x^2}}} = \frac{{D\left( {x + h} \right) + D\left( {x - h} \right) - 2D\left( x \right)}}{{{h^2}}}
\end{equation*}
where $x$ stands for $e_{2,0}$ or $i_{2,0}$, and $h$ is the associated step. Please refer to \citet{Daquin2023Detection} for more details about the definition, determination and applications of the indicator $\left\| {\Delta D} \right\|$.

Our recent works show that a normalised second-derivative-based index has a stronger ability to detect those structures with small strength \citep{lei2024dynamical,huang2024dynamical}. Following the same consideration, we introduce the normalised indicator $||\Delta D||$ as follows:
\begin{equation}
\left\| {\Delta D} \right\| = \frac{1}{D}\left( {\left| {\frac{{{\partial ^2}D}}{{\partial e_{2,0}^2}}} \right| + \left| {\frac{{{\partial ^2}D}}{{\partial i_{2,0}^2}}} \right|} \right)
\end{equation}
According to the definition of $\left\| {\Delta D} \right\|$, it measures the relative continuity of the distance metric $D$ in the considered phase space. Thus, it becomes possible to take $\left\| {\Delta D} \right\|$ to detect the separatrix crossing, meaning that the dynamical structures with $\left\| {\Delta D} \right\|$ as indicator could provide minute structures in the phase space. 

To produce dynamical maps with $||\Delta D||$ as chaotic index, the initial longitude of ascending node and argument of pericentre are assumed as $\Omega_{2,0} =\omega_{2,0}=\pi/2$. Such a choice of initial angles is based on the following considerations: (a) $\Omega_{2} = \pi/2$ is the centre of nodal librating trajectory (see Sect. \ref{Sect5}), (b) $\omega_{2} = \pi/2$ is the centre of inverse ZLK librating trajectory, and (c) it leads to $\varpi = \pi$ or $\varpi = 0$, which corresponds to an initial aligned or anti-aligned configuration. Thus, this choice can represent almost all types of trajectories in the considered phase space. By fixing the initial angles (this is similar to taking a snapshot for a beam of Hamiltonian flow), it becomes possible for us to explore dynamical structures in the action space. Our practical experiments indicate that dynamical structures are weakly related to planetary eccentricity $e_2$, thus without loss of generality we take $e_{2,0} = 0.1$ in simulations. The planetary inclination $i_{2}$ and binary's eccentricity $e_1$ are distributed in the domain $[0^{\circ},180^{\circ}] \times [0,0.9]$ with steps of $\delta i_2 = 0.2^{\circ}$ and $\delta e_1 = 0.01$. Numerical simulations are stopped if any one of the following conditions is satisfied: (a) the integration time is over $200 \pi$ normalised units of time ($t_{\max} \sim $100 period of $\Omega_2$); (b) the stability condition given by \citet{mardling2001tidal} is violated; (c) the time step of numerical integration is smaller than $1.0 \times 10^{-8}$. Almost all simulations are stopped due to condition (a).

\begin{figure*}
\centering
\includegraphics[width=\columnwidth]{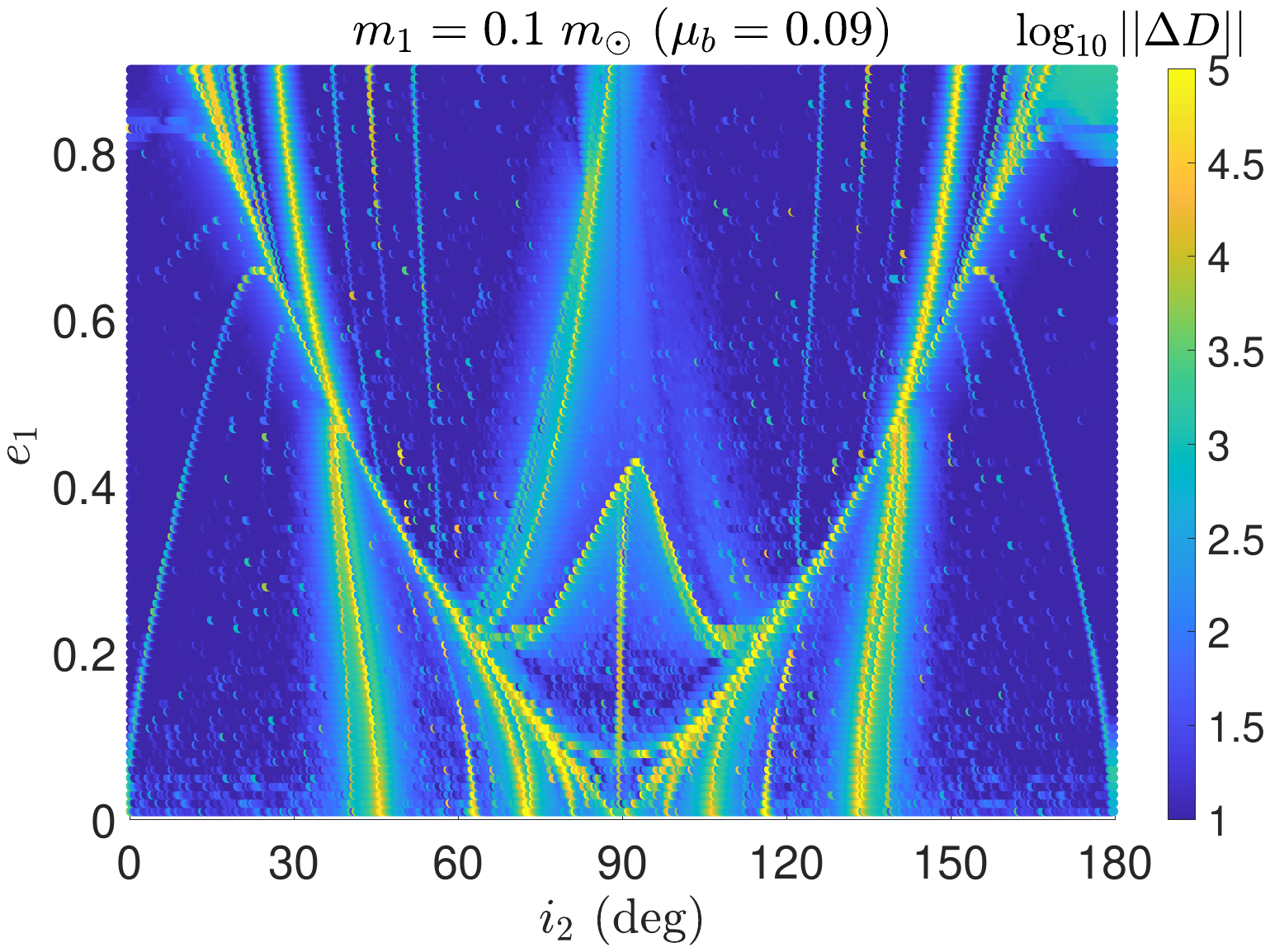}
\includegraphics[width=\columnwidth]{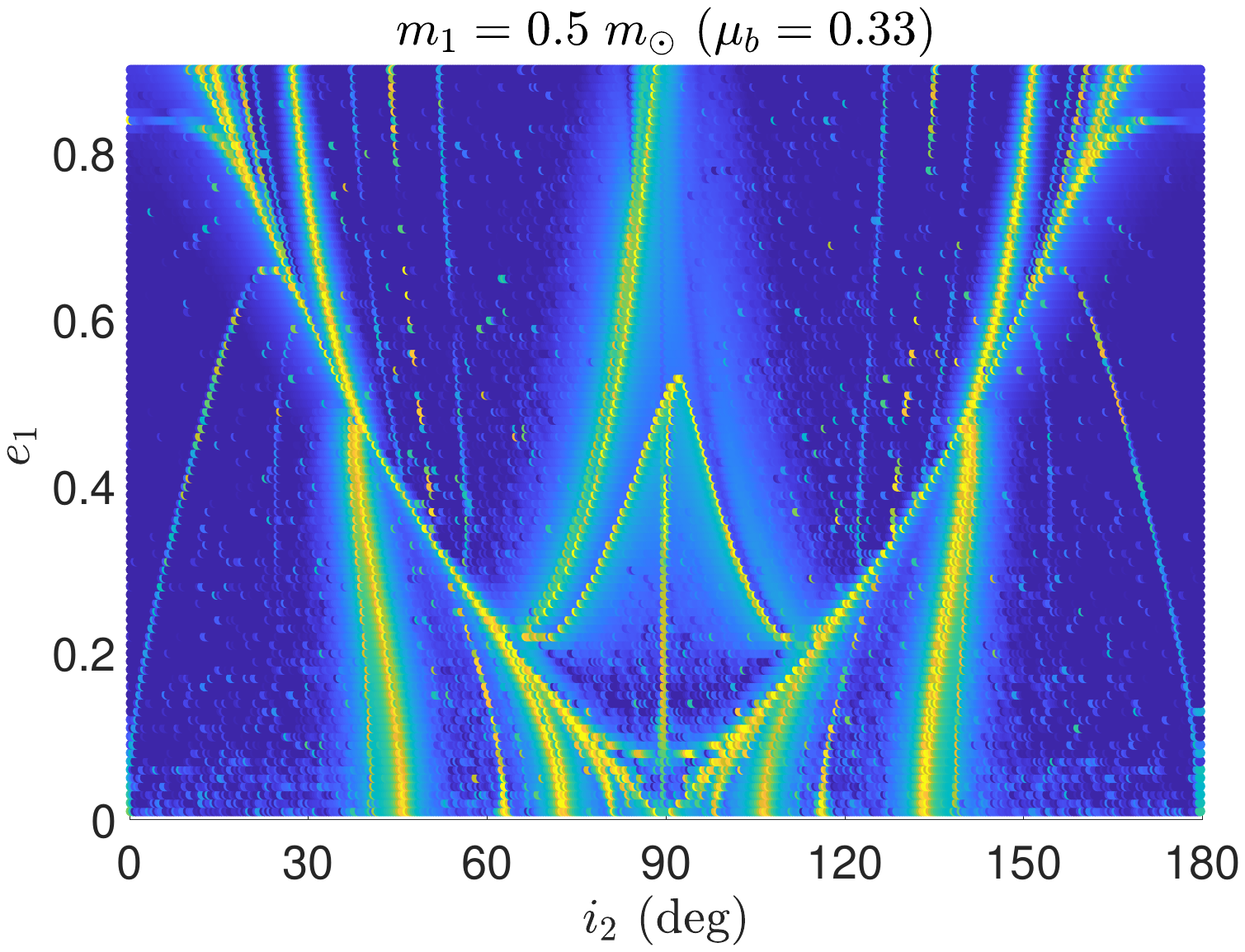}\\
\includegraphics[width=\columnwidth]{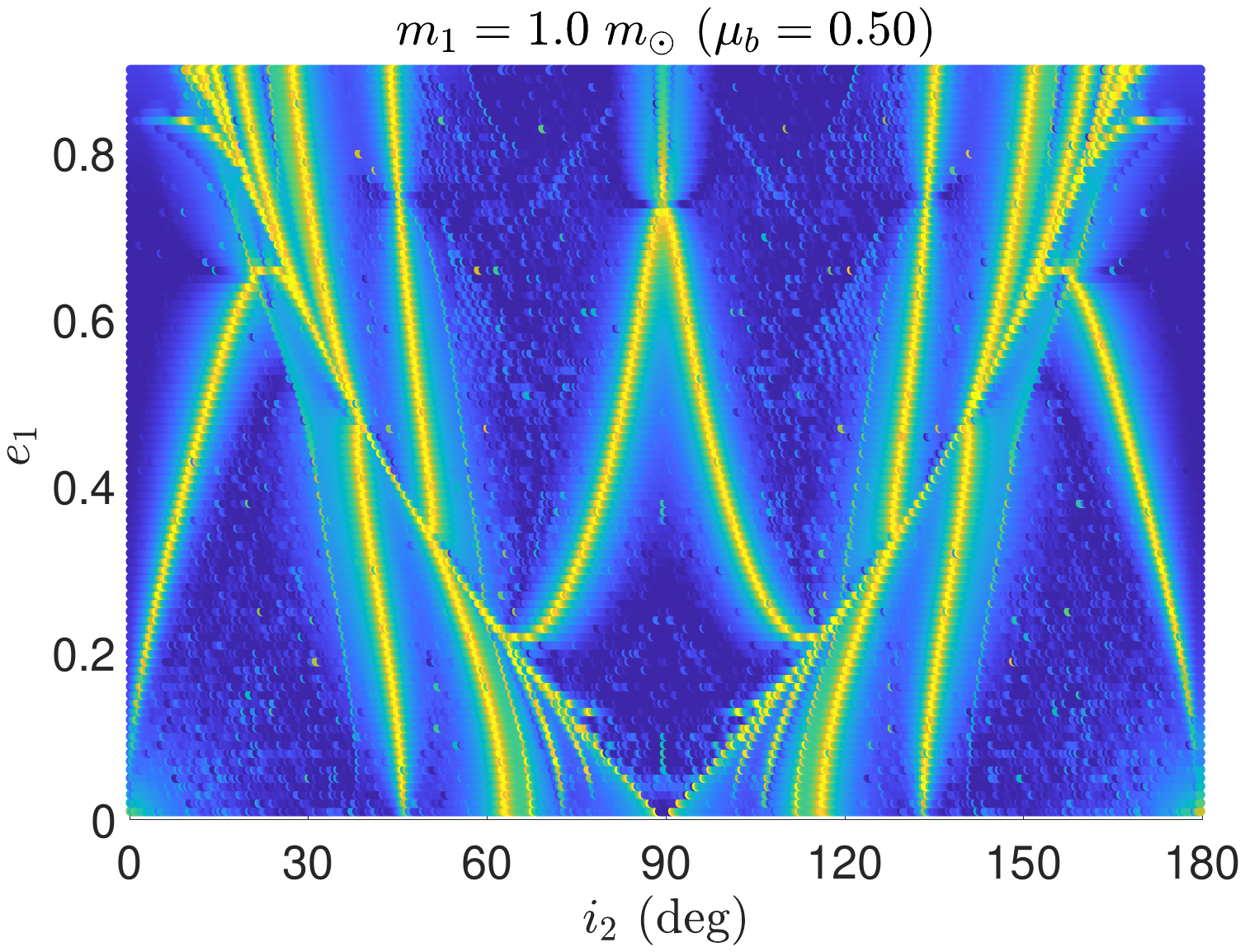}
\includegraphics[width=\columnwidth]{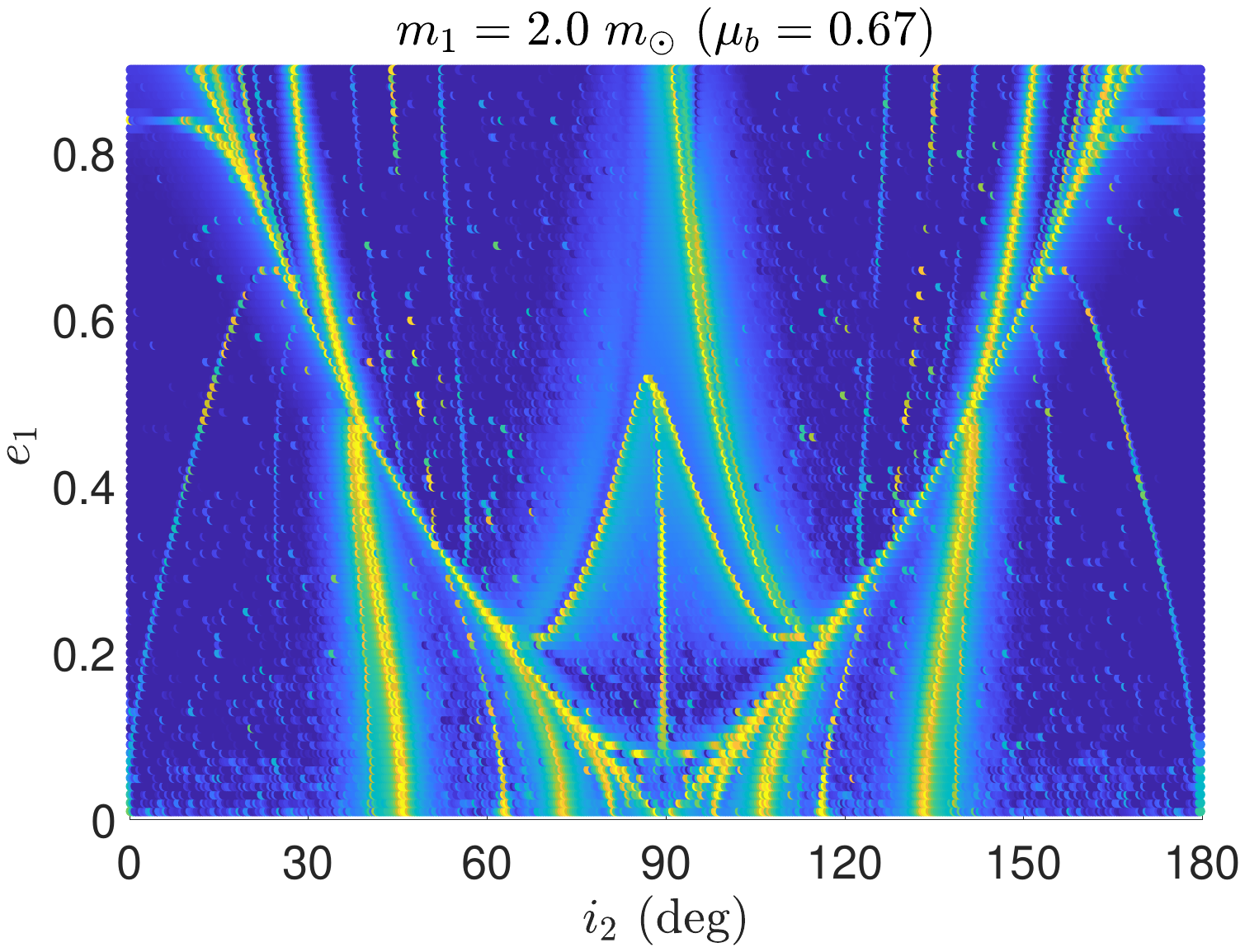}\\
\includegraphics[width=\columnwidth]{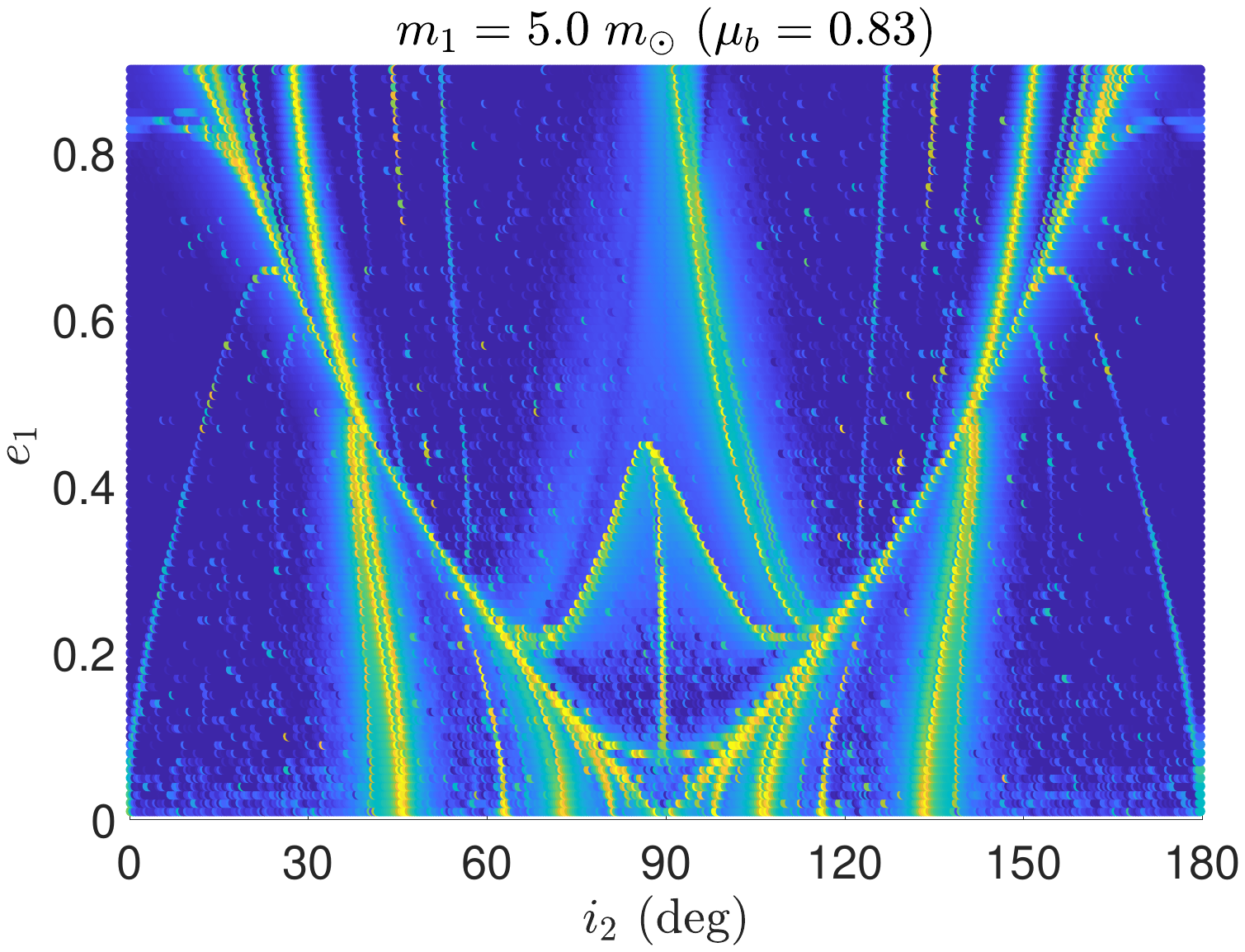}
\includegraphics[width=\columnwidth]{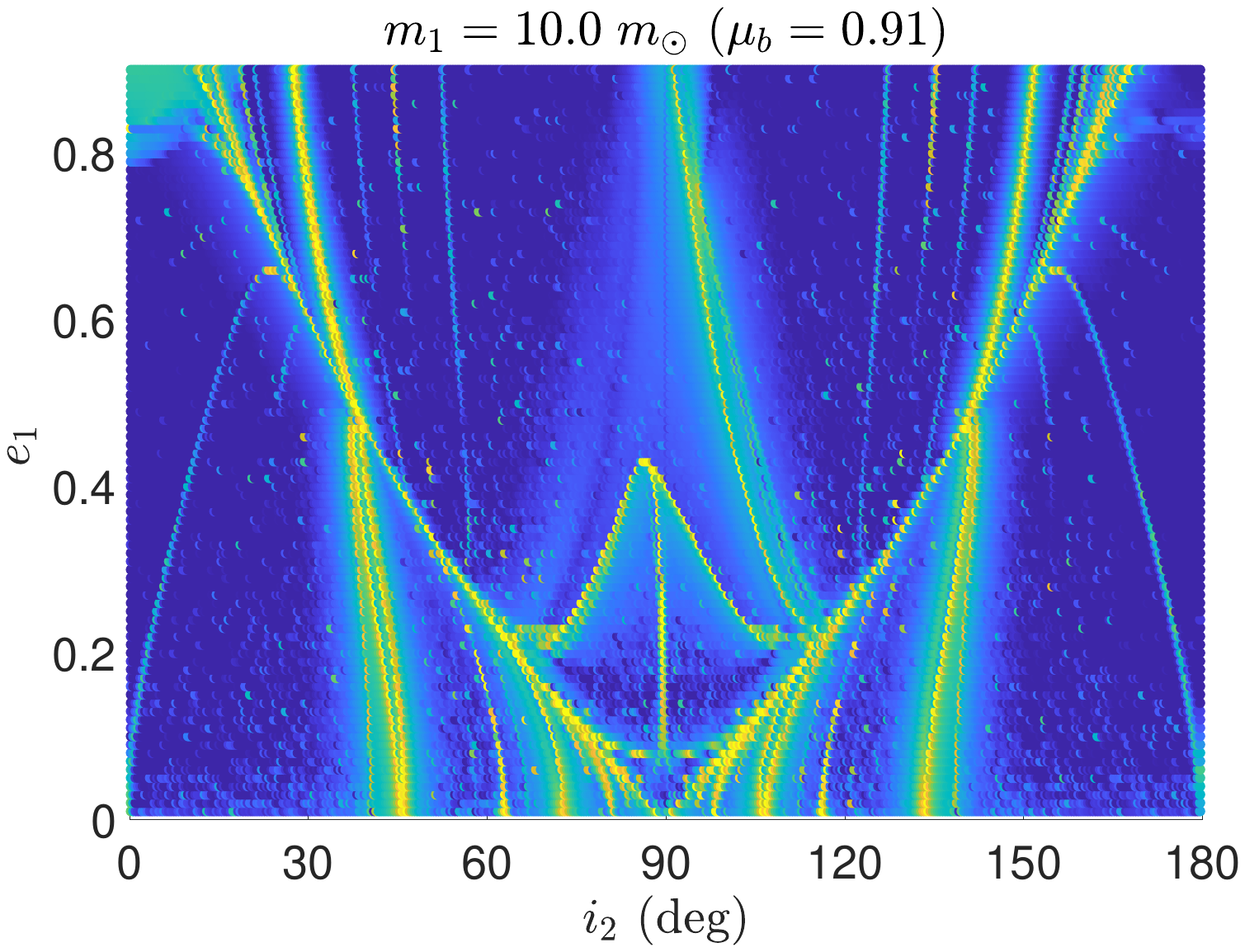}
\caption{Dynamical maps with the normalised second-derivative-based index $||\Delta D||$ shown in the $(i_2,e_1)$ parameter space for different mass parameters $\mu_b=m_1/(m_0+m_1)$ where $m_0$ is fixed as $m_0 = 1.0 m_{\odot}$ and $m_1$ is given on the top of each panel. To produce the dynamical maps, the initial eccentricity of test particle is assumed at $e_{2,0} = 0.1$ (additional experiments show that different initial eccentricities $e_{2,0}$ lead to similar maps), and the initial argument of perientre and longitude of ascending node are taken as $\Omega_{2,0}=\omega_{2,0}=\pi/2$ (please see the text for explanation why we choose such a pair of initial angles). For convenience of display, the magnitude of $||\Delta D||$ is limited to the range of $[10^1,10^5]$ and the index shown in the colour bar corresponds to the base 10 logarithm of $||\Delta D||$. Higher magnitude of $||\Delta D||$ indicates that the associated orbit is more chaotic.}
\label{Fig4}
\end{figure*}

Dynamical maps in the $(i_2, e_1)$ space are shown in Fig. \ref{Fig4} for different mass parameters $\mu_b = m_1/(m_0+m_1)$. In particular, the cases of $m_1=0.1 m_{\odot}$, $0.5 m_{\odot}$, $1.0 m_{\odot}$, $2.0 m_{\odot}$, $5.0 m_{\odot}$, $10.0 m_{\odot}$ are considered with the mass of the central object at $m_0 = 1.0 m_{\odot}$. For convenience of display, the magnitude of $||\Delta D||$ is limited to the range of $[10^1,10^5]$ and the index shown in the colour bar corresponds to the base 10 logarithm of $||\Delta D||$. The regions with higher $||\Delta D||$ are more chaotic.

From Fig. \ref{Fig4}, it is observed that (a) there are complex structures arising in the dynamical maps, (b) dynamical maps with different mass parameters hold slightly different structures, showing that dynamical structures are weakly dependent on the mass parameter, (c) in each panel there is a main V-shape structure which is symmetric with respect to the polar orbit, and (d) minute structures can be observed inside and outside the V-shape region, and they are no longer symmetric with respect to $i_2 = 90^{\circ}$ (i.e., symmetry breaking of minute structures). However, it is different for the case of $m_1 = 1.0 m_{\odot}$ (equal-mass binary), where the minute structures inside and/or outside the V-shape region are nearly symmetric with respect to $i_2 = 90^{\circ}$ and they hold almost the same strength. This is because, in the equal-mass case, the octupole-order contribution disappears (see the Hamiltonian developed in Sect. \ref{Sect2}), and all the substructures arising in the dynamical maps are attributed to the hexadecapolar-order terms, which are almost equal-strength.

Regarding dynamical maps, \citet{cuello2019planet} adopted the eccentricity variation $\Delta e_2$ of orbits as index (see their Fig. 6). In their dynamical maps, only the structures caused by 1:1 resonance can be observed but those minute structures caused by high-order and secondary resonances are not visible.

\begin{figure*}
\centering
\includegraphics[width=\columnwidth]{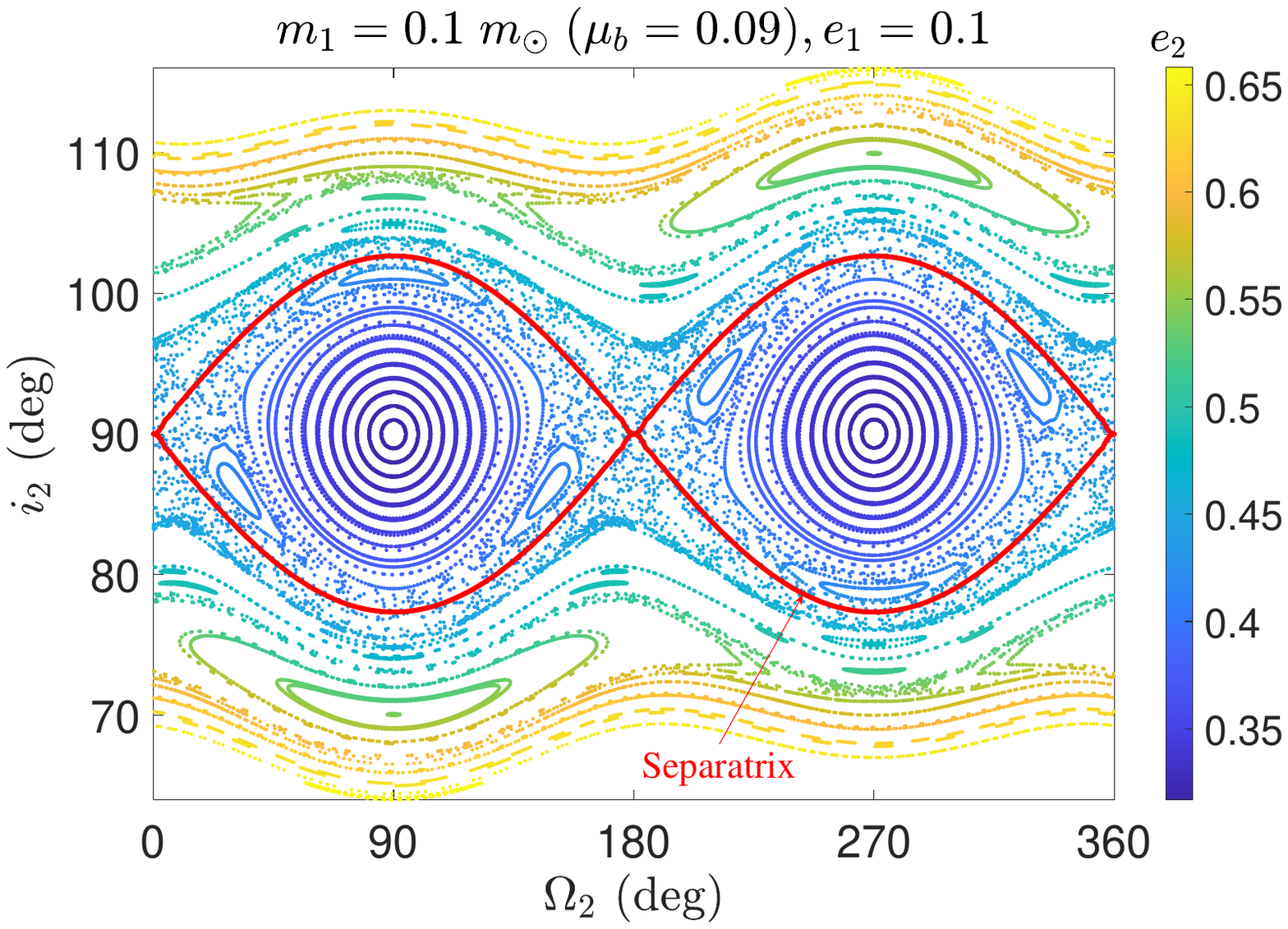}
\includegraphics[width=\columnwidth]{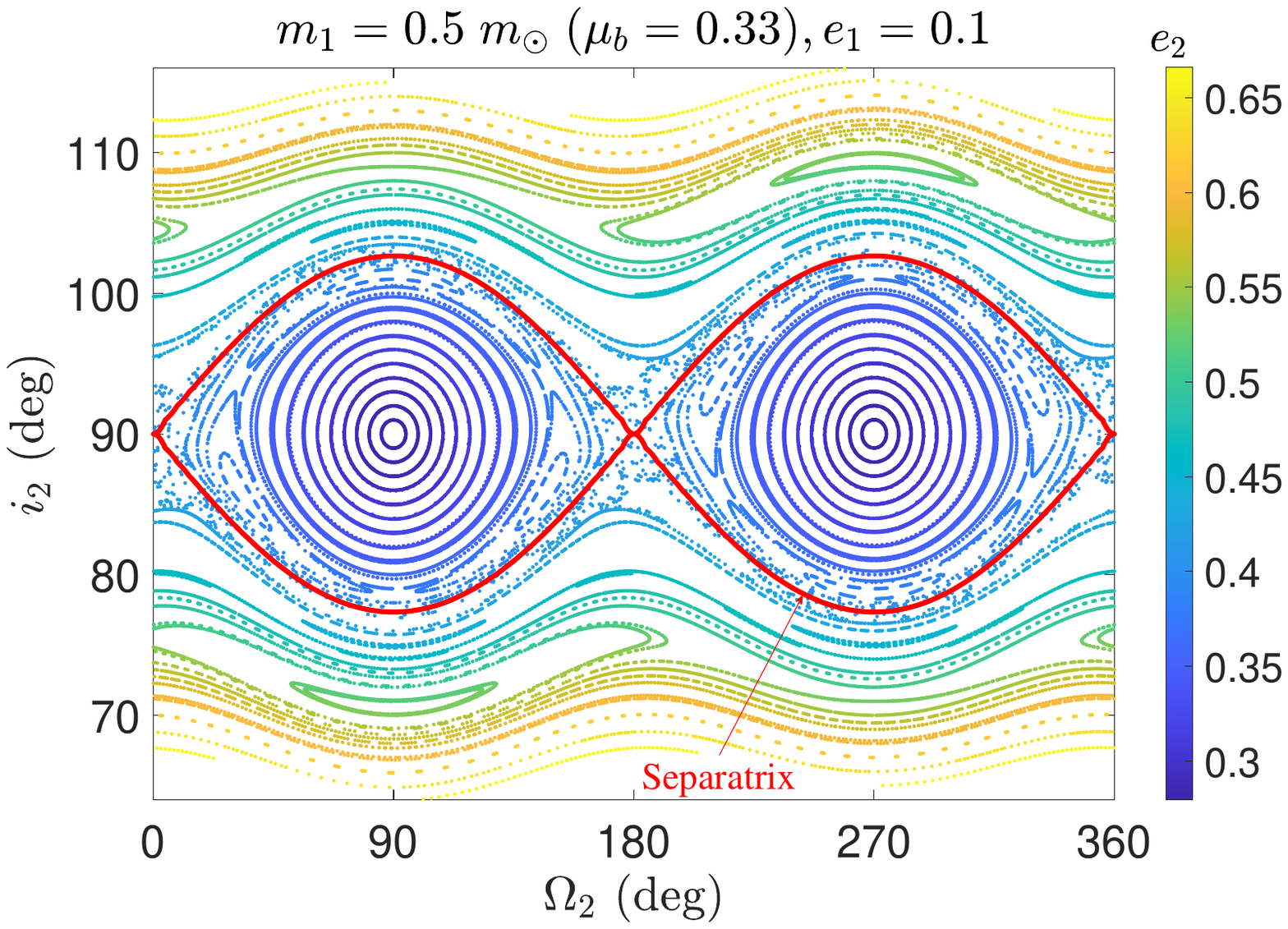}\\
\includegraphics[width=\columnwidth]{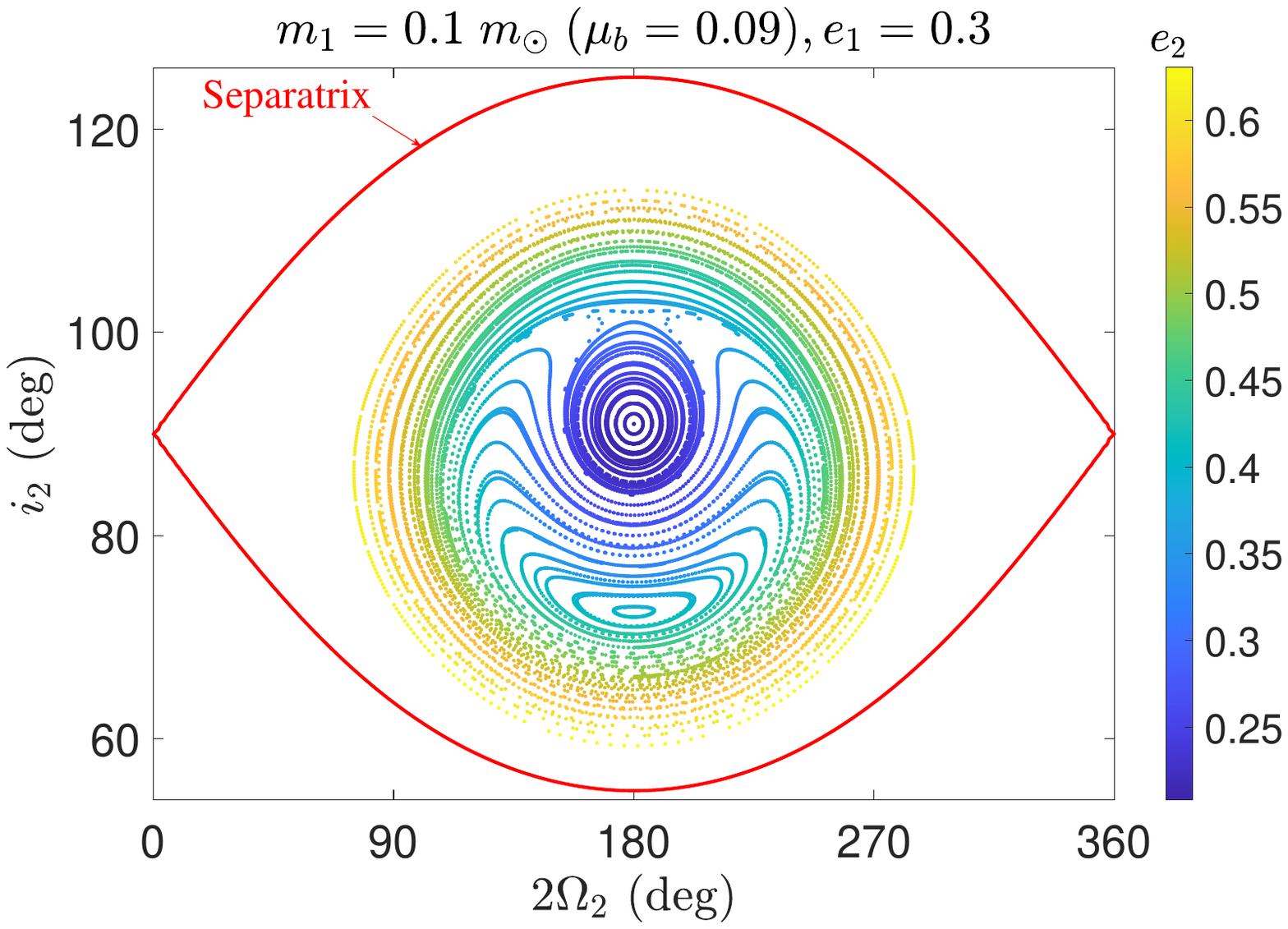}
\includegraphics[width=\columnwidth]{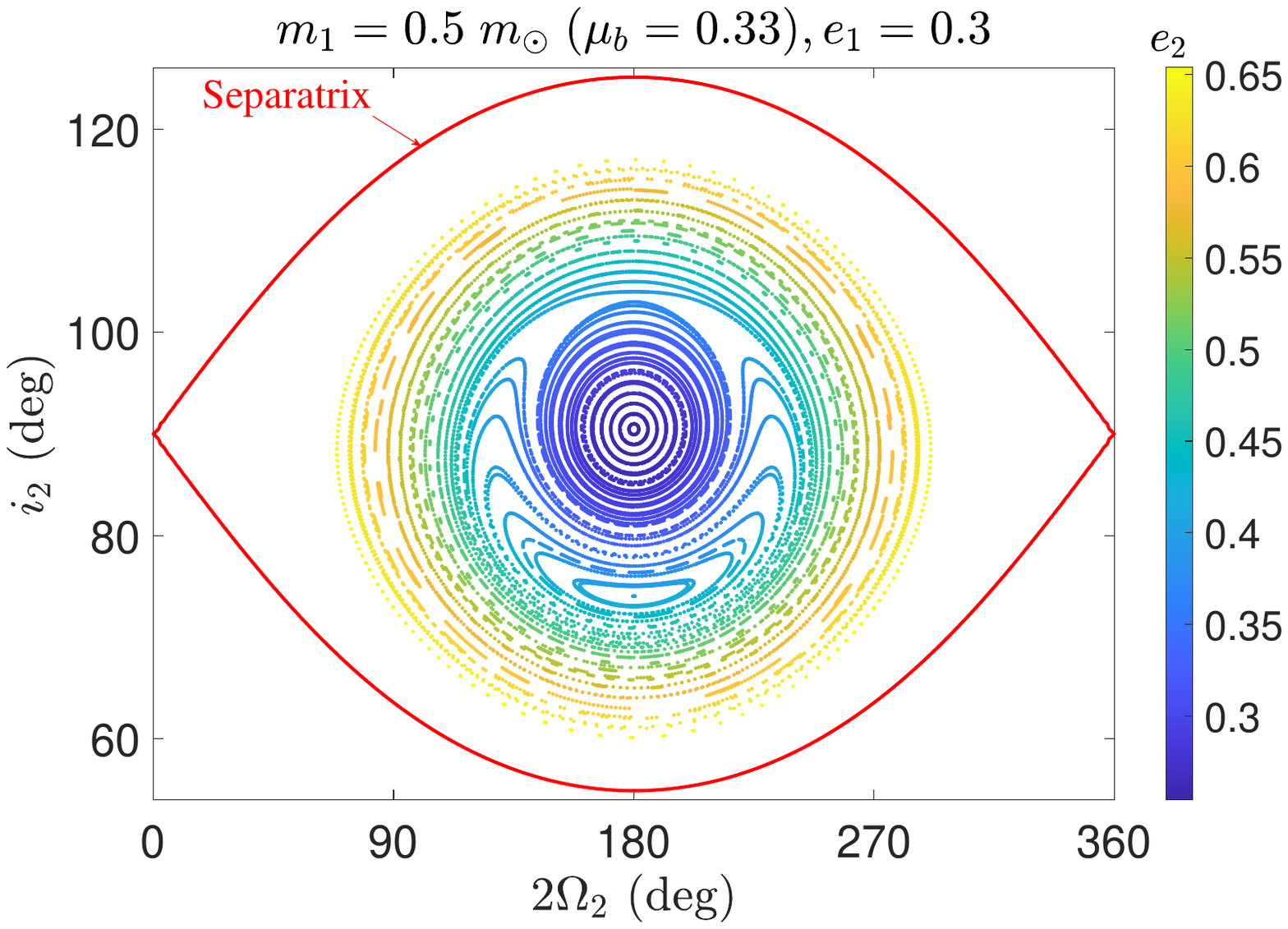}
\caption{Poincar\'e surfaces of section, defined by $g_2=\pi/2$ and $\dot g_2 > 0$. The index shown in the colour bar stands for the magnitude of $e_2$. For producing Poincar\'e sections, the stability condition given by \citet{mardling2001tidal} is satisfied. Hamiltonian is characterised by the critical inclination at $i_{2,c}=67^{\circ}$ (top-two panels), at $i_{2,c}=61^{\circ}$ (bottom-left panel) and at $i_{2,c}=63^{\circ}$ (bottom-right panel). In the case of $e_1 = 0.1$ (see the top-row panels), islands associated with high-order and secondary resonances can be observed. In the case of $e_1 = 0.3$ (see the bottom-row panels), there are two libration islands inside the nodal libration region. The dynamical separatrix marked by red lines are determined by the quadrupole-level Hamiltonian, which is to be discussed in Sect. \ref{Sect5}. The region bounded by dynamical separatrix is of nodal libration.}
\label{Fig5}
\end{figure*}

To understand the complex dynamical structures arising in Fig. \ref{Fig4}, we take the configurations with $e_1 = 0.1$ and $e_1=0.3$ as examples to produce Poincar\'e sections, defined by
\begin{equation}
\omega_2 = \pi/2,\quad {\dot \omega}_2 > 0.
\end{equation}
Under a given Hamiltonian, the points are recorded every time when CBPs pass through the section of $\omega_2 = \pi/2$ with positive angular velocity. For convenience, the Hamiltonian is characterised by the critical inclination $i_{2,c}$ in the form of ${\cal H}\left(\Omega_2=\omega_2=\pi/2,e_2 = 0.6, i_2 = i_{2,c}\right)$. See Fig. \ref{Fig5} for Poincar\'e sections shown in the $(\Omega_2,i_2)$ space for the cases of $m_1 = 0.1 m_{\odot}$ and $m_1 = 0.5 m_{\odot}$. The initial angles of CBPs are assumed at $\Omega_{2,0}=\omega_{2,0}=\pi/2$. The index shown in the colour bar stands for the magnitude of planetary eccentricity $e_2$, which is cut off to ensure the stability condition \citep{mardling2001tidal}. It should be mentioned that the physically allowed region in the $(\Omega_2,i_2)$ space is determined by the given Hamiltonian $\cal H$, mass parameter $\mu_b$, binary eccentricity $e_1$, semimajor axis ratio $\alpha = a_1/a_2$ and stability limit. Please refer to the caption of Fig. \ref{Fig5} for the detailed setting of $e_1$ and $i_{2,c}$.
To distinguish the regions of nodal libration and nodal circulation, dynamical separatrices under the quadrupole-order dynamical model are also shown in the Poincar\'e sections and they will be discussed in the next section.

When the eccentricity of the inner binary ($e_1$) is given, it is observed that Poincar\'e sections for the cases of $m_1 = 0.1 m_{\odot}$ and $m_1 = 0.5 m_{\odot}$ hold similar dynamical structures. However, when the mass parameter $m_1$ is given, dynamical structures arising in Poincar\'e sections are totally different for the cases of $e_1 = 0.1$ and $e_1 = 0.3$. It means that phase-space structures of CBPs are dominated by binary's eccentricity $e_1$.

For the case of $e_1 = 0.1$, chaotic layers can be observed around the dynamical separatrix of nodal resonance. Outside the nodal libration zone, islands associated with high-order resonances can be found and, inside the nodal libration zone, islands associated with secondary resonances can be observed. The primary nodal resonance (quadrupole-order resonance), high-order and secondary resonances (octupole- and hexadecapolar-order resonances) are together responsible for the entire phase-space structures arising in Poincar\'e sections. 

For the configuration of $e_1=0.3$, all the possible motion under the given Hamiltonian takes place inside the region bounded by the dynamical separatrix. In this case, islands associated with secondary 1:1 resonance can be observed in the section. The nodal resonance in combination with secondary 1:1 resonance dominates the dynamical structures arising in the section.

A question arises: how do the nodal, high-order and secondary resonances sculpt the dynamical structures arising in dynamical maps? To answer this question, we will analytically study the dynamics under the quadrupole-order model in Sect. \ref{Sect5}, applications to numerical structures in Sect. \ref{Sect6} and then discuss secondary resonances under the hexadecapolar-order Hamiltonian model in Sect. \ref{Sect7} by taking advantage of perturbative treatments.

\section{Dynamics of the quadrupole-order model}
\label{Sect5}

In this section, we concentrate on the dynamics under the quadrupole-order Hamiltonian model. 

\subsection{Phase portraits}
\label{Sect5-1}

At the quadrupole-level approximation, the Hamiltonian can be written as
\begin{equation}\label{Eq18}
{{\cal H}} =  - \frac{1}{{G_2^5}}\left[ {\left( {2 + 3e_1^2} \right)\left( {3H_2^2 - G_2^2} \right) + 15e_1^2\left( {G_2^2 - H_2^2} \right)\cos 2{h_2}} \right].
\end{equation}
The angle $g_2$ is a cyclic variable, thus its conjugate momentum $G_2$ is a motion integral, showing that the planetary eccentricity $e_2$ remains constant under the quadrupole-order model. Considering the fact that the coefficient $\frac{1}{{G_2^5}}$ is constant, we can normalise the quadrupole-level Hamiltonian represented by equation (\ref{Eq18}) as
\begin{equation}\label{Eq19}
{\cal H} =  - \left[ {\left( {2 + 3e_1^2} \right)\left( {3H_2^2 - G_2^2} \right) + 15e_1^2\left( {G_2^2 - H_2^2} \right)\cos 2{h_2}} \right].
\end{equation}
The resulting dynamical model is of one degree of freedom and thus it is integrable. In addition, the quadrupole-order Hamiltonian is independent on the mass parameter $\mu_b$, thus we could conclude that $\mu_b$ has no influence on the dynamics of the quadrupole-order model. This is the reason that dynamical maps with different $\mu_b$ shown in Fig. \ref{Fig4} hold similar main structures. Analytical expressions of the nodal librating and circulating cycles under the quadrupole-level Hamiltonian are discussed in \citet{li2014analytical}.

\begin{figure*}
\centering
\includegraphics[width=\columnwidth]{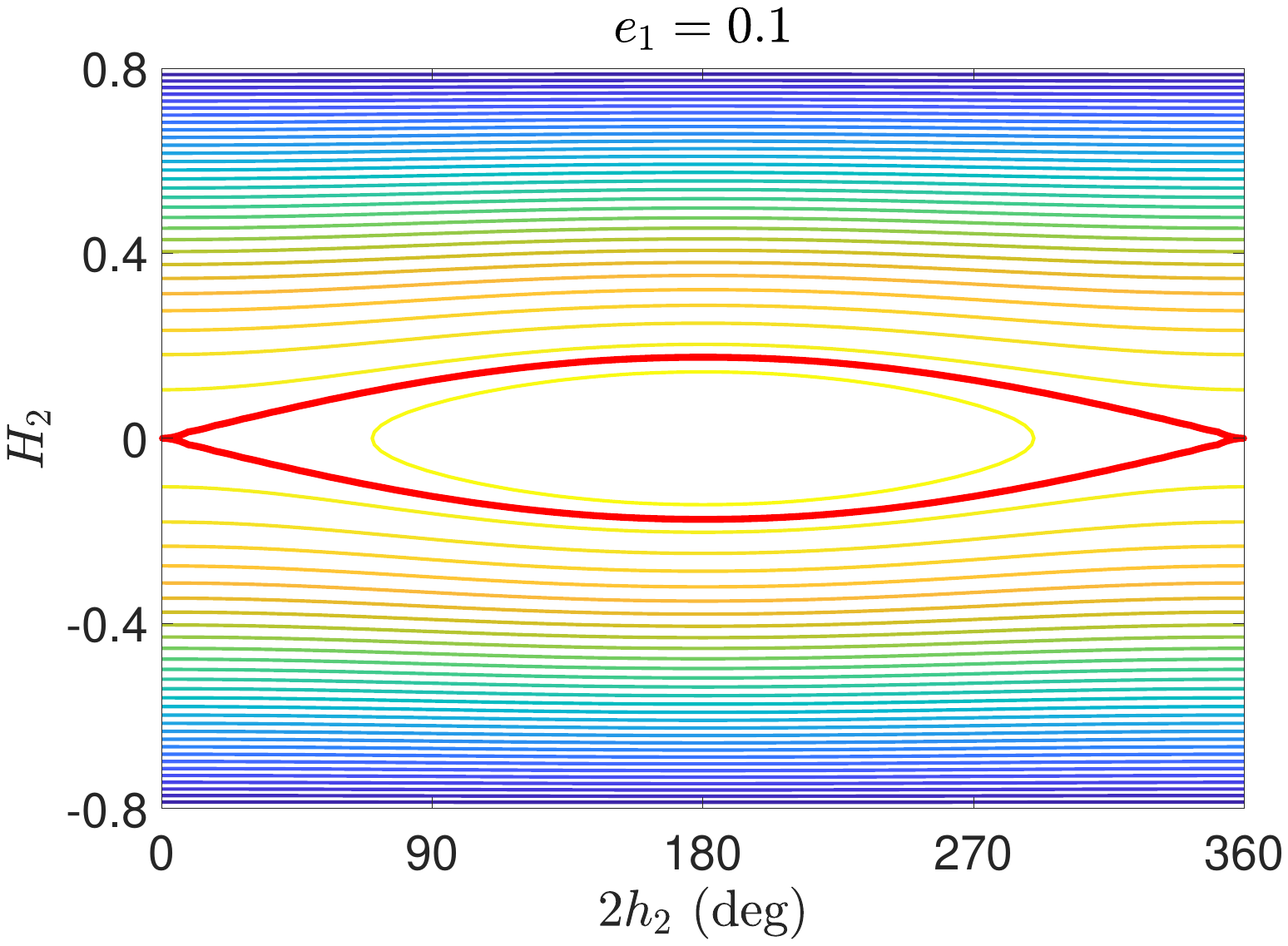}
\includegraphics[width=\columnwidth]{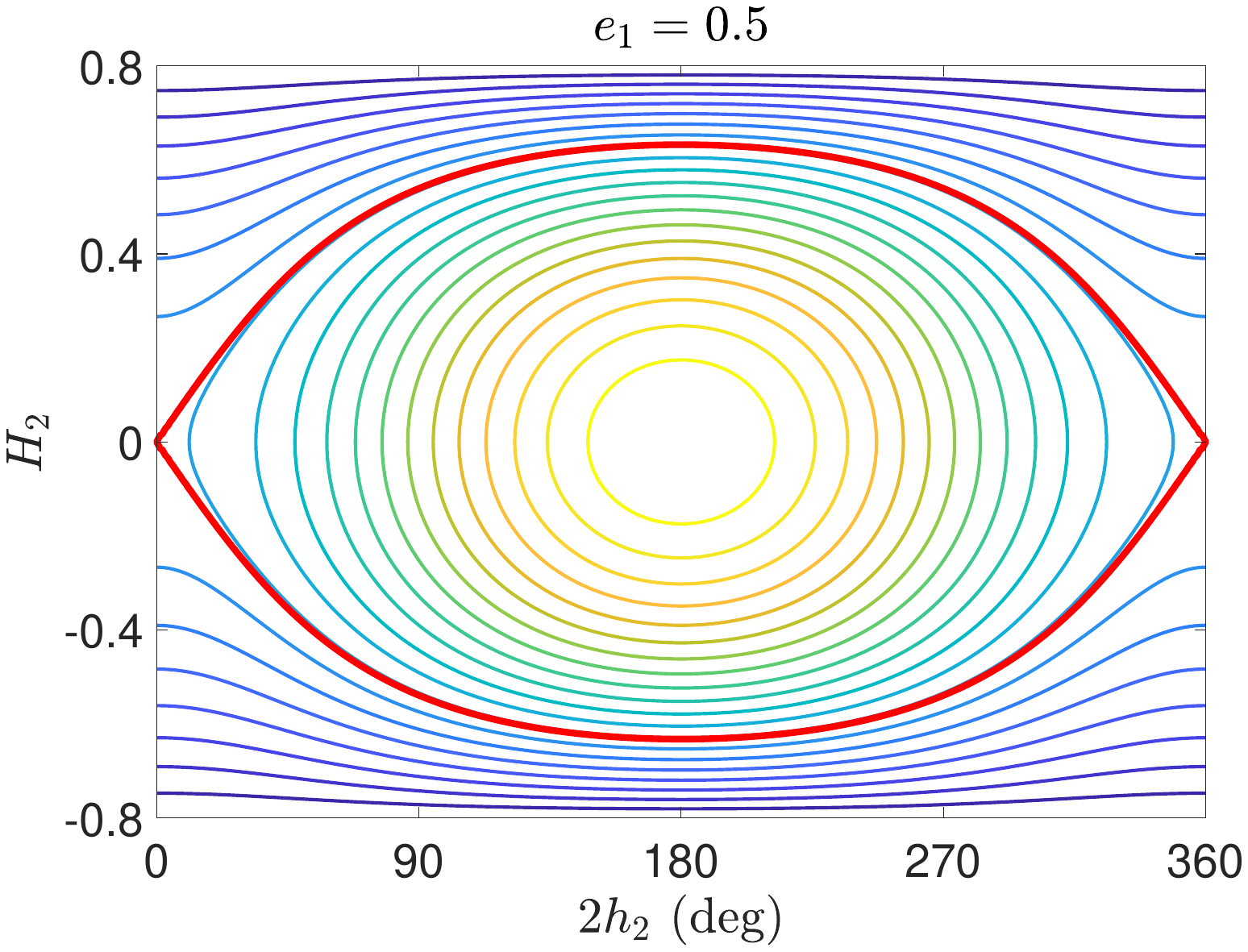}
\caption{Phase diagrams of the dynamical model at the quadrupole-level approximation (which is an integral model), shown in the $(2h_2, H_2)$ space. The left panel is for $e_1 = 0.1$ and the right one is for $e_1=0.5$. The red lines correspond to the dynamical separatrices, which divide the whole phase space into regions of circulation and libration. Inside the region bounded by the separatrix, the ascending node of CBPs is of libration. It is referred to as nodal resonance in this paper. Under the quadrupole-level approximation, the conventional ZLK resonance for inner test particles is replaced by the nodal resonance for outer test particles (i.e., the libration of argument of pericentre $\omega_1$ is replaced by the libration of longitude of ascending node $\Omega_2$).}
\label{Fig6}
\end{figure*}

Phase-space structures can be explored by plotting level curves of Hamiltonian (i.e., phase portraits). Please see Fig. \ref{Fig6} for phase portraits under the quadrupole-order Hamiltonian model in the cases of $e_1 = 0.1$ and $e_1=0.5$. It is observed that the nodal libration centre is located at $(2h_2 = \pi,H_2=0)$ and the saddle point is at $(2h_2 =0,H_2=0)$. The dynamical separatrix, corresponding to the level curves of Hamiltonian passing through the saddle point, is shown in red lines. Dynamical separatrix can be found in Poincar\'e section shown in Fig. \ref{Fig5}. Evidently, the separatrix plays a role in dividing the phase space into regions of nodal libration and circulation. The libration zone is surrounded by dynamical separatrix and the island of nodal libration is larger with a higher value of binary's eccentricity $e_1$.

\subsection{Orbit classifications}
\label{Sect5-2}

Under the quadrupole-order model, both the Hamiltonian $\cal H$ and the planetary eccentricity $e_2$ are integral of motion. Here let us classify the motion in the space spanned by conserved parameters. It is similar to the well-known Lidov triangle in the $(c_1,c_2)$ space \citep{lidov1962evolution, shevchenko2016lidov}. Similar discussions can be found in \citet{lei2022quadrupole}.

From the phase portraits shown in Fig. \ref{Fig6}, it is observed that the saddle point is located at $(2h_2=0,H_2=0)$. According to the definition of dynamical separatrix, we can get the Hamiltonian of separatrix
\begin{equation}\label{Eq20}
{{\cal H}_{\rm sep}} =  - 2\left( {6e_1^2 - 1} \right)\left( {1 - e_2^2} \right).
\end{equation}
${{\cal H}_{\rm sep}}$ is a decreasing function of $e_2$ if $e_1 < 1/\sqrt{6}$ and it is an increasing function of $e_2$ if $e_1 > 1/\sqrt{6}$. In particular, ${{\cal H}_{\rm sep}}$ is equal to zero when $e_1 = 1/\sqrt{6}$.

The minimum value of Hamiltonian takes place at $H_2 = G_2$ (i.e., $i_2 = 0^{\circ}$). Thus, the lower boundary of Hamiltonian can be expressed as
\begin{equation}\label{Eq21}
{{\cal H}_{\rm lower}} =  - \left( {6e_1^2 + 4} \right)\left( {1 - e_2^2} \right),
\end{equation}
which is an increasing function of $e_2$.

The maximum Hamiltonian happens at the libration centre $(2h_2 = \pi,H_2=0)$. As a result, the upper boundary of Hamiltonian can be given by
\begin{equation}\label{Eq22}
{{\cal H}_{\rm upper}} = \left( {18e_1^2 + 2} \right)\left( {1 - e_2^2} \right),
\end{equation}
which is a decreasing function of $e_2$. 

\begin{figure*}
\centering
\includegraphics[width=\columnwidth]{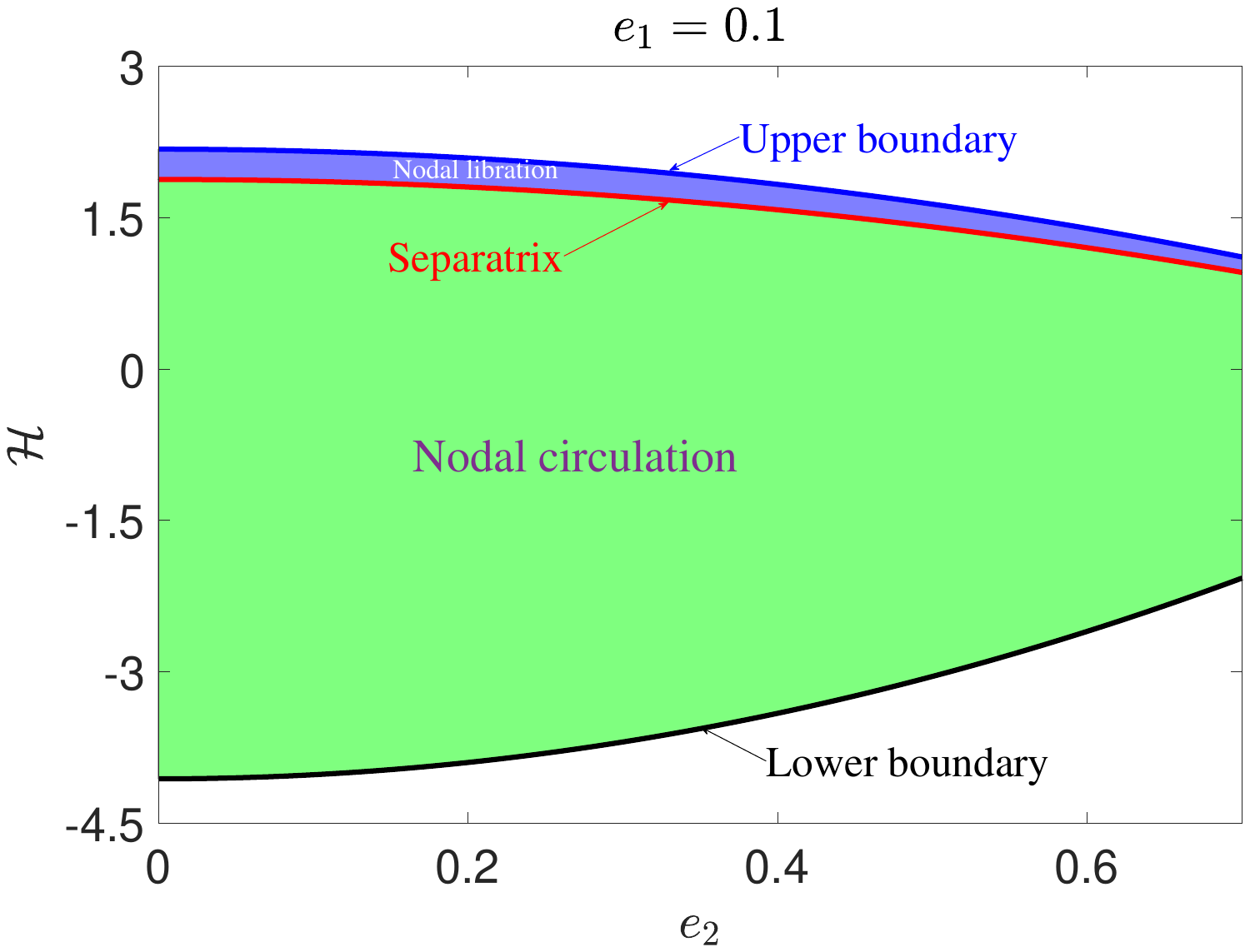}
\includegraphics[width=\columnwidth]{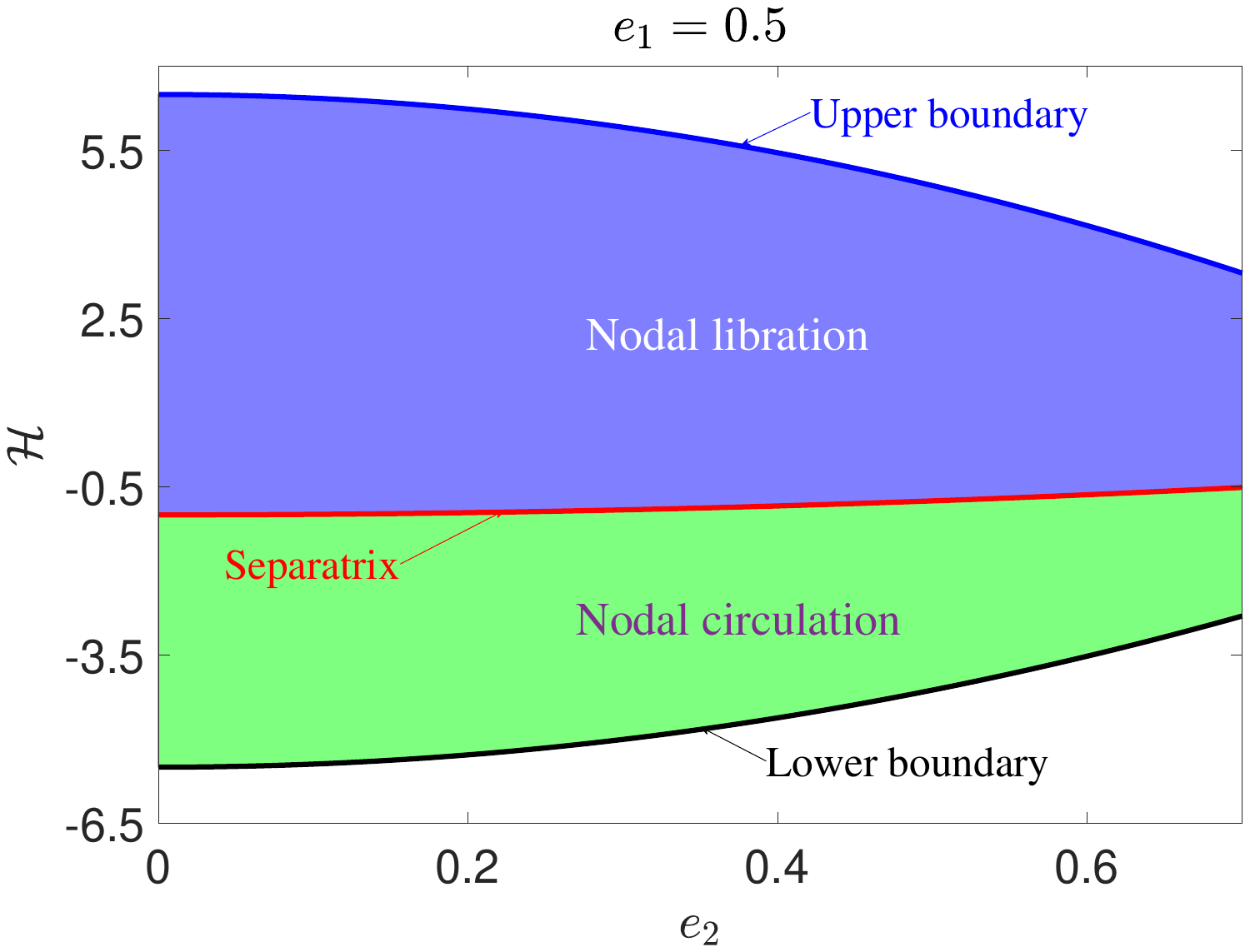}
\caption{Distribution of the nodal libration and circulation regions in the $(e_2,{\cal H})$ space. The left panel is for $e_1=0.1$ and the right panel is for $e_1=0.5$. In both panels, dynamical separatrix between the libration and circulation regions is marked by red line. The white space represents the physically forbidden region.}
\label{Fig7}
\end{figure*}

Fig. \ref{Fig7} shows the distribution of librating and circulating orbits in the $(e_2,{\cal H})$ space for the cases of $e_1=0.1$ (see the left panel) and $e_1=0.5$ (see the right panel). The lower boundary, separatrix and the upper boundary are marked by black line, red line and blue line, respectively. Centres of nodal libration occupy on the upper boundary. The green shaded regions bounded by the separatrix and the lower boundary are of nodal circulation, the blue shaded regions bounded by the separatrix and upper boundary are of nodal libration, and the white zones (below the lower boundary or above the upper boundary) stand for the physically forbidden regions. It is observed that the area of nodal libration region increases with the binary's eccentricity $e_1$. For a given $e_1$, the range of Hamiltonian decreases with planetary eccentricity $e_2$.

\subsection{Width of nodal resonance}
\label{Sect5-3}

According to equation (\ref{Eq20}), we can get the expression of separatrix in the form of $e_1$, $i_2$ and $h_2$ as follows: 
\begin{equation}\label{Eq23}
\left( {2 + 3e_1^2} \right){\cos ^2}{i_2} + 5e_1^2\left( {{{\sin }^2}{i_2}\cos 2{h_2} - 1} \right) = 0
\end{equation}
For a given $h_2$, there are two values of $i_2$ (see the phase portraits shown in Fig. \ref{Fig6}). Usually, resonant width is used to measure the size of libration island, which takes the maximum when it is evaluated at the libration centre (i.e., $2h_2 = \pi$). Taking $2h_2 = \pi$ in equation (\ref{Eq23}), we can get
\begin{equation*}
\left( {1 + 4e_1^2} \right){\cos ^2}{i_{2,\rm sep}} - 5e_1^2 = 0
\end{equation*}
which leads to
\begin{equation*}
\cos {i_{2,\rm sep}} =  \pm \sqrt {\frac{{5e_1^2}}{{1 + 4e_1^2}}}.
\end{equation*}
It shows that the inclinations of the lower and upper separatrices at $2h_2 = \pi$ are 
\begin{equation*}
\begin{aligned}
{i_{2,\rm low}} &= \arccos \left( {\sqrt {\frac{{5e_1^2}}{{1 + 4e_1^2}}} } \right),\\
{i_{2,\rm up}} &= \pi  - \arccos \left( {\sqrt {\frac{{5e_1^2}}{{1 + 4e_1^2}}} } \right),
\end{aligned}
\end{equation*}
which are consistent with the critical inclinations for alternation between nodal libration and circulation \citep{doolin2011dynamics, farago2010high, li2014analytical, chen2019orbital, martin2019polar}. It means that the critical inclinations ${i_{2,\rm low}}$ and ${i_{2,\rm up}}$ are symmetric with respect to $90^{\circ}$. As a result, the resonant width, denoted by $\Delta i_2$, can be obtained by
\begin{equation}\label{Eq22-1}
\Delta {i_2} ={i_{2,\rm up}} - {i_{2,\rm low}}= \pi  - 2\arccos \left( {\sqrt {\frac{{5e_1^2}}{{1 + 4e_1^2}}} } \right)
\end{equation}
which shows that the resonant width is only related to binary's eccentricity $e_1$ and it is an increasing function of $e_1$.

\begin{figure*}
\centering
\includegraphics[width=\columnwidth]{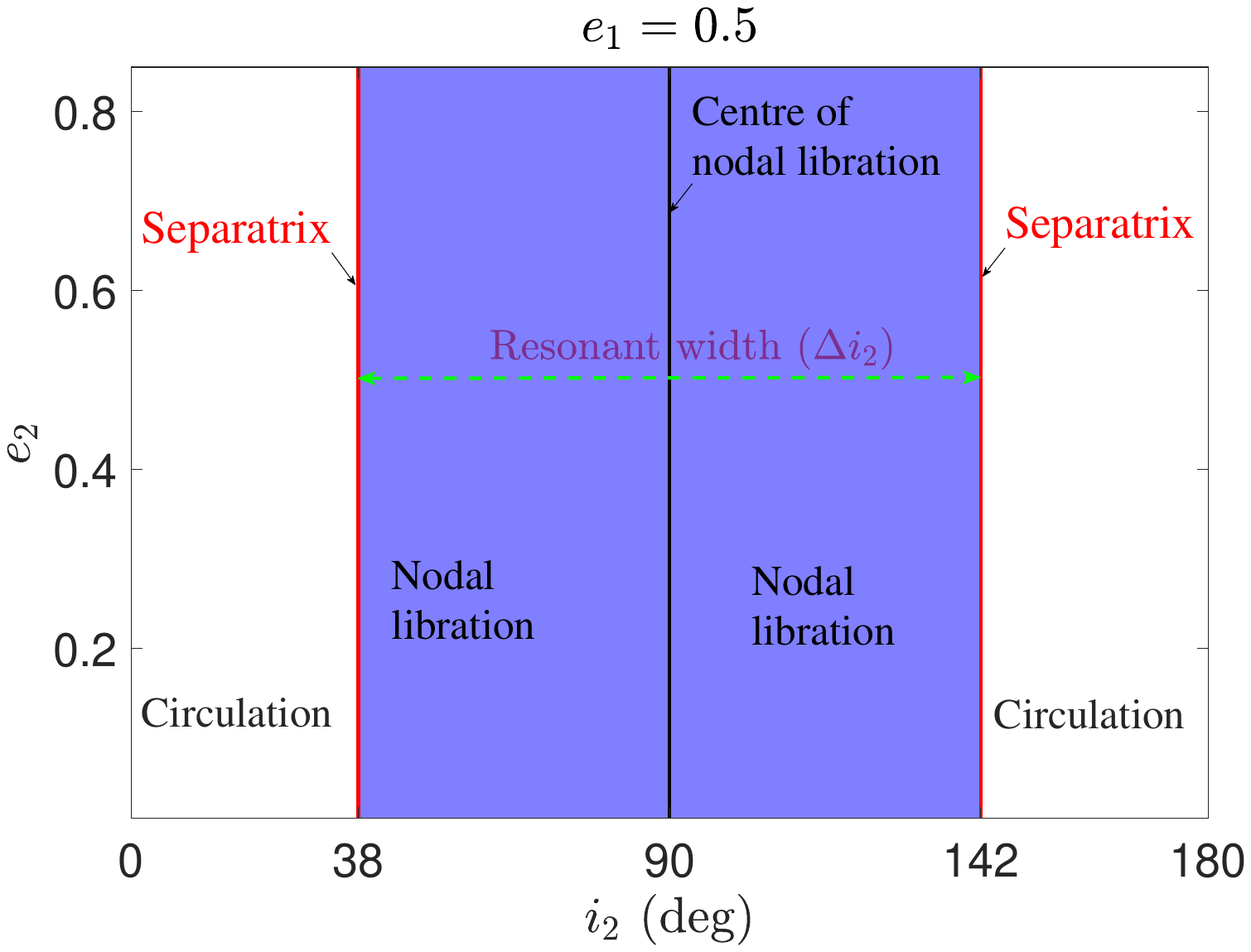}
\includegraphics[width=\columnwidth]{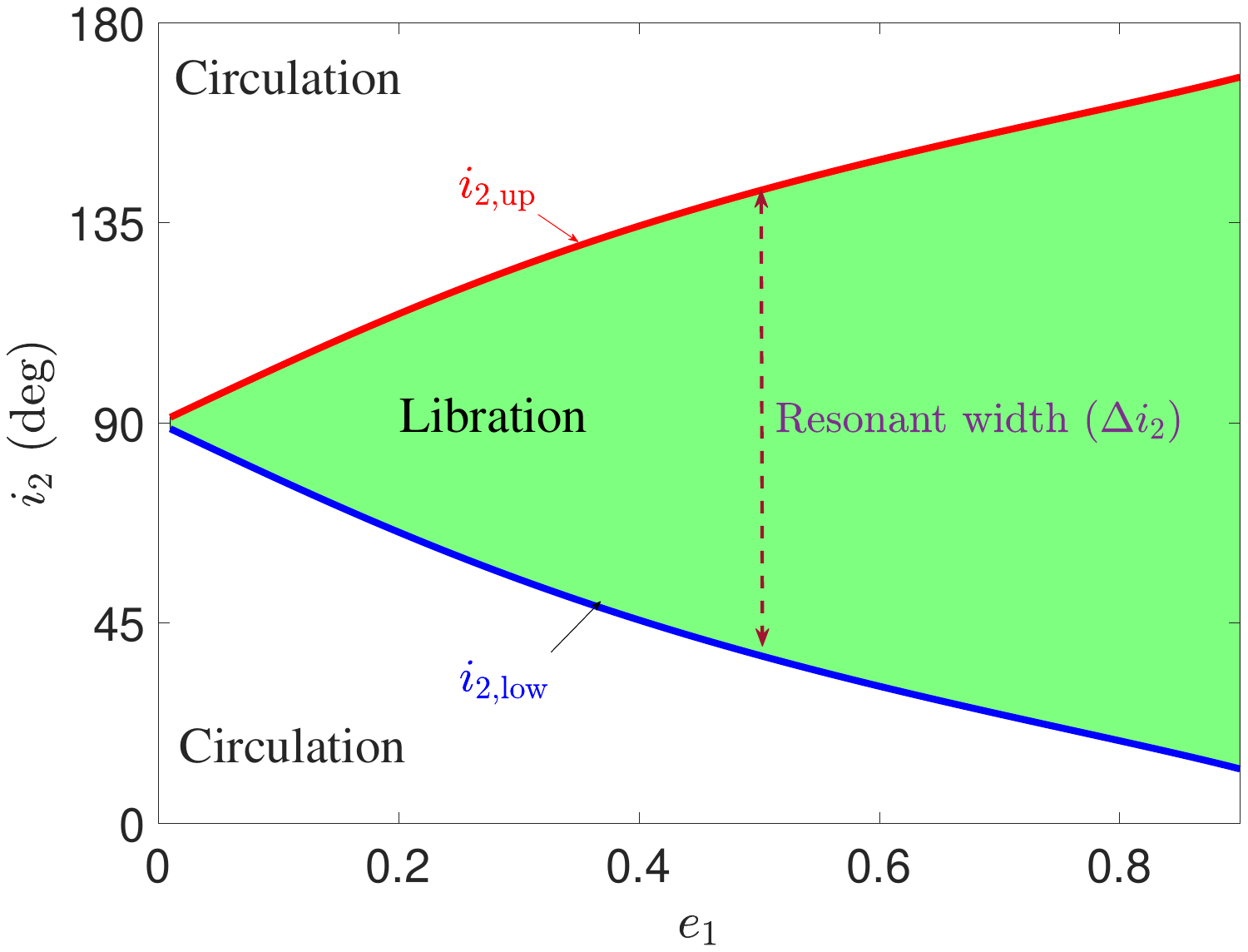}
\caption{Distribution of nodal libration and circulation regions in the $(i_2,e_2)$ space for the case of $e_1 = 0.5$ (left panel) and in the $(e_1,i_2)$ space for arbitrary planetary eccentricities (right panel) with the longitude of ascending node at $2h_2=\pi$ (libration centre). The shaded region is of libration and the white regions are of circulation. In both panels, the resonant width denoted by $\Delta i_2$ represents the distance of test particle's inclination between separatrices, evaluated at the libration centre. In the left panel, resonant width $\Delta i_2$ is plotted as a function of particle's eccentricity $e_2$ and in the right panel it is plotted as a function of the perturber's eccentricity $e_1$ (right panel). It is observed that the resonant width is independent on planetary eccentricity $e_2$ (see the left panel), while it is an increasing function of binary eccentricity $e_1$ (see the right panel). }
\label{Fig8}
\end{figure*}

Fig. \ref{Fig8} shows distribution of nodal libration and circulation regions in the $(i_2,e_2)$ space for the case of $e_1 = 0.5$ (see the left panel) and in the $(e_1,i_2)$ space for arbitrary binary's eccentricities (see the right panel). It is observed that the resonant width $\Delta i_2$ is independent on $e_2$ (see the left panel) and it is an increasing function of $e_1$ (see the right panel), which is in agreement with the prediction of equation (\ref{Eq22-1}). In particular, when the binary moves on a circular orbit ($e_1=0$), the nodal resonance disappears because of $\Delta i_2 = 0$ in this case.

It is interesting to note that the distribution of libration and circulation regions in the $(i_2,e_2)$ space (see the left panel of Fig. \ref{Fig8}) is similar to that of the conventional ZLK resonance of the inner test particles in the $(i_1,e_1)$ space \citep{lei2021structures}. However, the dynamics is different. For the case of inner test particles, the resonant argument is $2\omega_1$ (ZLK oscillations), the location of separatrix in the $(i_1,e_1)$ space is fixed at $39.8^{\circ}$ and $140.2^{\circ}$ and the motion integral is the $z$-component of particle's angular momentum $H=\sqrt{1-e_1^2}\cos{i_1}$ \citep{lidov1962evolution,kozai1962secular}. For the case of outer test particles (the topic in this work), the resonant argument is $2\Omega_2$ (nodal oscillation), the location of separatrix in the $(i_2,e_2)$ space is dependent on binary's eccentricity $e_1$ and the motion integral is the test particle's eccentricity $e_2$ \citep{naoz2017eccentric}.

To conclude, we can see that the dynamics of quadrupole-order model is determined by two parameters: the binary eccentricity $e_1$ and planetary inclination $i_2$. This is the reason that we discuss dynamical structures in the $(i_2,e_1)$ space in this work. 

\subsection{Web of resonances}
\label{Sect5-4}

Considering the quadrupole-order Hamiltonian model is integrable, we could introduce Arnold action-angle variables \citep{morbidelli2002modern}
\begin{equation}\label{Eq25}
\begin{aligned}
g_2^* &= {g_2} - {\rho _g}\left( {t,G_2^*,H_2^*} \right){\rm{ = }}g_2^*(0) + \frac{{2\pi }}{{{T_g}}}t,\\
G_2^* &= {G_2},\\
h_2^* &= {h_2} - {\rho _h}\left( {t,G_2^*,H_2^*} \right){\rm{ = }}h_2^*(0) + \frac{{2\pi }}{{{T_h}}}t,\\
H_2^* &= \frac{1}{{2\pi }}\oint {{H_2}{\rm d}{h_2}}
\end{aligned}
\end{equation}
which is a canonical transformation with the generating function,
\begin{equation*}
{\cal S}\left( {{g_2},{h_2},G_2^*,H_2^*} \right) = {h_2}H_2^* + \int {{H_2}{\rm d}{h_2}}
\end{equation*}
leading to the mutual transformation
\begin{equation*}
\begin{aligned}
{G_2} &= \frac{{\partial {\cal S}}}{{\partial {g_2}}},\quad g_2^* = \frac{{\partial {\cal S}}}{{\partial G_2^*}}  = \frac{\partial }{{\partial G_2^*}}\int {{H_2}{\rm d}{h_2}} ,\\
{H_2} &= \frac{{\partial {\cal S}}}{{\partial {h_2}}},\quad h_2^* = \frac{{\partial {\cal S}}}{{\partial H_2^*}} = {h_2} + \frac{\partial }{{\partial H_2^*}}\int {{H_2}{\rm d}{h_2}}.
\end{aligned}
\end{equation*}
In the transformation given by equation (\ref{Eq25}), $H_2^*$ is called Arnold action and it requires $h_2^*(0) = 0$. Without loss of generality, we assume that the state at the initial moment is $h_2(0) = \pi/2$ (it further requires $H_2(0)>0$ for librating cycles). $T_g$ and $T_h$ are periods of $g_2$ and $h_2$ under the quadrupole-order Hamiltonian model. ${\rho _g}\left( {t,G_2^*,H_2^*} \right)$ and ${\rho _h}\left( {t,G_2^*,H_2^*} \right)$ are periodic functions with zero average and have the same period of nodal cycle $T_h$ \citep{henrard1990semi}. It requires that the initial angles $g_2 (0)$ and $g_2^*(0)$ should be coincident, showing that ${\rho _g}\left( t=0 \right) = {\rho _g}\left( t=T_h \right) = 0$.

In order to compute Arnold action $H_2^*$, we could integrate the following set of differential equation over one period of $h_2$,
\begin{equation*}
\left\{ \begin{aligned}
{{\dot h}_2} &= \frac{{\partial {{\cal H}}}}{{\partial {H_2}}},\\
{{\dot H}_2} &=  - \frac{{\partial {{\cal H}}}}{{\partial {h_2}}},\\
\dot H_2^* &= \frac{1}{{2\pi }}{H_2}\frac{{\partial {{\cal H}}}}{{\partial {H_2}}}.
\end{aligned} \right.
\end{equation*}
starting from the initial condition at $(h_{2,0}=\pi/2,H_{2,0},H_{2,0}^* = 0)$.

\begin{figure*}
\centering
\includegraphics[width=\columnwidth]{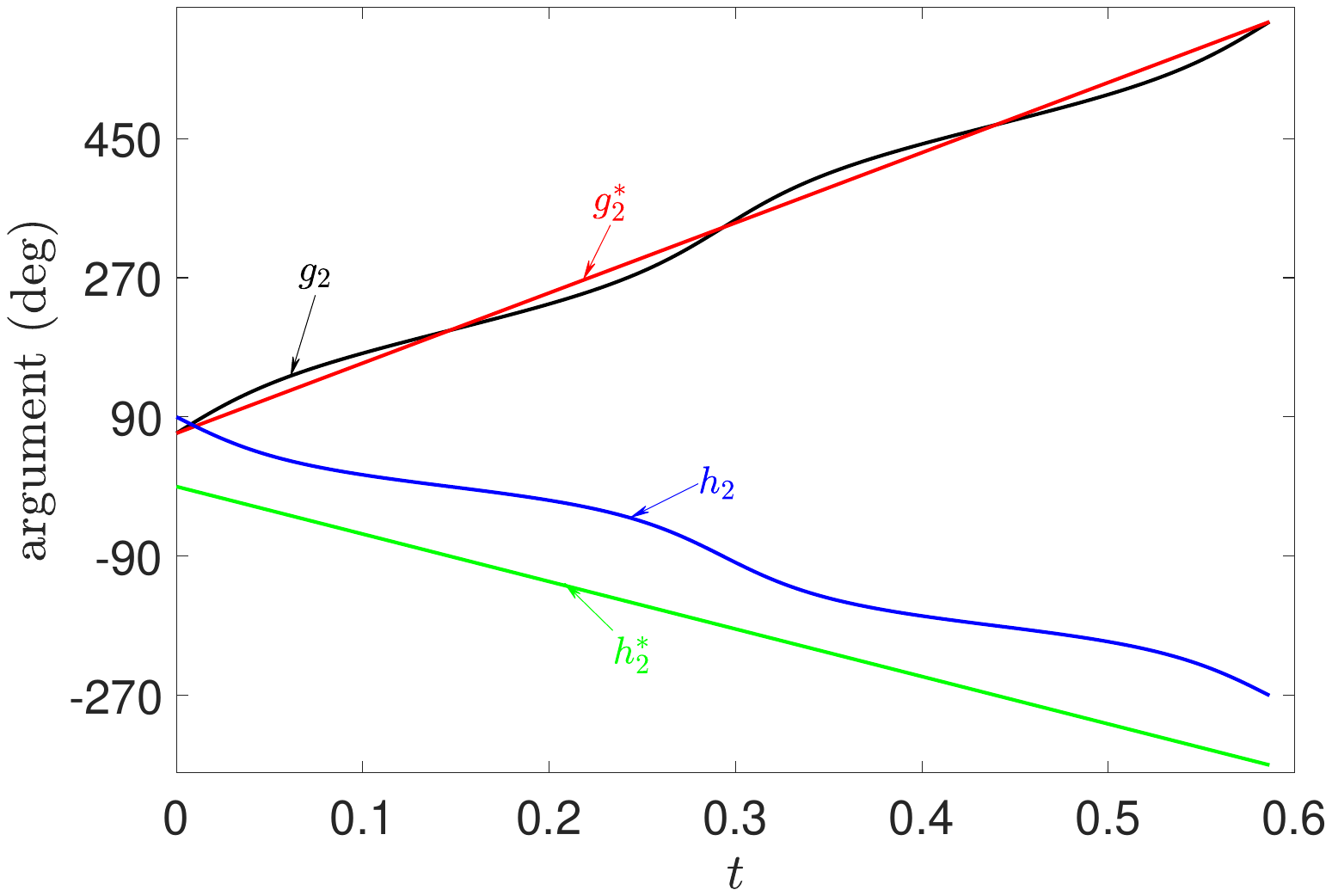}
\includegraphics[width=\columnwidth]{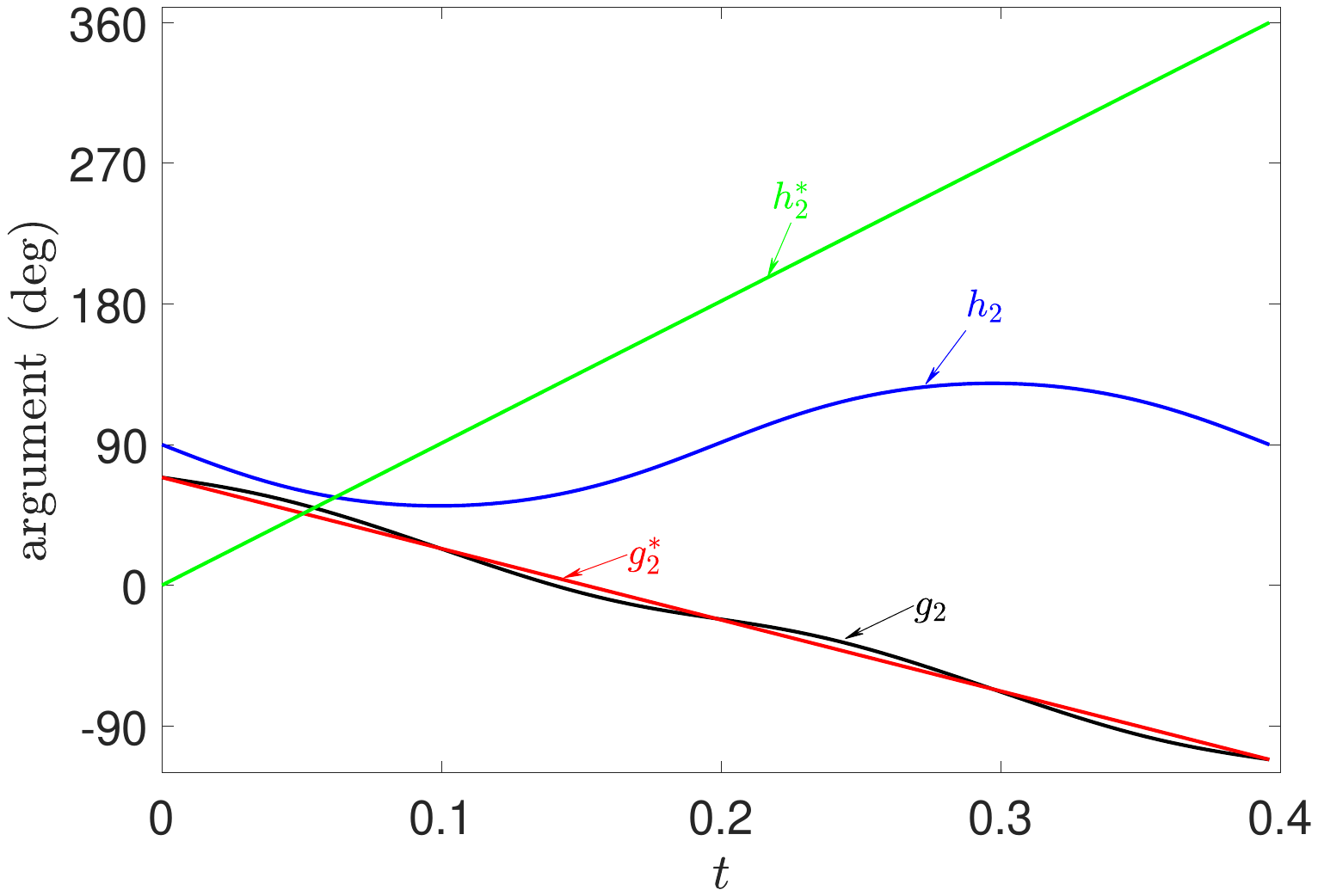}
\caption{Relationship between the original variables $g_2$ and $h_2$ as well as the associated transformed variables $g_2^*$ and $h_2^*$ for the nodal circulating case (left panel) and for the nodal librating case (right panel).}
\label{Fig9}
\end{figure*}

Performing the following transformation
\begin{equation*}
\left( {{g_2},{h_2},{G_2},{H_2}} \right) \leftrightarrow \left( {g_2^*,h_2^*,G_2^*,H_2^*} \right)
\end{equation*}
it is possible to write the quadrupole-level Hamiltonian as
\begin{equation}\label{Eq26}
{{\cal H}}\left( {{h_2},{G_2},{H_2}} \right) \Rightarrow  {{\cal H}}\left( {G_2^*,H_2^*} \right).
\end{equation}
The Hamiltonian, which is independent on action variables, is referred to as the normal form. Under the set of action-angle variables, the angles $(g_2^*,h_2^*)$ are cyclic variables, showing that the action variables $(G_2^*,H_2^*)$ are motion integral. As a result, $(g_2^*,h_2^*)$ are linear functions of time with fundamental frequencies of $2\pi/T_g$ and $2\pi/T_h$ (see equation \ref{Eq25}). Please refer to Fig. \ref{Fig9} for the relationship between $(g_2,h_2)$ and $(g_2^*,h_2^*)$ for the circulating cycle (see the left panel) and the librating cycle (see the right panel). As expected, the original angles are nonlinear functions of time, while the variables after transformation are linear functions of time. 

Applying the normal form of Hamiltonian to the canonical relation yields the fundamental frequencies as follows:
\begin{equation}\label{Eq27}
\begin{aligned}
\dot g_2^* = \frac{{\partial {{\cal H}}}}{{\partial G_2^*}},\\
\dot h_2^* = \frac{{\partial {{\cal H}}}}{{\partial H_2^*}}.
\end{aligned}
\end{equation}
The fundamental frequencies determines the nominal location of resonances, 
\begin{equation}\label{Eq28}
{k_1}\dot h_2^* + {k_2}\dot g_2^* = 0
\end{equation}
where $k_1 \in \mathbb{N}$ and $k_2 \in \mathbb{Z}$.  

\begin{figure*}
\centering
\includegraphics[width=2\columnwidth]{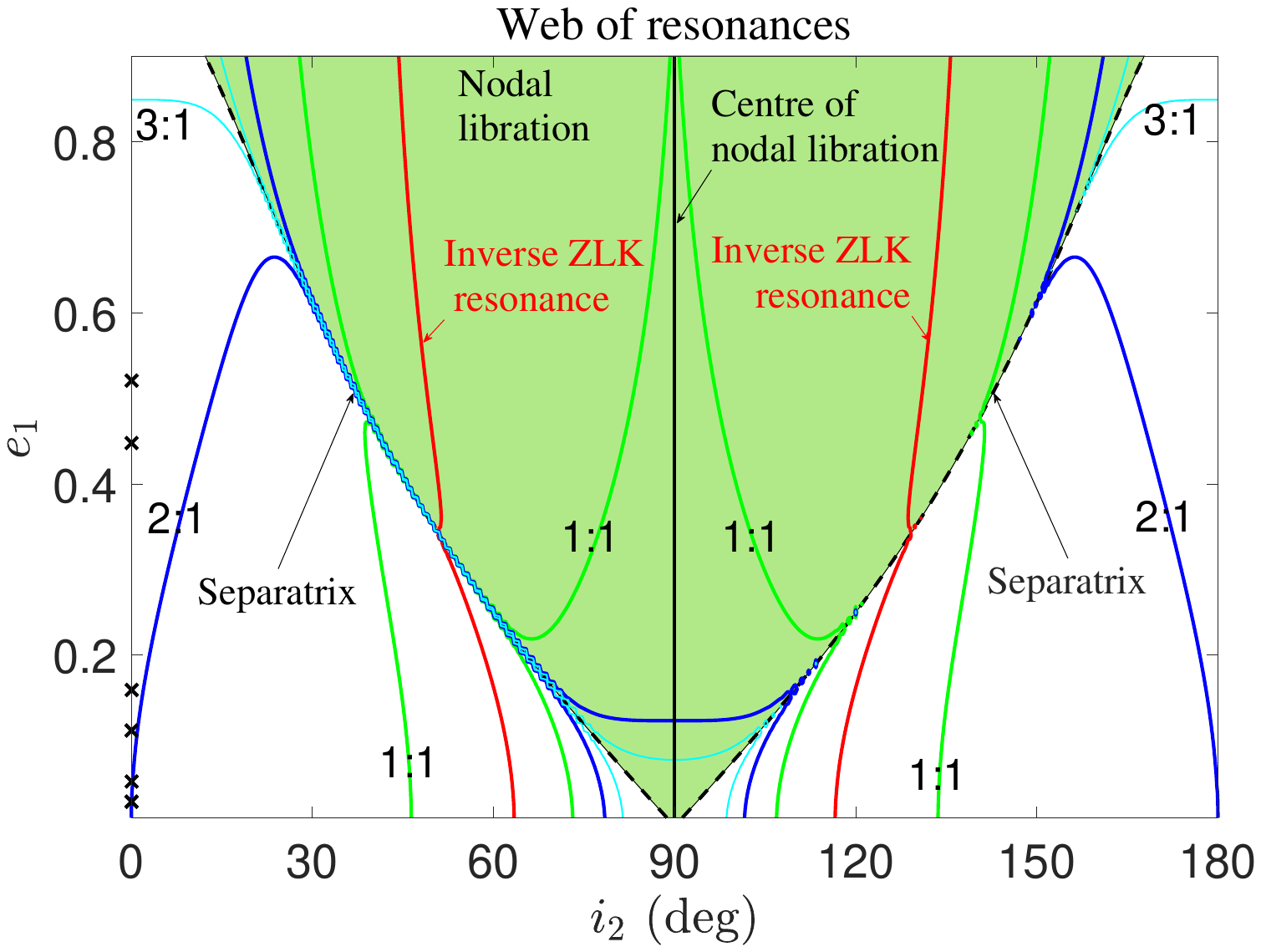}
\caption{Web of resonances, standing for the distribution of nominal location of high-order and secondary resonances shown in the $(i_2,e_1)$ space. The shaded region represents the parameter space where the nodal resonance takes place (it is similar to the right panel of Fig. \ref{Fig8}) and the white regions are of nodal circulation. Secondary resonances happen inside the region of nodal resonance and high-order resonances takes place outside the region of nodal resonance. The red lines stand for the inverse ZLK resonance. Black crosses represent the location of observed CBPs (see Table \ref{Table1} for their detailed parameters).}
\label{Fig10}
\end{figure*}

Fig. \ref{Fig10} presents the resonance curves with $|k_2|=1$ and $k_1=\{0,1,2,3\}$ in the $(i_2,e_1)$ space. The shaded zone stands for the nodal libration region, which is the same as the right panel of Fig. \ref{Fig8}. Inside the nodal libration region, secular resonances between $g_2^*$ and $h_2^*$ are referred to as secondary resonances and the ones outside the nodal libration region are called high-order resonances. In particular, the resonance with $k_1 = 0$ is the so-called inverse ZLK resonance with $\sigma = \omega_2$ as the resonant argument \citep{de2019inverse,vinson2018secular}. According to the Hamiltonian given by equation (\ref{Eq3}), we can see that the inverse ZLK resonance is of hexadecapolar order. In particular, when the binary's eccentricity is zero, we can see that the inclinations of inverse ZLK resonance are equal to $i_2 = 63.4^{\circ}$ and $i_2 = 116.6^{\circ}$, which is consistent with the critical inclinations derived by \citet{gallardo2012survey} and \citet{de2019inverse}.

It should be noted that the distribution of resonant curves shown here is representative because it is independent on the eccentricity of particles $e_2$, the mass parameter of the inner binary $\mu_b$ and semimajor axis ratio $\alpha = a_1/a_2$. Black crosses represent the location of observed CBPs with the semimajor axis ratio smaller than 1/3. See Table \ref{Table1} for their detailed parameters. We can see that all the observed CBPs are in nearly co-planar configurations, and thus they are weakly influenced by high-order and/or secondary resonances.

From Fig. \ref{Fig10}, we could know the nominal location of resonances. However, there is no resonant strength under the quadrupole-level approximation. Only if the octupole-order and hexacadepolar-order terms are included, the associated curves of high-order or secondary resonances may be replaced by a certain zone of libration around the nominal curve of resonance and the area of libration zone is proportional to the resonance strength. As expected, for the unequal-mass binary, the octupole-order resonances would have higher strength and the hexacadepolar-order resonances would hold weaker strength. However, it is different for the equal-mass case where all the resonances are of hexacadepolar order because of disappearance of octupole-order Hamiltonian, thus it is expected that all of them have comparable strength. This is in agreement with the dynamical maps shown in Fig. \ref{Fig4}.

\begin{table*}
\begin{center}	
\caption{Parameters of CBPs shown in Fig. \ref{Fig10}, where the semimajor axis ratio is smaller than 1/3. Here $i_2$ is the relative inclination of planets with respect to the orbit of the inner binary.} \label{Table1}
\vspace{0.1 cm}
{\begin{tabular}{c c c c c c c c c c}\hline\hline\\
System & $m_1 (m_{\odot})$ & $m_2 (m_{\odot})$ & $m_2 (m_J) $ & $i_2 (^{\circ})$ & $a_1(\rm au) $ & $a_2(\rm au)$ & $e_1$ & $e_2$ & Ref. \\
\hline
 Kepler-34  & 1.0479 & 1.0208 & 0.22 & 1.81 & 0.22882 & 1.0896 & 0.52087 & 0.182 & \citet{welsh2012transiting} \\
 Kepler-47 (c) & 0.957 & 0.342 & 0.05984 & 1.165 & 0.08145 & 0.6992 & 0.0288 & 0.024 & \citet{orosz2012kepler} \\
 Kepler-47 (d)  & 0.957 & 0.342 & 0.00997 & 1.38 & 0.08145 & 0.9638 & 0.0288 & 0.044 & \citet{orosz2019discovery}\\
 Kepler-453   & 0.944 & 0.1951 & 0.05 & 2.258 & 0.18539 & 0.7903 & 0.0524 & 0.0359 & \citet{welsh2015kepler} \\
 Kepler-1647  & 1.21 & 0.975 & 1.52 & 2.9855 & 0.1276 & 2.7205 & 0.1593 & 0.0581 & \citet{kostov2016kepler} \\
 Kepler-1661  & 0.841 & 0.262 & 0.053 & 0.93 & 0.187 & 0.633 & 0.112 & 0.057 & \citet{socia2020kepler} \\
 TIC 172900988  & 1.2388 & 1.2023 & 2.74 & 1.45 & 0.191928 & 0.89809 & 0.44793 & 0.088 & \citet{kostov2021tic} \\
\hline
\end{tabular}}
\end{center}
\end{table*}

\section{Applications to numerical structures}
\label{Sect6}

In this section, resonance curves associated with high-order and secondary resonances between $g_2^*$ and $h_2^*$ discussed in the previous section are utilised to explain dynamical structures produced under the hexadecapolar-order Hamiltonian model.

\subsection{Nodal and Inverse ZLK resonances}
\label{Sect6-1}

The nodal resonance with argument of $\sigma = 2\Omega_2$ is of quadrupole order and, without doubt, it dominates the main structure. The inverse ZLK resonance has been studied in \citet{vinson2018secular} and \citet{de2019inverse} under the hexadecapolar-level approximation. The resonant argument is $\sigma = \omega_2$ and its strength is of hexadecapolar order. It is much weaker than the conventional ZLK resonance which is of quadrupole order in strength \citep{lidov1962evolution,kozai1962secular}.

\begin{figure*}
\centering
\includegraphics[width=\columnwidth]{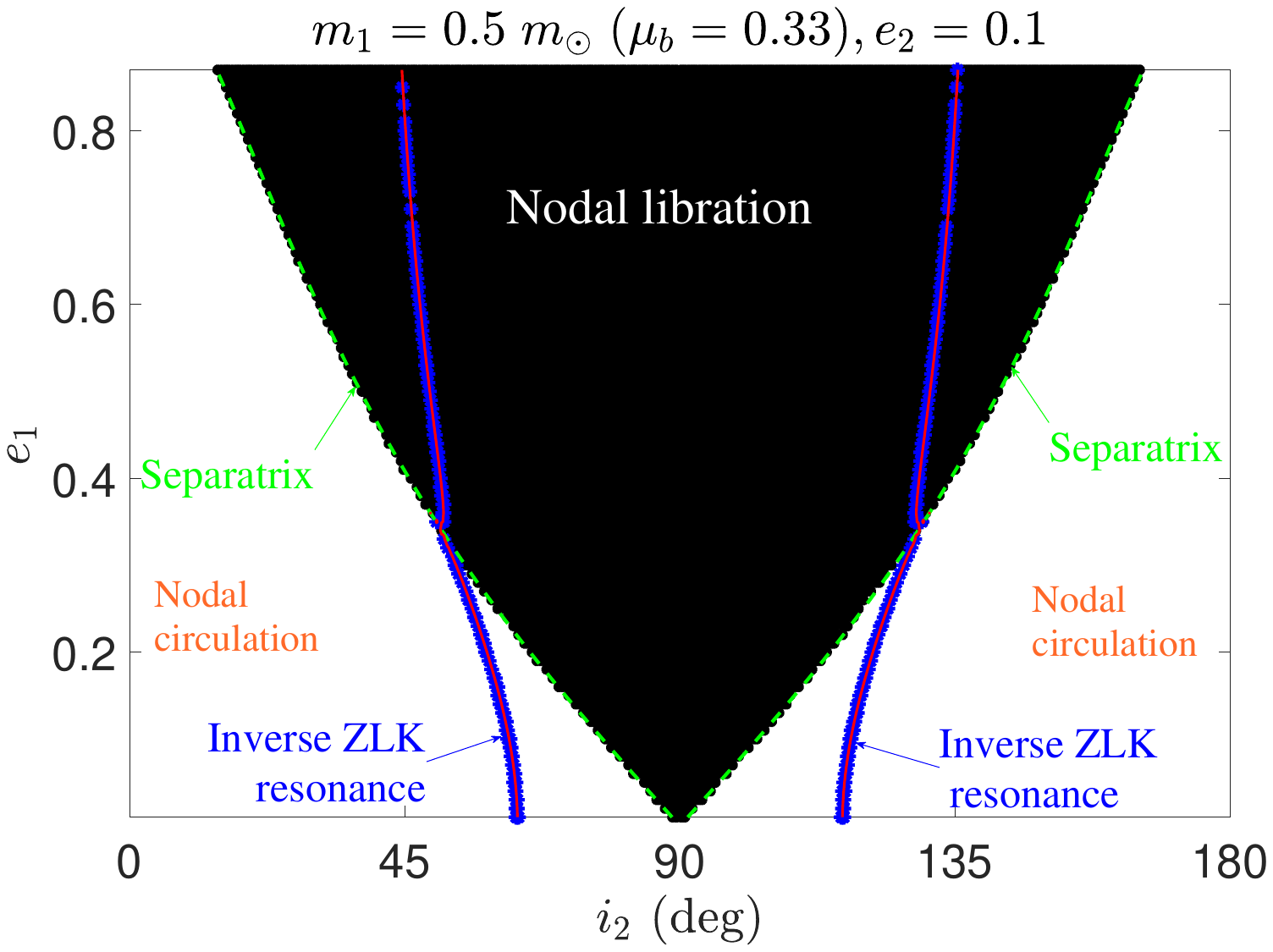}
\includegraphics[width=\columnwidth]{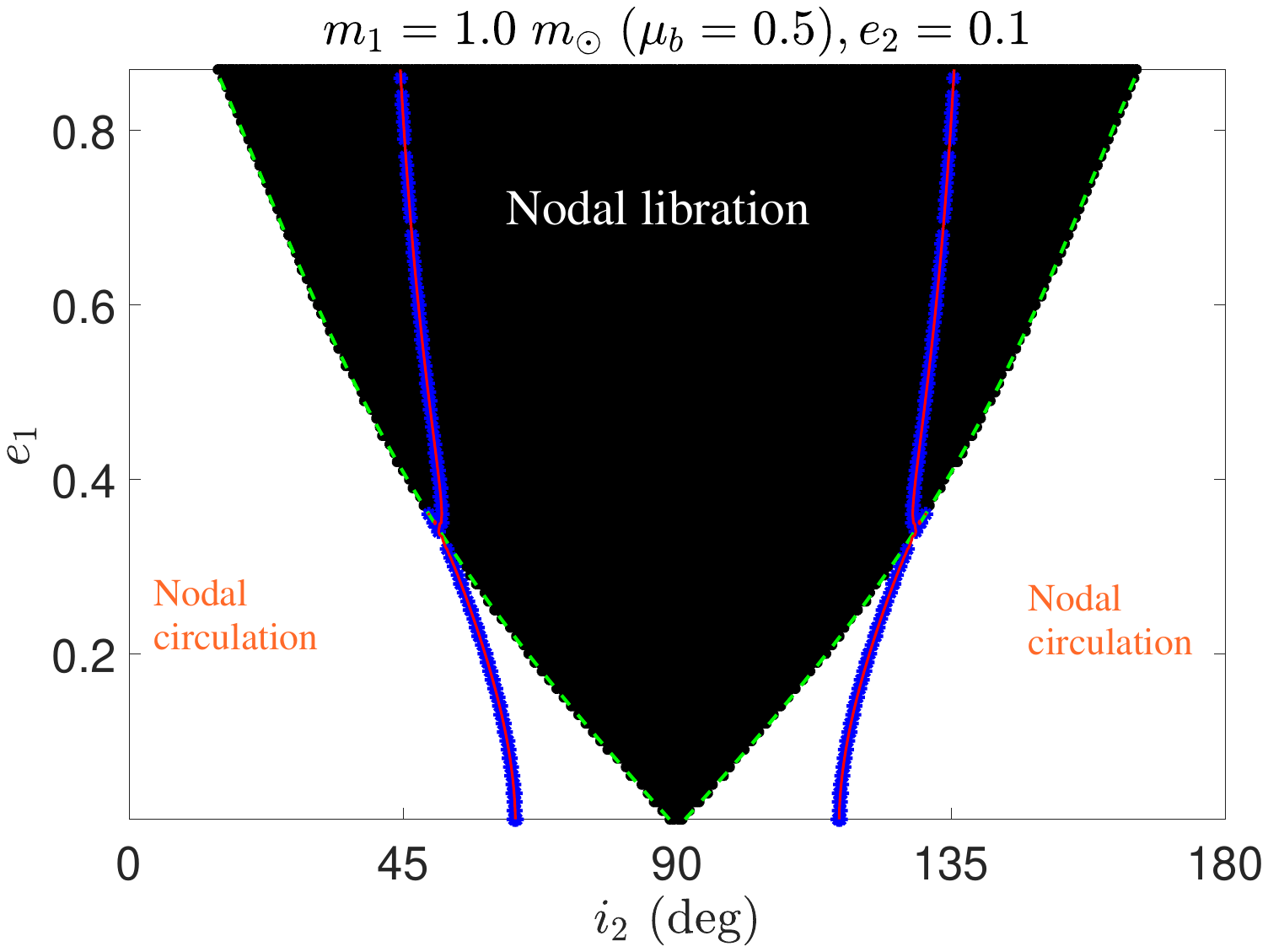}
\caption{Numerical distribution of the nodal resonance (black dots) and inverse ZLK resonance (blue dots), determined under the hexadecapolar-order dynamical model. The left panel is for $m_1 = 0.5m_{\odot}$ representing the case of unequal-mass binary and the right panel is for $m_1 = 1.0m_{\odot}$ representing the case of equal-mass binary. The red lines stand for the nominal location of inverse ZLK resonance (similar to Fig. \ref{Fig6}), and the green dashed lines stand for the dynamical separatrices of nodal resonance.}
\label{Fig11}
\end{figure*}

Here we numerically explore dynamical structures caused by nodal resonance and inverse ZLK resonance. To produce the numerical structures, the initial angles are assumed at $g_{2,0} = \pi/2$ and $h_{2,0}=\pi/2$ and the secular equations of motion at the hexadecapolar-level approximation are numerically integrated over 10 nodal periods. The points with libration of $h_2$ and/or $g_2$ are recorded. Numerical structures are shown in Fig. \ref{Fig11}. In practical simulations, $m_1 = 0.5 m_{\odot}$ is taken for the unequal-mass case, and $m_1 = 1.0 m_{\odot}$ is taken for the equal-mass case.

In Fig. \ref{Fig11}, black dots stand for nodal resonance and blue dots represent the inverse ZLK resonance. For comparison, the nominal location of inverse ZLK resonance and dynamical separatrix of nodal resonance are also plotted. See the green dashed lines for dynamical separatrix and the solid red lines for nominal location of inverse ZLK resonance. It is observed that (a) there are qualitatively similar numerical structures caused by nodal resonance and inverse ZLK resonance with different parameters of $\mu_b$, (b) numerical structures are nearly symmetric with respect to $i_2 = 90^{\circ}$, meaning that nodal and inverse ZLK resonances will not lead to symmetry breaking, (c) the distribution of nodal resonance in the $(i_2,e_1)$ space presents the V-shape structure bounded by the dynamical separatrix of nodal resonance, which is dominated by the quadrupole-order dynamics discussed in the previous section, (d) the inverse ZLK resonance can happen in almost the entire $e_1$ space, which is in agreement with the discussion given by \citet{de2019inverse}, (e) for a given $e_1$ the libration zone of inverse ZLK resonance is very narrow because it is of hexadecapolar order, (f) the inverse ZLK resonance happens outside the nodal libration region if the binary's eccentricity $e_1$ is smaller than $\sim$$0.35$, else it happens inside the nodal libration region, and (g) excellent agreement can be observed between the analytical and numerical results for the structures.

\subsection{Dynamical structures}
\label{Sect6-2}

\begin{figure*}
\centering
\includegraphics[width=\columnwidth]{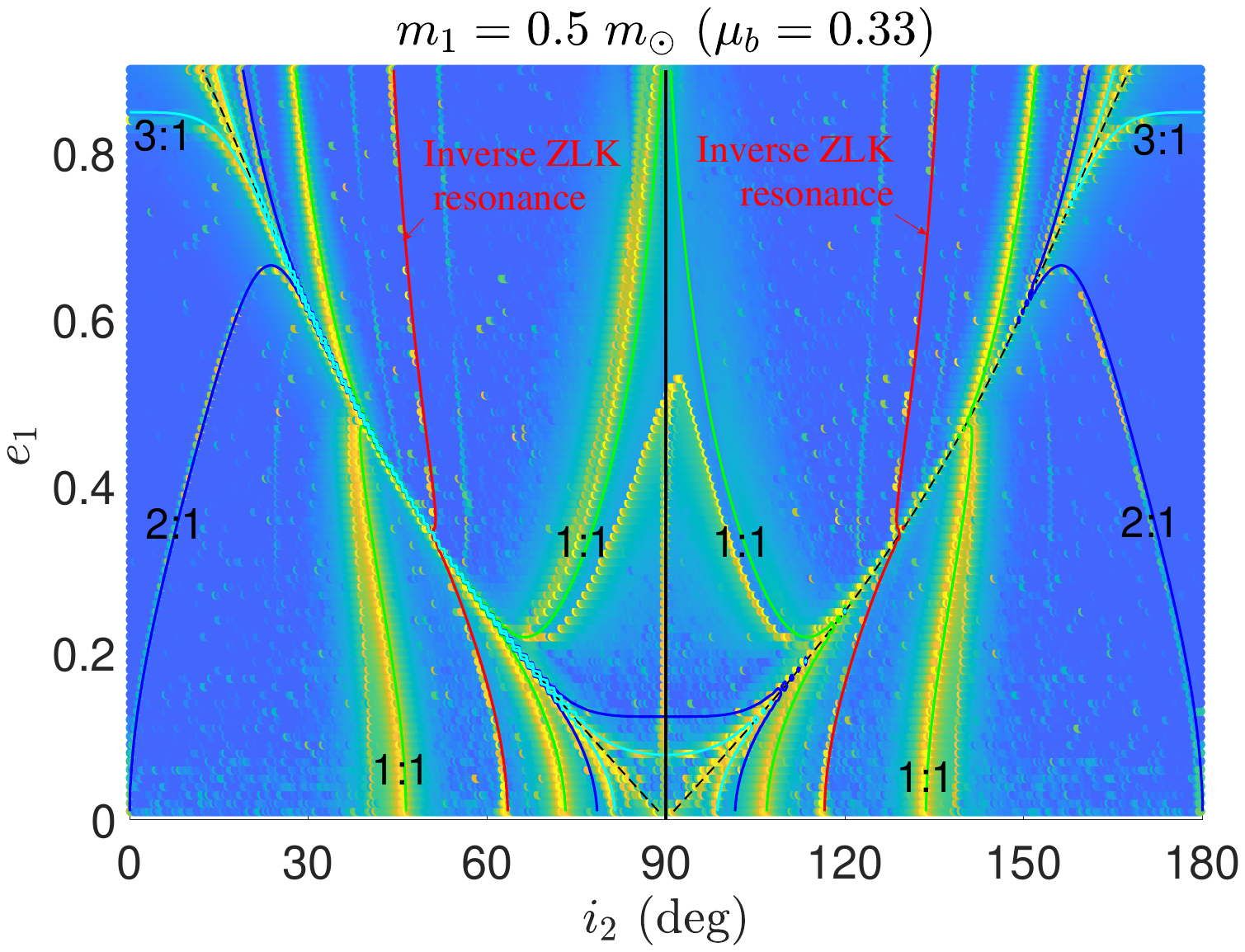}
\includegraphics[width=\columnwidth]{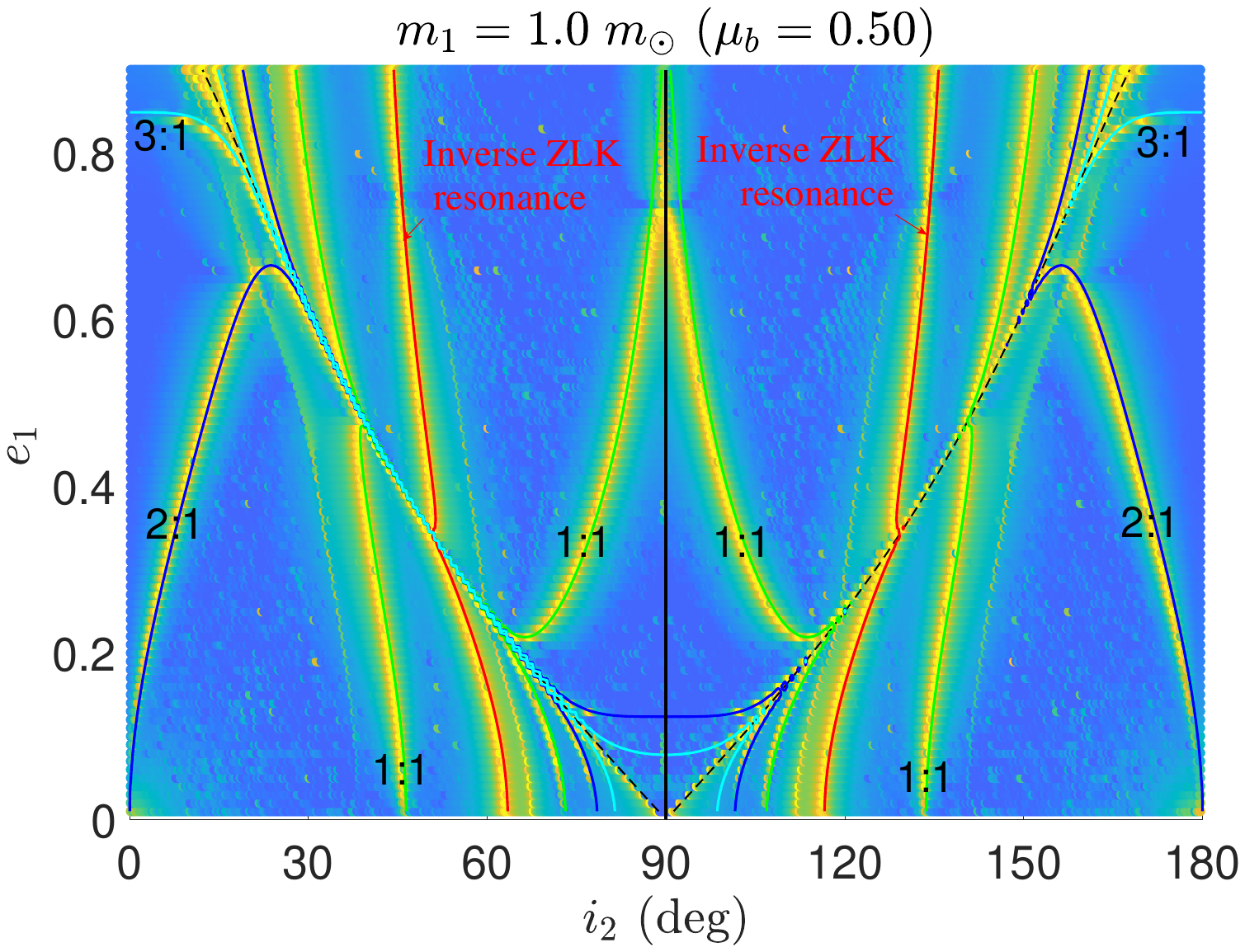}
\caption{The nominal location of nodal resonance, high-order resonance as well as secondary resonances are plotted together with dynamical maps. The left panel is for the case of $m_1=0.5 m_{\odot}$ and the right panel is for the case of $m_1=1.0 m_{\odot}$. Excellent agreement can be observed between the numerical structures arising in dynamical maps and resonance curves determined under the quadrupole-level approximation.}
\label{Fig12}
\end{figure*}

Dynamical maps corresponding to $m_1=0.5 m_{\odot}$ (standing for unequal-mass configuration) and $m_1=1.0 m_{\odot}$ (standing for equal-mass configuration) together with web of resonances are shown in Fig. \ref{Fig12}. For convenience, dynamical separatrix of nodal resonance is marked by black dashed lines, which form the main V-shape structure in the $(i_2,e_1)$ space. 

Fig. \ref{Fig12} shows that there is a perfect correspondence between dynamical structures and resonance curves. It is observed that (a) the main V-shape structure is dominated by the nodal resonance which is of quadrupole order in strength, (b) the structures inside the nodal libration region are governed by secondary resonances, and (c) the structures outside the nodal libration region are dominated by high-order secular resonances. In the unequal-mass case, the 1:1 resonance is of octupole order and the remaining resonances are of hexadecapolar order. In the equal-mass case, all the high-order and secondary resonances are of hexadecapolar order. 

As shown in Table \ref{Table1}, CBPs currently detected so far are almost co-planar to the binary orbit. This is due to the fact that misaligned CBPs are more difficult to detect. However, in recent years polar-aligned circumbinary gas disk (HD 98800 B) and debris disk (99 Herculis) have been observed around eccentric binaries \citep{kennedy2019circumbinary, kennedy201299, smallwood2020formation}, indicating that it is possible for formation of polar planets. In addition, it is more likely to form CBPs on polar orbits than on co-planar orbits, predicting that polar-aligned terrestrial planets may be found in the future \citep{childs2021formation}. In the polar region, an unexpected symmetry breaking is found between prograde and retrograde configurations around eccentric binary \citep{cuello2019planet}. 

If we focus on the structures inside the polar region, we can further find from Fig. \ref{Fig12} that the dynamics is dominated by the nodal resonance when $e_1 < 0.22$ and it is governed by the nodal resonance in combination with the secondary 1:1 resonance when $e_1 > 0.22$. In addition, we can observe that the culprit causing symmetry breaking of dynamical structures inside polar region is the secondary 1:1 resonance.

\section{Secondary resonances}
\label{Sect7}

It is observed from Fig. \ref{Fig5} that (a) in the configuration of $e_1 = 0.1$ there are islands of libration associated with high-order resonances (including 1:1, 2:1 and 3:1 resonances located outside the nodal libration region) as well as secondary 3:1 resonance (inside the nodal libration region) and (b) in the configuration of $e_1=0.3$ all the motion happen inside the nodal libration region and there are two islands of libration, corresponding to the 1:1 bifurcation. In this section, we will analytically study secondary resonances inside the nodal libration zone by means of perturbative treatments. 

\subsection{Perturbative treatment}
\label{Sect7-1}

Under the set of action-angle variables $(g_2^*,h_2^*,G_2^*,H_2^*)$ defined by equation (\ref{Eq25}), the hexadecapolar-order Hamiltonian can be expressed as
\begin{equation}\label{Eq29}
{\cal H}\left( {g_2^*,h_2^*,G_2^*,H_2^*} \right) = {{\cal H}_0}\left( {G_2^*,H_2^*} \right) + \varepsilon {{\cal H}_1}\left( {g_2^*,h_2^*,G_2^*,H_2^*} \right)
\end{equation}
which is composed of the unperturbed Hamiltonian ${{\cal H}_0}\left( {G_2^*,H_2^*} \right)$ corresponding to the quadrupole-order component and the perturbation Hamiltonian ${{\cal H}_1}\left( {g_2^*,h_2^*,G_2^*,H_2^*} \right)$ corresponding to the octupole- and hexadecapolar-order components. In general, the perturbation part is much smaller than the unperturbed term, showing that such a Hamiltonian model can be treated by means of perturbation theory. In this section, we will take advantage of perturbation theories proposed by \citet{wisdom1985perturbative} and \citet{henrard1986perturbation} to study the resonant dynamics. It should be mentioned that similar approach of Henrard's method is adopted by \citet{li2014analytical} to study dynamics of low-inclination and nearly polar CBPs. In particular, excitation of planetary eccentricity in near coplanar configurations is analytically discussed in \citet{moriwaki2004planetesimal}, \citet{leung2013analytic} and \citet{li2014analytical}.

According to Sect. \ref{Sect6}, it is shown that the unperturbed Hamiltonian ${{\cal H}_0}\left( {G_2^*,H_2^*} \right)$ produces the fundamental frequencies and their integer commensurability determines the nominal location of resonances between $g_2^*$ and $h_2^*$. To study the dynamics of $k_1$:$k_2$ resonance, it is convenient for us to introduce the following transformation,
\begin{equation}\label{Eq30}
\begin{aligned}
{\sigma _1} &= h_2^* + \frac{{{k_2}}}{{{k_1}}}g_2^*,\quad {\Sigma _1} = H_2^*,\\
{\sigma _2} &= g_2^*,\quad {\Sigma _2} = G_2^* - \frac{{{k_2}}}{{{k_1}}}H_2^*
\end{aligned}
\end{equation}
which is canonical with the generating function
\begin{equation*}
{\cal S}\left( {g_2^*,h_2^*,{\Sigma _1},{\Sigma _2}} \right) = h_2^*{\Sigma _1} + g_2^*\left( {\frac{{{k_2}}}{{{k_1}}}{\Sigma _1} + {\Sigma _2}} \right).
\end{equation*}
In terms of the canonical variables $(\sigma_1,\sigma_2,\Sigma_1,\Sigma_2)$, the hexadecapolar-order Hamiltonian can be written as
\begin{equation}\label{Eq31}
{\cal H}\left( {{\sigma _1},{\sigma _2},{\Sigma _1},{\Sigma _2}} \right) = {{\cal H}_0}\left( {{\Sigma _1},{\Sigma _2}} \right) + \varepsilon {{\cal H}_1}\left( {{\sigma _1},{\sigma _2},{\Sigma _1},{\Sigma _2}} \right).
\end{equation}
When a circumbinary planet is located inside the $k_1$:$k_2$ resonance, it holds $k_1 \dot h_2^* + k_2 \dot g_2^* \approx 0$, meaning that $\dot \sigma_1 \approx 0$. As a result, the time derivatives of $\sigma_1$ and $\sigma_2$ satisfy the hierarchical relation $\dot \sigma_1 \ll  \dot \sigma_2$, showing that equation (\ref{Eq31}) determines a separable Hamiltonian model \citep{henrard1990semi}. In particular, $(\sigma_1,\Sigma_1)$ stands for the slow degree of freedom and $(\sigma_2,\Sigma_2)$ stands for the fast degree of freedom. According to Wisdom's perturbation theory \citep{wisdom1985perturbative}, during the timescale of the fast degree of freedom (for example, one period of $\sigma_2$), the slow variables $(\sigma_1,\Sigma_1)$ have negligible change and thus they can be approximated as parameters. Under the assumption with $(\sigma_1,\Sigma_1)$ as parameters, equation (\ref{Eq31}) reduces to an integrable Hamiltonian model with $(\sigma_2,\Sigma_2)$ as phase-space variables. Regarding the integrable Hamiltonian, it is possible to introduce action-angle variables as follows \citep{morbidelli2002modern}:
\begin{equation}\label{Eq32}
\sigma _2^* = \frac{{2\pi }}{T}t,\quad \Sigma _2^* = \frac{1}{{2\pi }}\int\limits_0^{2\pi } {{\Sigma _2}{\rm d}{\sigma _2}}
\end{equation}
where $T$ is the period of $\sigma_2$. Under the set of action-angle variables $(\sigma_2^*,\Sigma_2^*)$, the hexadecapolar-order Hamiltonian becomes
\begin{equation}\label{Eq33}
{\cal H}\left( {{\sigma _1},{\Sigma _1},{\Sigma _2^*}} \right) = {{\cal H}_0}\left( {{\Sigma _1},{\Sigma _2^*}} \right) + \varepsilon {{\cal H}_1}\left( {{\sigma _1},{\Sigma _1},{\Sigma _2^*}} \right)
\end{equation}
where $\sigma _2^*$ is a cyclic variable, indicating that its conjugate momentum $\Sigma _2^*$ is a motion integral, which remains constant during the long-term evolution. Such a motion integral $\Sigma _2^*$ is usually referred to as the adiabatic invariant \citep{henrard1990semi}. For convenience, we denote the adiabatic invariant as
\begin{equation}\label{Eq34}
S = \Sigma _2^* = \frac{1}{{2\pi }}\int\limits_0^{2\pi } {{\Sigma _2}{\rm d}{\sigma _2}}
\end{equation}
which measures the path integration of the phase curve in the $(\sigma_2,\Sigma_2)$ space (divided by $2\pi$).
  
For the two-degree-of-freedom model governed by the Hamiltonian (\ref{Eq31}), there are two conserved quantities during the long-term evolution: (a) the Hamiltonian $\cal H$ itself and (b) the adiabatic invariant $S$. As a result, the dynamical model becomes approximately integrable.

\subsection{Phase-space structures}
\label{Sect7-2}

Regarding the integrable two-degree-of-freedom Hamiltonian model, the phase-space structures can be be explored by analysing phase portraits: (a) plotting level curves of the Hamiltonian $\cal H$ under a given adiabatic invariant $S$ or (b) plotting level curves of the adiabatic invariant $S$ under a given Hamiltonian $\cal H$. Both versions of phase portraits are equivalent. The former definition of phase portraits is adopted in \citet{saillenfest2016long} and \citet{saillenfest2020long}, and the latter definition of phase portraits is taken in \citet{lei2022systematic} and \citet{lei2022zeipel}.

For the current problem, the Hamiltonian $\cal H$ is easy to calculate because it has an explicit expression, while the adiabatic invariant is relatively difficult to determine. Thus, in this work we produce phase portraits by plotting level curves of the adiabatic invariant $S$ under a given Hamiltonian $\cal H$ (i.e., the second definition). As mentioned in Sect. \ref{Sect4}, the Hamiltonian is characterised by the critical inclination $i_{2,c}$. 

\begin{figure*}
\centering
\includegraphics[width=\columnwidth]{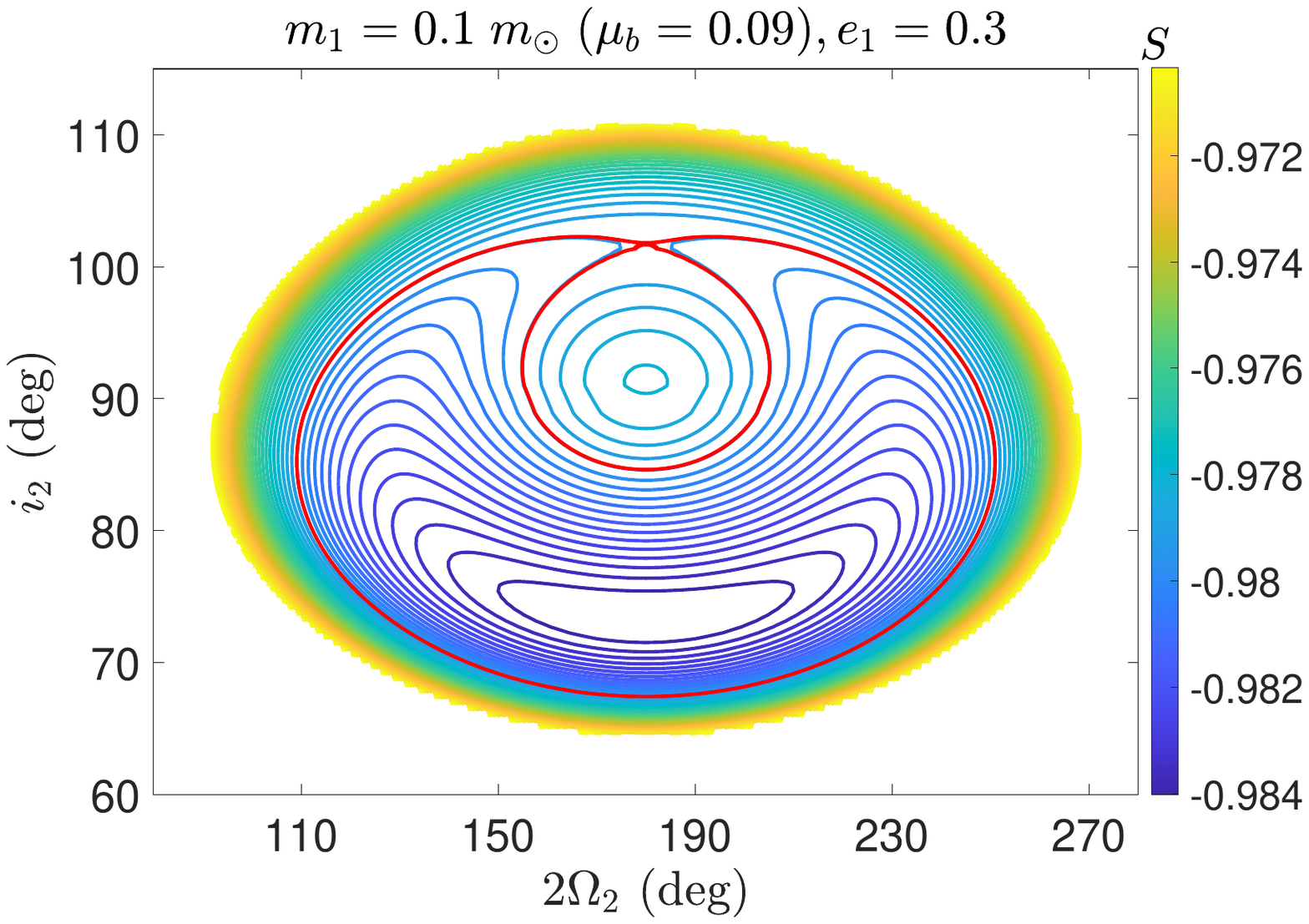}
\includegraphics[width=\columnwidth]{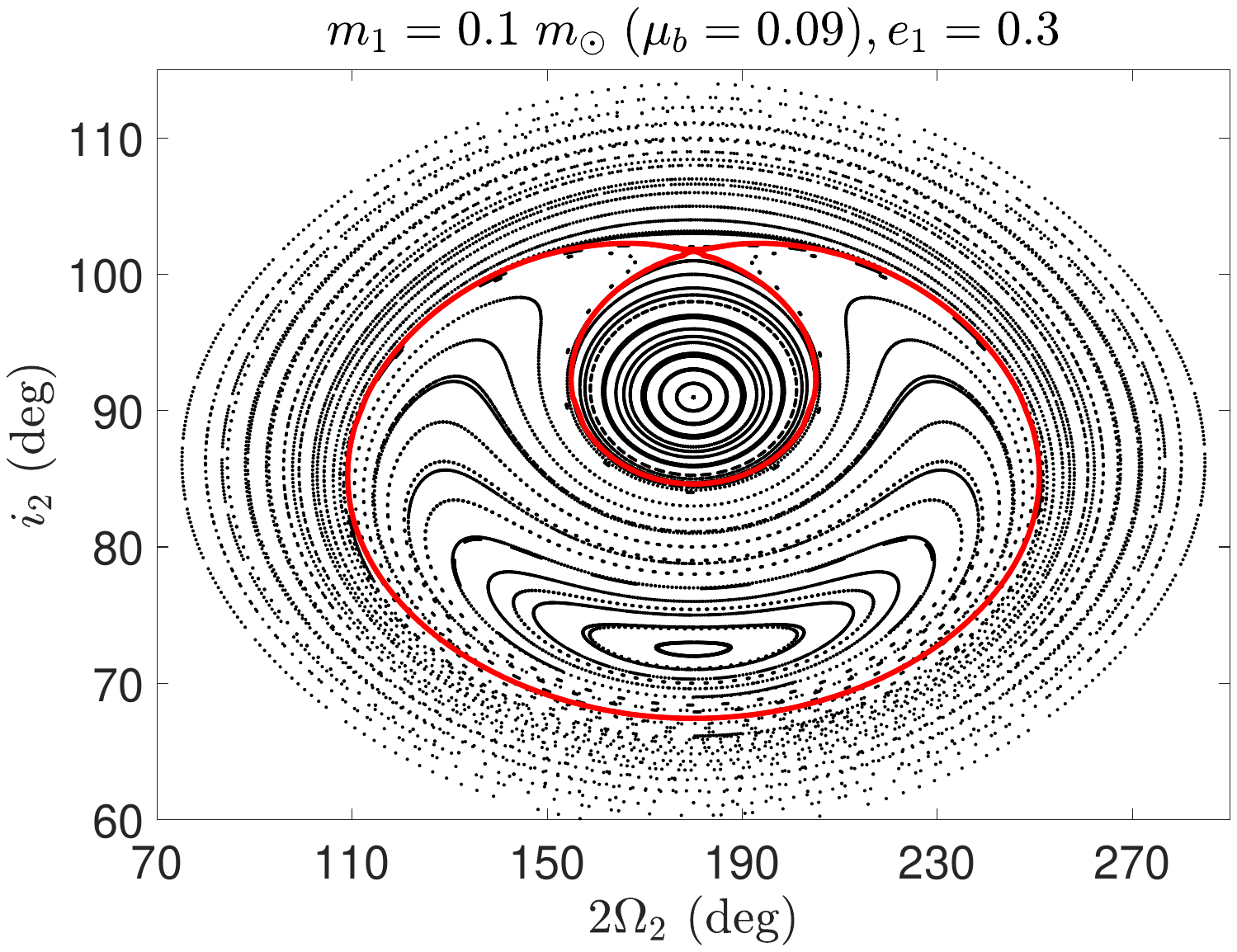}
\caption{Dynamical structures of the 1:1 secondary resonances inside the nodal resonance (left panel), as well as the associated numerical structures arising in Poincaré sections (right panel) for CBPs with system parameters of $m_1 = 0.1 m_{\odot}$, $e_1 = 0.3$ and $i_{2,c} = 61^{\circ}$. The index shown in the colour bar represents the magnitude of the motion integral $S$. In both panels, red line stands for the dynamical separatrix arising in phase portrait. There is an excellent agreement between the analytical and numerical structures.}
\label{Fig13}
\end{figure*}

\begin{figure*}
\centering
\includegraphics[width=\columnwidth]{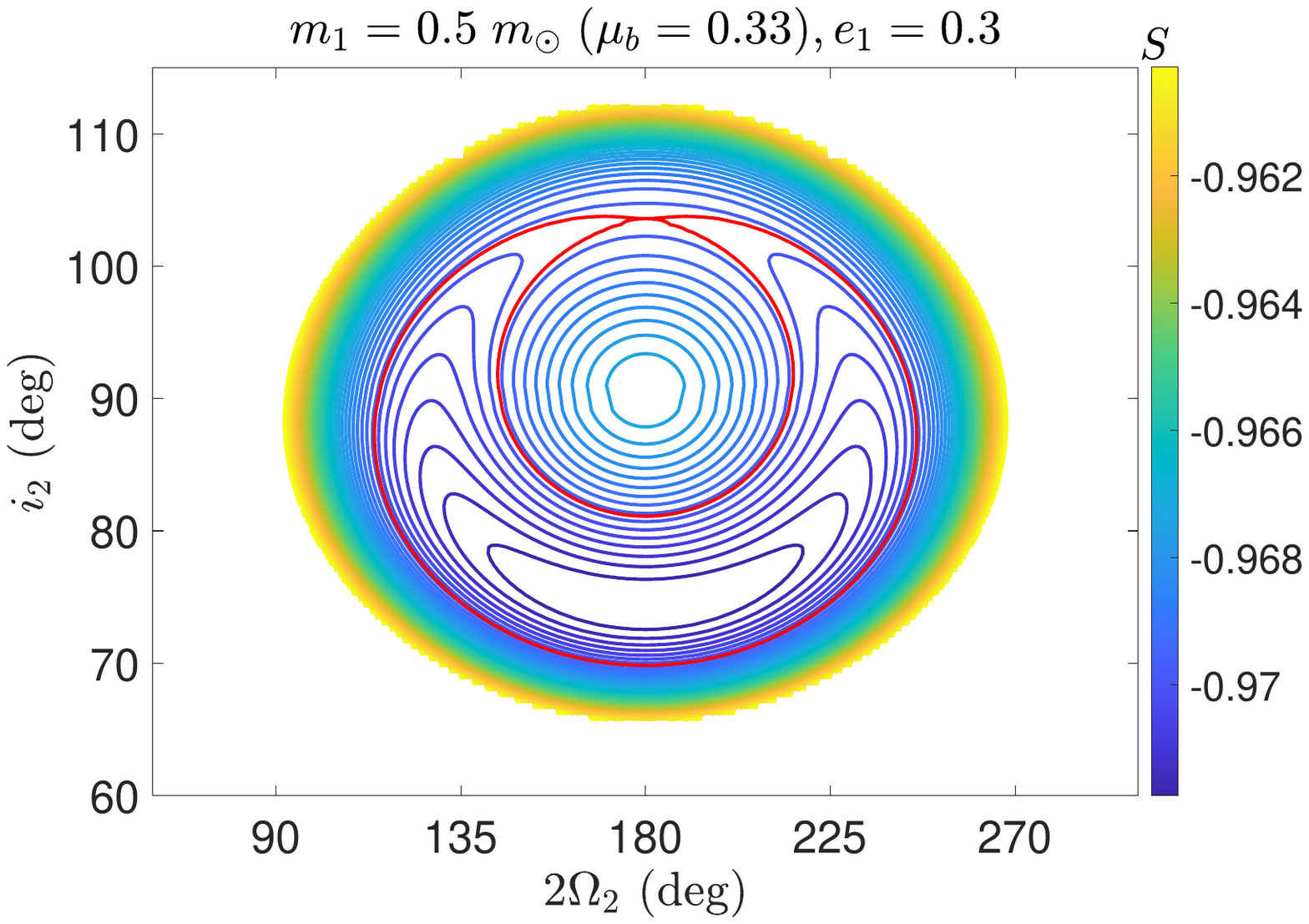}
\includegraphics[width=\columnwidth]{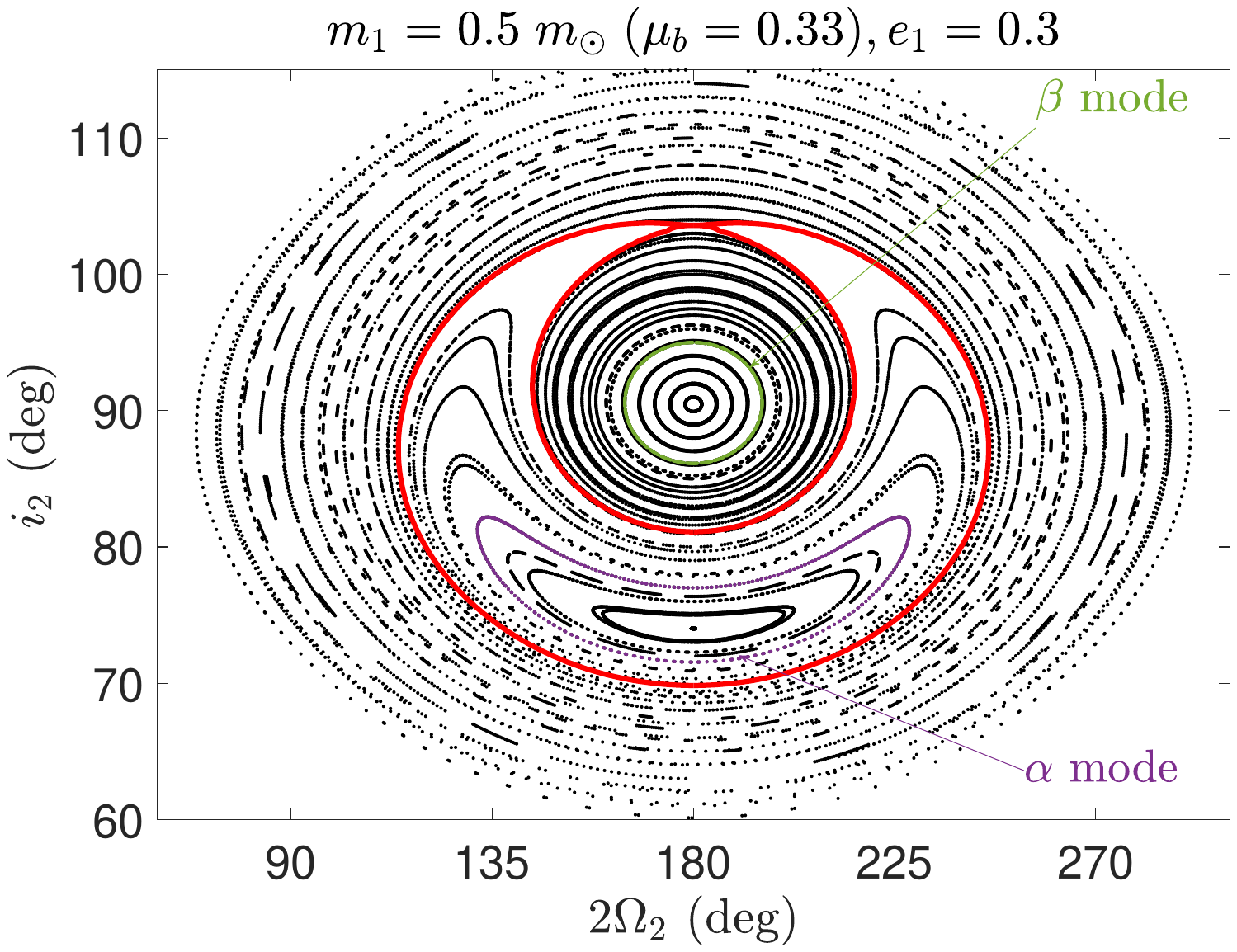}
\caption{Similar to Fig. \ref{Fig13} but for the parameters of $m_1 = 0.5 m_{\odot}$, $e_1 = 0.3$ and $i_{2,c} = 63^{\circ}$. In the right panel, the bottom island of libration  is denoted as $\alpha$ mode and the upper island of libration is denoted by $\beta$ mode.}
\label{Fig14}
\end{figure*}

Fig. \ref{Fig13} presents phase-space structures of the 1:1 secondary resonances under the configurations with $m_1 = 0.1 m_{\odot}$, $e_1 = 0.3$ and $i_{2,c} = 61^{\circ}$, and Fig. \ref{Fig14} corresponds to the case of $m_1 = 0.5 m_{\odot}$, $e_1 = 0.3$ and $i_{2,c} = 63^{\circ}$. Red lines stand for the dynamical separatrices, which play a role in dividing regions of libration and circulation associated with secondary resonance. For comparison, the associated Poincar\'e sections are provided in right panels. It is observed that (a) there is a perfect agreement between the analytical structures arising in phase portraits and numerical structures arising in Poincar\'e sections, (b) there are two islands of libration, which are separated by the dynamical separatrix, and (c) the 1:1 secondary resonance is indeed the culprit that leads to symmetry breaking of dynamical structures.

\begin{figure*}
\centering
\includegraphics[width=\columnwidth]{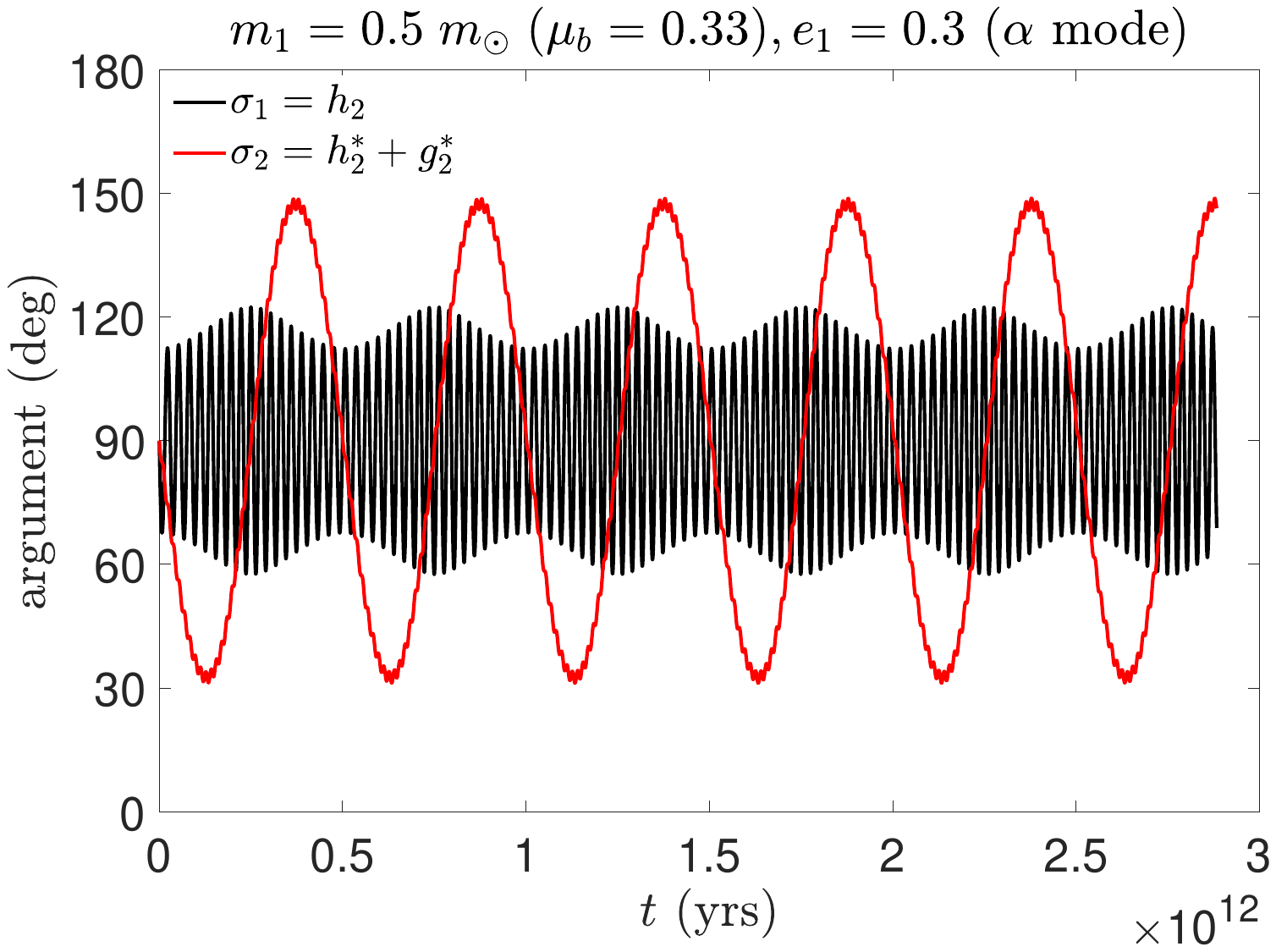}
\includegraphics[width=\columnwidth]{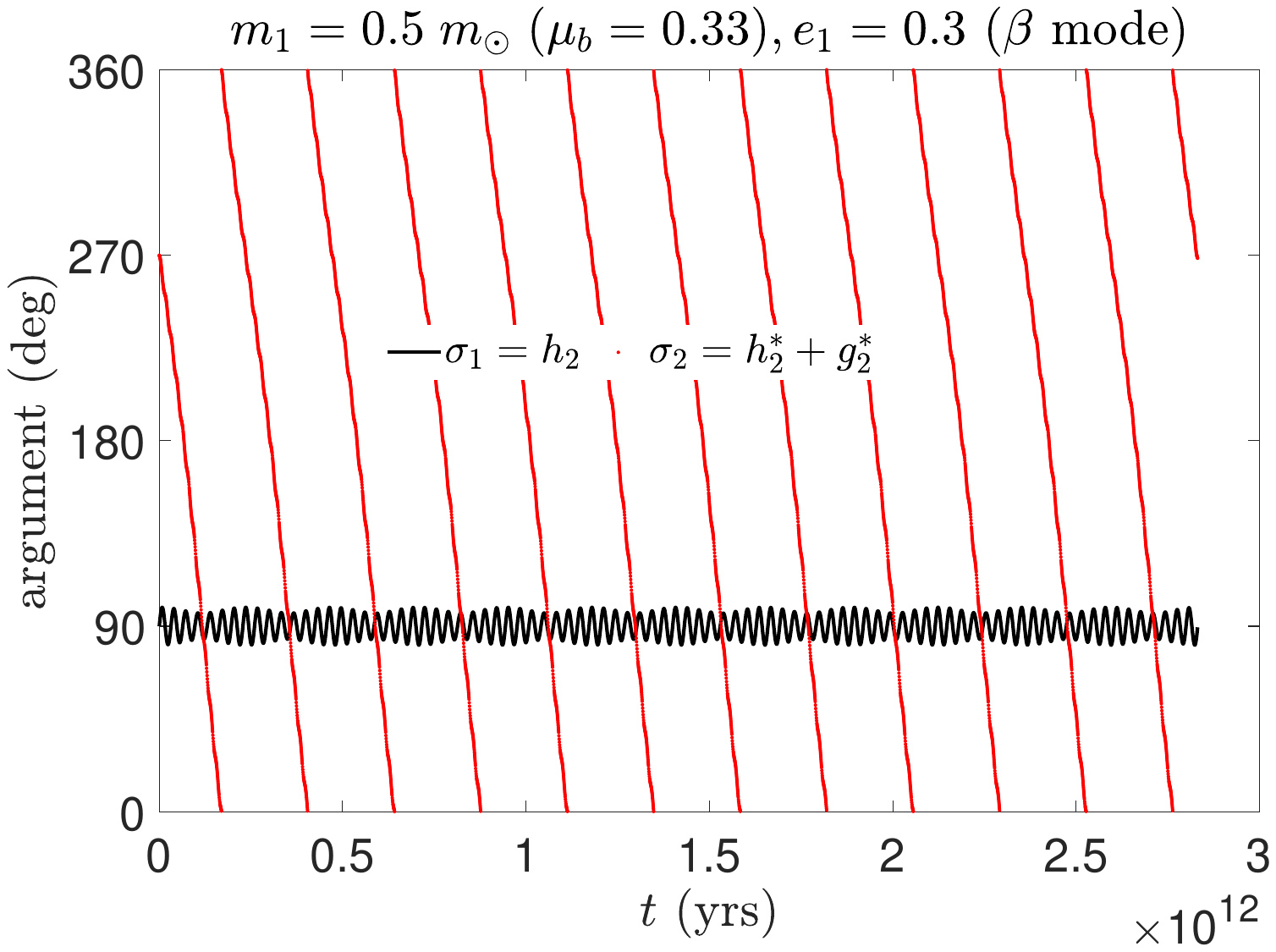}
\caption{Time histories of the arguments associated with the nodal resonance $\sigma = h_2$ (black lines) and the secondary resonance $\sigma = h_2^* + g_2^*$ (red lines) for the trajectories marked in the right panel of Fig. \ref{Fig14}. The left panel is for the $\alpha$ model and the right panel is for the $\beta$ mode. Trajectories in both modes are inside the nodal resonance (i.e., $\Omega_2$ of both trajectories is of libration). The initial condition of the trajectory in $\alpha$ mode is $e_{2,0} = 0.363$, $i_{2,0} = 77^{\circ}$ and $\Omega_{2,0}=\omega_{2,0}=\pi/2$, and the initial condition of the trajectory in $\beta$ mode is $e_{2,0} = 0.278$, $i_{2,0} = 95^{\circ}$ and $\Omega_{2,0}=\omega_{2,0}=\pi/2$.}
\label{Fig15}
\end{figure*}

In the Poincar\'e section, there are two islands of libration associated with 1:1 secondary resonance. For convenience, we denote the bottom island of libration as $\alpha$ mode and the upper island of libration as $\beta$ mode. Please see the right panel of Fig. \ref{Fig14}. In $\alpha$ and $\beta$ modes, we take a representative trajectory in each mode as an example and show its time evolution of resonant arguments including $\sigma = h_2$ (for nodal resonance) and $\sigma = h_2^* + g_2^*$ (for secondary resonance) in Fig. \ref{Fig15}. It is observed that (a) for the trajectory in $\alpha$ mode both the arguments $\sigma = h_2$ and $\sigma = h_2^* + g_2^*$ are of libration and (b) for the trajectory in $\beta$ mode the argument of $\sigma = h_2$ is of libration while the argument of $\sigma = h_2^* + g_2^*$ is of circulation. Thus, secondary 1:1 resonance happens inside the $\alpha$-mode island of libration. 

It should be mentioned that although the simulations are performed for the secondary 1:1 resonance, which is the strongest among secondary resonances, the perturbation method adopted in this work is applicable for other resonances shown in Fig. \ref{Fig10}.

\section{Conclusions}
\label{Sect8}

In this work, secular dynamics of CBPs is systematically investigated under the hierarchical three-body configuration, where the semimajor axis ratio $\alpha =a_1/a_2$ is a small parameter. The phase-averaged Hamiltonian truncated up to hexadecapolar order in $\alpha$ is derived and it is in agreement with the Hamiltonian form given in \citet{naoz2017eccentric}, \citet{vinson2018secular} and \citet{de2019inverse}. Numerical simulations show that the secular approximation at the hexadecapolar level could catch the long-term behaviours of orbit elements.

Initially, different versions of stability criteria are briefly introduced and compared for P-type configurations. The stability criteria given by \citet{dvorak1989stability}, \citet{holman1999long} and \citet{quarles2018stability} are suitable for low-inclination, low-eccentricity planetary orbits, the one presented by \citet{adelbert2023stability} is suitable for low-inclination planetary orbits with arbitrary eccentricities, and the ones developed in \cite{mardling2001tidal} and \citet{georgakarakos2024empirical} can apply to inclined and eccentric planetary orbits. With a certain set of parameter, there exists different degrees of difference among the stability curves. In practice, we adopt the stability condition given by \cite{mardling2001tidal} as a criterion to stop numerical integration.

Then, global structures of CBPs in the phase space are revealed by analysing dynamical maps, with the second-derivative-based indicator $||\Delta D||$ as index (see Fig. \ref{Fig4}). It is found that structures arising in dynamical maps are weakly related to the mass parameter $\mu_b$, semimajor axis ratio $\alpha=a_1/a_2$, planetary eccentricity $e_2$. In the $(i_2,e_1)$ space, there is a main V-shape structure and inside or outside the V-shape region there are complex minute structures. Dynamical maps could provide us a powerful tool to predict possible regions where the inner binary could host planets from the viewpoint of long-term stability. In the case of unequal-mass binary, significant symmetry breaking of dynamical structures is observed inside the polar region. However, the symmetry breaking is relatively weak for the case of equal-mass binary, because in this case the octupole-order contribution disappears. Furthermore, Poincar\'e surfaces of sections indicate that these dynamical structures are caused by nodal resonance, high-order and secondary resonances.

To understand the structures arising in dynamical maps and Poincar\'e sections, the dynamics under the quadrupole-order Hamiltonian is systematically explored. The quadrupole-level Hamiltonian determines an integrable model, where the phase-space structures can be revealed by plotting phase portraits. For convenience, we refer the quadrupole-order resonance as the nodal resonance with argument of $\sigma = 2\Omega_2$. It is found that the island of nodal libration increases with binary eccentricity $e_1$. Orbit classifications are made in the $({\cal H}, e_2)$ space, where ${\cal H}$ and $e_2$ are motion integral for the quadrupole-order Hamiltonian model. Expression of resonant width, denoted by $\Delta i_2$, is explicitly provided and it is only related to the binary eccentricity $e_1$ (an increasing function of $e_1$). 

At the quadrupole-level approximation, the dynamics is distinctively different for inner and outer test particles. In particular, the quadrupole-order dynamics is determined by the binary eccentricity for the case of outer test particles, while it is independent on binary's eccentricity for the case of inner test particles (this observation is evident from their respective expression of quadrupole-level Hamiltonian). For the case of inner test particles, the quadrupole-order resonance is the conventional ZLK resonance with argument of $\sigma=2\omega_1$ librating around $\pi$, the motion integral is the $z$-component of angular momentum $H={\sqrt{1-e_1^2}}\cos{i_1}$, and boundaries of inclination where the quadrupole-order resonance happens are $39.2^{\circ}$ and $140.8^{\circ}$ \citep{kozai1962secular,lidov1962evolution}. When ZLK resonance takes place, it is possible to stimulate large oscillations in eccentricity and inclination, conserving the motion integral $H$. For the case of outer test particles, the quadrupole-order resonance is the nodal resonance with argument of $\sigma=2\Omega_2$ librating around $\pi$, the motion integral is planetary eccentricity $e_2$, and the boundaries of inclination where the quadrupole-order resonance takes place are dependent on the binary's eccentricity $e_1$ (see Fig. \ref{Fig8}). The dynamical response is to excite planetary inclination by conserving the motion integral $e_2$ (planetary eccentricity remains unchanged), which is distinctively different from the case of inner test particles. 

Regarding the integrable dynamical model determined by the quadrupole-order Hamiltonian, a canonical transformation is performed to introduce the set of action-angle variables. The quadrupole-level Hamiltonian can be expressed in a normal form, which determines the fundamental frequencies $(\dot g_2^*, \dot h_2^*)$. Consequently, a web of resonance, satisfying the relation $k_1 \dot h_2^* + k_2 \dot g_2^* = 0$, can be produced (see Fig. \ref{Fig10} for those dominant resonances). The curves of quadrupole-order resonance (nodal resonance), high-order resonances and secondary resonances are very important because they constitute the backbone dominating varieties of dynamical structures in phase space. It is observed that the CBPs detected so far (see Table \ref{Table1}) are located in almost co-planar configurations and thus they are not influenced by the resonances. Resonance curves determined under the quadrupole-order Hamiltonian can well explain numerical structures (see Figs. \ref{Fig11} and \ref{Fig12}). It is concluded that (a) the main V-shape structure is dominated by the quadrupole-order resonance (nodal resonance), (b) the dynamical structures inside the nodal libration region are governed by secondary resonances (as well as inverse ZLK resonance), (c) those structures outside the nodal libration region are sculpted by high-order resonances and (d) the secondary 1:1 resonance is the culprit that leads to symmetry breaking of dynamical structures inside polar region.

At last, perturbative treatments developed by \citet{wisdom1985perturbative} and \citet{henrard1986perturbation} are adopted to study secondary resonances. In particular, a set of action-angle variables is introduced by means of canonical transformation. When the circumbinary planet is located inside a certain secondary resonance, the hexadecapolar-order Hamiltonian determines a separable dynamical model, where the frequencies are hierarchical in magnitude. According to Wisdom's theory \citep{wisdom1985perturbative}, an adiabatic invariant (denoted by $S$) measuring the area bounded by the phase trajectory of the fast degree of freedom can be introduced. The conservation of Hamiltonian $\cal H$ and adiabatic invariant $S$ makes the current two-degree-of-freedom dynamical model be integrable. The dynamical structures can be explored by analysing phase portraits, corresponding to level curves of adiabatic invariant $S$ with a given Hamiltonian $\cal H$. The secondary 1:1 resonance, which is the strongest one among secondary resonances, is taken as an example. Results show that (a) there is an excellent agreement between analytical structures arising in phase portraits and numerical structures arising in Poincar\'e sections, (b) the secondary 1:1 resonance corresponds to the bifurcation of the nodal resonance, and (c) the secondary 1:1 resonance is indeed the culprit causing symmetry breaking (with respect to $i_2 = 90^{\circ}$) of dynamical structures.

The results of this work provide a global picture about the dynamics of CBPs in a large range of parameters. In addition, our solutions in this work have potential applications to asteroids in Kuiper belt, debris discs around binary stars, circumbinary planets, and stars around binary supermassive black holes.

\section*{Acknowledgements}
We wish to thank an anonymous referee for providing us helpful comments that improved the quality of the manuscript. This work is financially supported by the National Natural Science Foundation of China (Nos. 12073011, 12073019 and 12233003) and the National Key R\&D Program of China (No. 2019YFA0706601). 

\section*{Data availability}
The analysis and codes are available upon request.

\bibliographystyle{mn2e}
\bibliography{mybib}

\begin{thebibliography}{}

\bibitem[\protect\citeauthoryear{Adelbert, Penzlin, Sch{\"a}fer, Kley, Quarles
  \& Sfair}{Adelbert et~al.}{2023}]{adelbert2023stability}
Adelbert S.,  Penzlin A.~B.,  Sch{\"a}fer C.~M.,  Kley W.,  Quarles B.,
  Sfair R.,  2023, A\&A, 680, A29

\bibitem[\protect\citeauthoryear{Borkovits, Hajdu, Sztakovics, Rappaport,
  Levine, B{\'\i}r{\'o} \& Klagyivik}{Borkovits
  et~al.}{2016}]{borkovits2016comprehensive}
Borkovits T.,  Hajdu T.,  Sztakovics J.,  Rappaport S.,  Levine A.,
  B{\'\i}r{\'o} I.~B.,    Klagyivik P.,  2016, MNRAS, 455, 4136

\bibitem[\protect\citeauthoryear{Brinch, J{\o}rgensen, Hogerheijde, Nelson \&
  Gressel}{Brinch et~al.}{2016}]{brinch2016misaligned}
Brinch C.,  J{\o}rgensen J.~K.,  Hogerheijde M.~R.,  Nelson R.~P.,    Gressel
  O.,  2016, ApJL, 830, L16

\bibitem[\protect\citeauthoryear{Cazzoletti, Ricci, Birnstiel \&
  Lodato}{Cazzoletti et~al.}{2017}]{cazzoletti2017testing}
Cazzoletti P.,  Ricci L.,  Birnstiel T.,    Lodato G.,  2017, A\&A, 599, A102

\bibitem[\protect\citeauthoryear{Chen, Franchini, Lubow \& Martin}{Chen
  et~al.}{2019}]{chen2019orbital}
Chen C.,  Franchini A.,  Lubow S.~H.,    Martin R.~G.,  2019, MNRAS, 490, 5634

\bibitem[\protect\citeauthoryear{Chen, Lubow \& Martin}{Chen
  et~al.}{2020}]{chen2020polar}
Chen C.,  Lubow S.~H.,    Martin R.~G.,  2020, MNRAS, 494, 4645

\bibitem[\protect\citeauthoryear{Childs \& Martin}{Childs \&
  Martin}{2021}]{childs2021formation}
Childs A.~C.,  Martin R.~G.,  2021, ApJL, 920, L8

\bibitem[\protect\citeauthoryear{Childs \& Martin}{Childs \&
  Martin}{2022}]{childs2022misalignment}
Childs A.~C.,  Martin R.~G.,  2022, ApJL, 927, L7

\bibitem[\protect\citeauthoryear{Cuello \& Giuppone}{Cuello \&
  Giuppone}{2019}]{cuello2019planet}
Cuello N.,  Giuppone C.~A.,  2019, A\&A, 628, A119

\bibitem[\protect\citeauthoryear{Czekala, Chiang, Andrews, Jensen, Torres,
  Wilner, Stassun \& Macintosh}{Czekala et~al.}{2019}]{czekala2019degree}
Czekala I.,  Chiang E.,  Andrews S.~M.,  Jensen E.~L.,  Torres G.,  Wilner
  D.~J.,  Stassun K.~G.,    Macintosh B.,  2019, ApJ, 883, 22

\bibitem[\protect\citeauthoryear{Daquin \& Charalambous}{Daquin \&
  Charalambous}{2023}]{Daquin2023Detection}
Daquin J.,  Charalambous C.,  2023, Celest. Mech. Dyn. Astron., 135, 31

\bibitem[\protect\citeauthoryear{Daquin, P{\'e}denon-Orlanducci, Agaoglou,
  Garc{\'\i}a-S{\'a}nchez \& Mancho}{Daquin et~al.}{2022}]{daquin2022global}
Daquin J.,  P{\'e}denon-Orlanducci R.,  Agaoglou M.,  Garc{\'\i}a-S{\'a}nchez
  G.,    Mancho A.~M.,  2022, Physica D: Nonlinear Phenomena, 442, 133520

\bibitem[\protect\citeauthoryear{de Elia, Zanardi, Dugaro \& Naoz}{de~Elia
  et~al.}{2019}]{de2019inverse}
de Elia G.~C.,  Zanardi M.,  Dugaro A.,    Naoz S.,  2019, A\&A, 627, A17

\bibitem[\protect\citeauthoryear{Doolin \& Blundell}{Doolin \&
  Blundell}{2011}]{doolin2011dynamics}
Doolin S.,  Blundell K.~M.,  2011, MNRASy, 418, 2656

\bibitem[\protect\citeauthoryear{Dvorak, Froeschl{\'e} \& Froeschle}{Dvorak
  et~al.}{1989}]{dvorak1989stability}
Dvorak R.,  Froeschl{\'e} C.,    Froeschle C.,  1989, A\&A, 226, 335

\bibitem[\protect\citeauthoryear{Farago \& Laskar}{Farago \&
  Laskar}{2010}]{farago2010high}
Farago F.,  Laskar J.,  2010, MNRAS, 401, 1189

\bibitem[\protect\citeauthoryear{Ford, Kozinsky \& Rasio}{Ford
  et~al.}{2000}]{ford2000secular}
Ford E.~B.,  Kozinsky B.,    Rasio F.~A.,  2000, ApJ, 535, 385

\bibitem[\protect\citeauthoryear{Gallardo, Hugo \& Pais}{Gallardo
  et~al.}{2012}]{gallardo2012survey}
Gallardo T.,  Hugo G.,    Pais P.,  2012, Icarus, 220, 392

\bibitem[\protect\citeauthoryear{Georgakarakos, Eggl, Ali-Dip \&
  Dobbs-Dixon}{Georgakarakos et~al.}{2024}]{georgakarakos2024empirical}
Georgakarakos N.,  Eggl S.,  Ali-Dip M.,    Dobbs-Dixon I.,  2024, arXiv
  preprint arXiv:2404.13746

\bibitem[\protect\citeauthoryear{Getley, Carter, King \& O’Toole}{Getley
  et~al.}{2017}]{getley2017evidence}
Getley A.,  Carter B.,  King R.,    O’Toole S.,  2017, MNRAS, 468, 2932

\bibitem[\protect\citeauthoryear{Guzzo, Lega \& Froeschl{\'e}}{Guzzo
  et~al.}{2002}]{guzzo2002numerical}
Guzzo M.,  Lega E.,    Froeschl{\'e} C.,  2002, Physica D: Nonlinear Phenomena,
  163, 1

\bibitem[\protect\citeauthoryear{Harrington}{Harrington}{1968}]{harrington1968dynamical}
Harrington R.~S.,  1968, AJ, 73, 190

\bibitem[\protect\citeauthoryear{Harrington}{Harrington}{1969}]{harrington1969stellar}
Harrington R.~S.,  1969, Celest. Mech., 1, 200

\bibitem[\protect\citeauthoryear{Henrard}{Henrard}{1990}]{henrard1990semi}
Henrard J.,  1990, Celest. Mech. Dyn. Astron., 49, 43

\bibitem[\protect\citeauthoryear{Henrard \& Lemaitre}{Henrard \&
  Lemaitre}{1986}]{henrard1986perturbation}
Henrard J.,  Lemaitre A.,  1986, Celest. Mech., 39, 213

\bibitem[\protect\citeauthoryear{Holman \& Wiegert}{Holman \&
  Wiegert}{1999}]{holman1999long}
Holman M.~J.,  Wiegert P.~A.,  1999, AJ, 117, 621

\bibitem[\protect\citeauthoryear{Huang \& Lei}{Huang \&
  Lei}{2024}]{huang2024dynamical}
Huang X.,  Lei H.,  2024, AJ, 167, 234

\bibitem[\protect\citeauthoryear{Katz, Dong \& Malhotra}{Katz
  et~al.}{2011}]{katz2011long}
Katz B.,  Dong S.,    Malhotra R.,  2011, Physical Review Letters, 107, 181101

\bibitem[\protect\citeauthoryear{Kennedy, Wyatt, Sibthorpe, Duch{\^e}ne, Kalas,
  Matthews, Greaves, Su \& Fitzgerald}{Kennedy et~al.}{2012}]{kennedy201299}
Kennedy G.,  Wyatt M.,  Sibthorpe B.,  Duch{\^e}ne G.,  Kalas P.,  Matthews B.,
   Greaves J.,  Su K.,    Fitzgerald M.,  2012, MNRAS, 421, 2264

\bibitem[\protect\citeauthoryear{Kennedy, Matr{\`a}, Facchini, Milli,
  Pani{\'c}, Price, Wilner, Wyatt \& Yelverton}{Kennedy
  et~al.}{2019}]{kennedy2019circumbinary}
Kennedy G.~M.,  Matr{\`a} L.,  Facchini S.,  Milli J.,  Pani{\'c} O.,  Price
  D.,  Wilner D.~J.,  Wyatt M.~C.,    Yelverton B.~M.,  2019, Nature Astronomy,
  3, 230

\bibitem[\protect\citeauthoryear{Kostov, Orosz, Feinstein, Welsh, Cukier,
  Haghighipour, Quarles, Martin, Montet, Torres et~al.,}{Kostov
  et~al.}{2020}]{kostov2020toi}
Kostov V.~B.,  Orosz J.~A.,  Feinstein A.~D.,  Welsh W.~F.,  Cukier W.,
  Haghighipour N.,  Quarles B.,  Martin D.~V.,  Montet B.~T.,  Torres G.,
  et~al., 2020, AJ, 159, 253

\bibitem[\protect\citeauthoryear{Kostov, Orosz, Welsh, Doyle, Fabrycky,
  Haghighipour, Quarles, Short, Cochran, Endl et~al.,}{Kostov
  et~al.}{2016}]{kostov2016kepler}
Kostov V.~B.,  Orosz J.~A.,  Welsh W.~F.,  Doyle L.~R.,  Fabrycky D.~C.,
  Haghighipour N.,  Quarles B.,  Short D.~R.,  Cochran W.~D.,  Endl M.,
  et~al., 2016, ApJ, 827, 86

\bibitem[\protect\citeauthoryear{Kostov, Powell, Orosz, Welsh, Cochran,
  Collins, Endl, Hellier, Latham, MacQueen et~al.,}{Kostov
  et~al.}{2021}]{kostov2021tic}
Kostov V.~B.,  Powell B.~P.,  Orosz J.~A.,  Welsh W.~F.,  Cochran W.,  Collins
  K.~A.,  Endl M.,  Hellier C.,  Latham D.~W.,  MacQueen P.,    et~al., 2021,
  AJ, 162, 234

\bibitem[\protect\citeauthoryear{Kozai}{Kozai}{1962}]{kozai1962secular}
Kozai Y.,  1962, AJ, 67, 591

\bibitem[\protect\citeauthoryear{Lei}{Lei}{2020}]{lei2020dynamical}
Lei H.,  2020, Astrodyn, 4, 57

\bibitem[\protect\citeauthoryear{Lei}{Lei}{2021}]{lei2021structures}
Lei H.,  2021, Celest. Mech. Dyn. Astron., 133, 40

\bibitem[\protect\citeauthoryear{Lei}{Lei}{2022}]{lei2022systematic}
Lei H.,  2022, AJ, 163, 214

\bibitem[\protect\citeauthoryear{Lei}{Lei}{2024}]{lei2024dynamical}
Lei H.,  2024, AJ, 167, 121

\bibitem[\protect\citeauthoryear{Lei \& Huang}{Lei \&
  Huang}{2022}]{lei2022quadrupole}
Lei H.,  Huang X.,  2022, MNRAS, 515, 1086

\bibitem[\protect\citeauthoryear{Lei, Li, Huang \& Li}{Lei
  et~al.}{2022}]{lei2022zeipel}
Lei H.,  Li J.,  Huang X.,    Li M.,  2022, AJ, 164, 74

\bibitem[\protect\citeauthoryear{Lei, Ortore \& Circi}{Lei
  et~al.}{2022}]{lei2022secular}
Lei H.,  Ortore E.,    Circi C.,  2022, Astrodyn, 6, 357

\bibitem[\protect\citeauthoryear{Lepp, Martin \& Childs}{Lepp
  et~al.}{2022}]{lepp2022radial}
Lepp S.,  Martin R.~G.,    Childs A.~C.,  2022, ApJL, 929, L5

\bibitem[\protect\citeauthoryear{Leung \& Lee}{Leung \&
  Lee}{2013}]{leung2013analytic}
Leung G.~C.,  Lee M.~H.,  2013, The Astrophysical Journal, 763, 107

\bibitem[\protect\citeauthoryear{Li, Zhou \& Zhang}{Li
  et~al.}{2014}]{li2014analytical}
Li D.,  Zhou J.,    Zhang H.,  2014, MNRAS, 437, 3832

\bibitem[\protect\citeauthoryear{Lidov}{Lidov}{1962}]{lidov1962evolution}
Lidov M.,  1962, P\&SS, 9, 719

\bibitem[\protect\citeauthoryear{Lithwick \& Naoz}{Lithwick \&
  Naoz}{2011}]{lithwick2011eccentric}
Lithwick Y.,  Naoz S.,  2011, ApJ, 742, 94

\bibitem[\protect\citeauthoryear{Liu, Mu{\~n}oz \& Lai}{Liu
  et~al.}{2015}]{liu2015suppression}
Liu B.,  Mu{\~n}oz D.~J.,    Lai D.,  2015, MNRAS, 447, 747

\bibitem[\protect\citeauthoryear{Mardling \& Aarseth}{Mardling \&
  Aarseth}{2001}]{mardling2001tidal}
Mardling R.~A.,  Aarseth S.~J.,  2001, MNRAS, 321, 398

\bibitem[\protect\citeauthoryear{Martin}{Martin}{2018}]{martin2018populations}
Martin D.~V.,  2018, arXiv preprint arXiv:1802.08693

\bibitem[\protect\citeauthoryear{Martin \& Triaud}{Martin \&
  Triaud}{2014}]{martin2014planets}
Martin D.~V.,  Triaud A.~H.,  2014, A\&A, 570, A91

\bibitem[\protect\citeauthoryear{Martin \& Lubow}{Martin \&
  Lubow}{2017}]{martin2017polar}
Martin R.~G.,  Lubow S.~H.,  2017, ApJL, 835, L28

\bibitem[\protect\citeauthoryear{Martin \& Lubow}{Martin \&
  Lubow}{2019}]{martin2019polar}
Martin R.~G.,  Lubow S.~H.,  2019, MNRAS, 490, 1332

\bibitem[\protect\citeauthoryear{Martin, Lubow, Vallet, Anugu \& Gies}{Martin
  et~al.}{2023}]{martin2023ac}
Martin R.~G.,  Lubow S.~H.,  Vallet D.,  Anugu N.,    Gies D.~R.,  2023, ApJL,
  957, L28

\bibitem[\protect\citeauthoryear{Morbidelli}{Morbidelli}{2002}]{morbidelli2002modern}
Morbidelli A.,  2002, Modern celestial mechanics: aspects of solar system
  dynamics.
Taylor \& Francis, London and New York

\bibitem[\protect\citeauthoryear{Moriwaki \& Nakagawa}{Moriwaki \&
  Nakagawa}{2004}]{moriwaki2004planetesimal}
Moriwaki K.,  Nakagawa Y.,  2004, The Astrophysical Journal, 609, 1065

\bibitem[\protect\citeauthoryear{Murray \& Dermott}{Murray \&
  Dermott}{1999}]{murray1999solar}
Murray C.~D.,  Dermott S.~F.,  1999, Solar system dynamics.
Cambridge university press

\bibitem[\protect\citeauthoryear{Naoz}{Naoz}{2016}]{naoz2016eccentric}
Naoz S.,  2016, ARA\&A, 54, 441

\bibitem[\protect\citeauthoryear{Naoz, Farr, Lithwick, Rasio \&
  Teyssandier}{Naoz et~al.}{2013}]{naoz2013secular}
Naoz S.,  Farr W.~M.,  Lithwick Y.,  Rasio F.~A.,    Teyssandier J.,  2013,
  MNRAS, 431, 2155

\bibitem[\protect\citeauthoryear{Naoz, Li, Zanardi, De~El{\'\i}a \&
  Di~Sisto}{Naoz et~al.}{2017}]{naoz2017eccentric}
Naoz S.,  Li G.,  Zanardi M.,  De~El{\'\i}a G.~C.,    Di~Sisto R.~P.,  2017,
  AJ, 154, 18

\bibitem[\protect\citeauthoryear{Orosz, Welsh, Carter, Fabrycky, Cochran, Endl,
  Ford, Haghighipour, MacQueen, Mazeh et~al.,}{Orosz
  et~al.}{2012}]{orosz2012kepler}
Orosz J.~A.,  Welsh W.~F.,  Carter J.~A.,  Fabrycky D.~C.,  Cochran W.~D.,
  Endl M.,  Ford E.~B.,  Haghighipour N.,  MacQueen P.~J.,  Mazeh T.,
  et~al., 2012, Science, 337, 1511

\bibitem[\protect\citeauthoryear{Orosz, Welsh, Haghighipour, Quarles, Short,
  Mills, Satyal, Torres, Agol, Fabrycky et~al.,}{Orosz
  et~al.}{2019}]{orosz2019discovery}
Orosz J.~A.,  Welsh W.~F.,  Haghighipour N.,  Quarles B.,  Short D.~R.,  Mills
  S.~M.,  Satyal S.,  Torres G.,  Agol E.,  Fabrycky D.~C.,    et~al., 2019,
  AJ, 157, 174

\bibitem[\protect\citeauthoryear{Quarles, Satyal, Kostov, Kaib \&
  Haghighipour}{Quarles et~al.}{2018}]{quarles2018stability}
Quarles B.,  Satyal S.,  Kostov V.,  Kaib N.,    Haghighipour N.,  2018, ApJ,
  856, 150

\bibitem[\protect\citeauthoryear{Saillenfest}{Saillenfest}{2020}]{saillenfest2020long}
Saillenfest M.,  2020, Celest. Mech. Dyn. Astron., 132, 1

\bibitem[\protect\citeauthoryear{Saillenfest, Fouchard, Tommei \&
  Valsecchi}{Saillenfest et~al.}{2016}]{saillenfest2016long}
Saillenfest M.,  Fouchard M.,  Tommei G.,    Valsecchi G.~B.,  2016, Celest.
  Mech. Dyn. Astron., 126, 369

\bibitem[\protect\citeauthoryear{Shevchenko}{Shevchenko}{2016}]{shevchenko2016lidov}
Shevchenko I.~I.,  2016, The Lidov-Kozai effect-applications in exoplanet
  research and dynamical astronomy.
Vol.~441, Springer

\bibitem[\protect\citeauthoryear{Smallwood, Franchini, Chen, Becerril, Lubow,
  Yang \& Martin}{Smallwood et~al.}{2020}]{smallwood2020formation}
Smallwood J.~L.,  Franchini A.,  Chen C.,  Becerril E.,  Lubow S.~H.,  Yang
  C.-C.,    Martin R.~G.,  2020, MNRAS, 494, 487

\bibitem[\protect\citeauthoryear{Socia, Welsh, Orosz, Cochran, Endl, Quarles,
  Short, Torres, Windmiller \& Yenawine}{Socia et~al.}{2020}]{socia2020kepler}
Socia Q.~J.,  Welsh W.~F.,  Orosz J.~A.,  Cochran W.~D.,  Endl M.,  Quarles B.,
   Short D.~R.,  Torres G.,  Windmiller G.,    Yenawine M.,  2020, AJ, 159, 94

\bibitem[\protect\citeauthoryear{Sybilski, Konacki \& Koz{\l}owski}{Sybilski
  et~al.}{2010}]{sybilski2010detecting}
Sybilski P.,  Konacki M.,    Koz{\l}owski S.,  2010, MNRAS, 405, 657

\bibitem[\protect\citeauthoryear{Tokovinin}{Tokovinin}{1997}]{tokovinin1997multiplicity}
Tokovinin A.,  1997, Astronomy Letters, 23, 727

\bibitem[\protect\citeauthoryear{Verhoeff, Min, Pantin, Waters, Tielens, Honda,
  Fujiwara, Bouwman, Van~Boekel, Dougherty et~al.,}{Verhoeff
  et~al.}{2011}]{verhoeff2011complex}
Verhoeff A.,  Min M.,  Pantin E.,  Waters L.,  Tielens A.,  Honda M.,  Fujiwara
  H.,  Bouwman J.,  Van~Boekel R.,  Dougherty S.,    et~al., 2011, A\&A, 528,
  A91

\bibitem[\protect\citeauthoryear{Vinson \& Chiang}{Vinson \&
  Chiang}{2018}]{vinson2018secular}
Vinson B.~R.,  Chiang E.,  2018, MNRAS, 474, 4855

\bibitem[\protect\citeauthoryear{Welsh, Orosz, Carter, Fabrycky, Ford,
  Lissauer, Pr{\v{s}}a, Quinn, Ragozzine, Short et~al.,}{Welsh
  et~al.}{2012}]{welsh2012transiting}
Welsh W.~F.,  Orosz J.~A.,  Carter J.~A.,  Fabrycky D.~C.,  Ford E.~B.,
  Lissauer J.~J.,  Pr{\v{s}}a A.,  Quinn S.~N.,  Ragozzine D.,  Short D.~R.,
  et~al., 2012, Nature, 481, 475

\bibitem[\protect\citeauthoryear{Welsh, Orosz, Short, Cochran, Endl, Brugamyer,
  Haghighipour, Buchhave, Doyle, Fabrycky et~al.,}{Welsh
  et~al.}{2015}]{welsh2015kepler}
Welsh W.~F.,  Orosz J.~A.,  Short D.~R.,  Cochran W.~D.,  Endl M.,  Brugamyer
  E.,  Haghighipour N.,  Buchhave L.~A.,  Doyle L.~R.,  Fabrycky D.~C.,
  et~al., 2015, ApJ, 809, 26

\bibitem[\protect\citeauthoryear{Wisdom}{Wisdom}{1985}]{wisdom1985perturbative}
Wisdom J.,  1985, Icarus, 63, 272

\bibitem[\protect\citeauthoryear{Zhang \& Fabrycky}{Zhang \&
  Fabrycky}{2019}]{zhang2019distinguishing}
Zhang Z.,  Fabrycky D.~C.,  2019, ApJ, 879, 92

\end{thebibliography}


\bsp
\label{lastpage}
\end{document}